\documentclass[10pt,prd,preprintnumbers,floatfix,aps,nofootinbib,showpacs,twocolumn,superscriptaddress,nopacs]{revtex4}

\usepackage{epsfig}
\usepackage{url}
\usepackage{hyperref}

\usepackage[normalem]{ulem} 

\usepackage{latexsym}
\usepackage{epsfig}
\usepackage{amsmath}
\usepackage{amssymb}
\usepackage{wasysym}
\usepackage{graphicx}
\usepackage{verbatim}
\usepackage{enumerate,mdwlist}
\usepackage[titletoc]{appendix}
\usepackage{amsfonts}
\usepackage{tikz} 

\usepackage[normalem]{ulem}

\newcommand{\ba}{\begin{eqnarray}}
\newcommand{\ea}{\end{eqnarray}}
\newcommand{\be}{\begin{equation}}
\newcommand{\ee}{\end{equation}}

\newcommand{\M}{\mathcal{M}}

\definecolor{grey}{rgb}{0.4,0.4,0.4}
\definecolor{dullmagenta}{rgb}{0.4,0,0.4}
\definecolor{darkblue}{rgb}{0,0,0.4}
\definecolor{midblue}{rgb}{0,0,0.5}
\definecolor{midred}{rgb}{0.5,0,0}
\definecolor{orange}{rgb}{1,0.5,0}
\definecolor{lightbrown}{rgb}{0.75,0.5,0.25}
\definecolor{tan}{cmyk}{0.14,0.42,0.56,0}
\definecolor{djunglegreen}{cmyk}{0.99,0,0.52,0}
\definecolor{lightgreen}{rgb}{0,1,0}
\definecolor{olivegreen}{cmyk}{0.64,0,0.95,0.40}
\definecolor{midgreen}{rgb}{0.0,0.675,0.0}
\definecolor{darkgreen}{rgb}{0,0.5,0}

\usepackage{multirow}

\usepackage[all]{xy} 
\usepackage{amsfonts}

\begin{document}

\title{Gravity in the Era of Equality: \\
Towards solutions to the Hubble problem without fine-tuned initial conditions
}

\author{Miguel Zumalac\'arregui}
\email{miguel.zumalacarregui@aei.mpg.de}
\affiliation{Max Planck Institute for Gravitational Physics (Albert Einstein Institute) \\
Am M\"uhlenberg 1, D-14476 Potsdam-Golm, Germany}
\affiliation{Berkeley Center for Cosmological Physics, LBNL and University of California at Berkeley, \\
Berkeley, California 94720, USA}
\affiliation{Institut de Physique Th\' eorique, Universit\'e  Paris Saclay 
CEA, CNRS, 91191 Gif-sur-Yvette, France}

\begin{abstract}
Discrepant measurements of the Universe's expansion rate ($H_0$) may signal physics beyond the standard cosmological model. 
Here I describe \textit{two early modified gravity} mechanisms that reconcile $H_0$ value by increasing the expansion rate in the era of matter-radiation equality. 
These mechanisms, based on viable Horndeski theories, require significantly less fine-tuned initial conditions than early dark energy with oscillating scalar fields.
In \textit{Imperfect Dark Energy at Equality} (IDEE), the initial energy density dilutes slower than radiation but faster than matter, naturally peaking around the era of equality.
The minimal IDEE model, a cubic Galileon, is too constrained by the cosmic microwave background (Planck) and baryon acoustic oscillations (BAO) to relieve the $H_0$ tension.
In \textit{Enhanced Early Gravity} (EEG), the scalar field value modulates the cosmological strength of gravity. 
The minimal EEG  model, an exponentially coupled cubic Galileon, gives a Planck+BAO value $H_0=68.7 \pm 1.5$ (68\% c.l.), reducing the tension with SH0ES from $4.4\sigma$ to $2.6\sigma$.
Additionally, Galileon contributions to cosmic acceleration may reconcile $H_0$ via \textit{Late-Universe Phantom Expansion} (LUPE). Combining LUPE, EEG  and $\Lambda$ reduces the tension between Planck, BAO and SH0ES to $2.5\sigma$.
I will also describe additional tests of coupled Galileons based on local gravity tests, primordial element abundances and gravitational waves.
While further model building is required to fully resolve the $H_0$ problem and satisfy all available observations, these examples show the wealth of possibilities to solve cosmological tensions beyond Einstein's General Relativity.
\end{abstract}

\date{\today}



 
\maketitle

\tableofcontents

\section{Introduction}

Observational cosmology has established a simple and successful standard model of the Universe. $\Lambda$CDM is named after the dominant components: the cosmological constant ($\Lambda$) accelerates the expansion at late times and cold dark matter (CDM) drives the formation of large-scale structure (LSS). In addition, the model includes other matter species known from Earthly experiments (atoms, photons, neutrinos) and assumes the validity of Einstein's general relativity (GR).
This remarkably simple model successfully describes most cosmological observations in terms of a handful of parameters \cite{Aghanim:2018eyx}.
But despite $\Lambda$CDM's success, several datasets interpreted within the standard model are in conflict \cite{Raveri:2015maa}. 

The most significant tension involves the Universe's expansion rate. 
Late-universe measurements of $H_0$ clash with observations of early-universe processes interpreted within $\Lambda$CDM. 
Late probes include distance ladder \cite{Riess:2019cxk} and lensing time delays \cite{Wong:2019kwg,Shajib:2019toy}. They are direct and largely independent of the cosmological model.
Probes based on early-Universe processes (or early probes) rely on Planck's cosmic microwave background (CMB) plus baryon acoustic oscillations (BAO) data \cite{Aghanim:2018eyx}. Early probes are indirect and rely on the predictions of the $\Lambda$CDM model.
Unless the Hubble problem is due to unknown systematics, its significance demands physics beyond the simple $\Lambda$CDM model \cite{Freedman:2017yms,Verde:2019ivm}.

New-physics solutions to the Hubble problem reflect the conflict between the early and late universe.
Late-universe solutions rely on new astrophysical effects \cite{Desmond:2019ygn} or dark energy (DE) beyond $\Lambda$ \cite{Barreira:2013jma,Barreira:2014jha,Renk:2017rzu,DiValentino:2017iww,Sola:2017znb,Poulin:2018zxs,Yang:2018euj,Vattis:2019efj,DiValentino:2019ffd}.
Adjusting $H_0$ for fixed CMB angular scale requires that the density of DE grows in time instead of remaining constant, i.e. a \textit{phantom equation of state}
\begin{equation}
 w_\phi = 
 \frac{\mathcal{P}}{\mathcal{E}} 
 < -1\,,
\end{equation}
where $\mathcal{E},\mathcal{P}$ are the energy and pressure density of DE, respectively. 
While disfavoured by combining BAO and type Ia supernovae (SNe) \cite{Bernal:2016gxb,Poulin:2018zxs,Raveri:2019mxg,Mortsell:2018mfj}, other analyses favour late-time solutions to the Hubble problem \cite{Wong:2019kwg,Millon:2019slk,Vagnozzi:2019ezj,Krishnan:2020obg,Suyu:2020opl}.

Galileon gravity \cite{Nicolis:2008in} once provided a late-universe solution to the Hubble problem.
Simple models with $\Lambda=0$ were compatible with Planck and BAO \cite{Barreira:2014jha} and \textit{unambiguously predicted} a value of $H_0$ in agreement with distance ladder, well before the Hubble problem was troubling.
Latter investigations showed that the only Galileon compatible with CMB$\times$LSS cross-correlation 
modify the speed of gravitational waves (GWs)\cite{Renk:2017rzu}. The observation of coincident gravitational and electromagnetic radiation from the neutron star merger GW170817 \cite{Monitor:2017mdv} swiftly ruled out Galileons as a solution to the Hubble problem, along with many other theories of gravity
\cite{Ezquiaga:2017ekz,Creminelli:2017sry,Baker:2017hug,Sakstein:2017xjx}.

\begin{figure*}
 \includegraphics[width=0.69\textwidth]{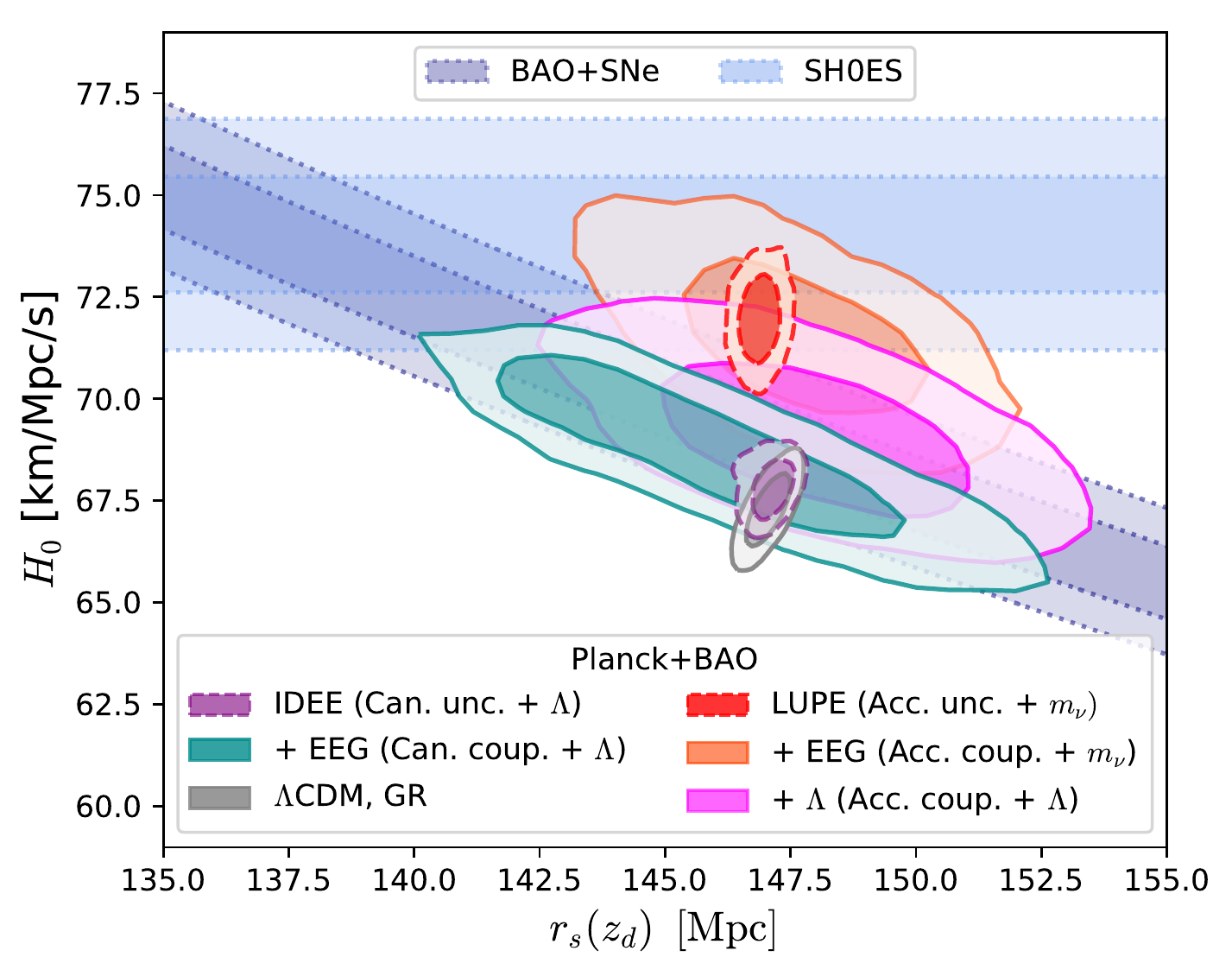} 
 \caption{Galileons, early modified gravity and the Hubble problem. Model-independent constraints on $H_0$ (dotted bands) prefer a lower acoustic scale $r_s$ than $\Lambda$CDM.
 Filled contours show the model-dependent Planck+BAO constraints for Galileon models implementing IDEE, EEG  and LUPE. 
 In IDEE-only models (dashed) the stringent constraints limit the impact on $r_s$.
 Coupled EEG  models (solid) relax the bounds considerably, extending the degeneracy across the BAO+SNe direction. 
 Uncoupled/coupled LUPE models with $\Lambda=0$ (red/orange) predict a high central value of $H_0$ compared to the canonical $\Lambda\neq 0$ cases (purple, dark green), but have a worse fit and` are ruled out by other observations.
 LUPE models with $\Lambda\neq0$ (magenta) provide an intermediate case.
 (Figure adapted from \cite{Knox:2019rjx}).}\label{fig:master_plot}
\end{figure*}

Early-universe solutions invoke new physics before recombination to ``re-calibrate'' the comoving acoustic scale 
\begin{equation}\label{eq:acoustic_scale}
r_s=\int^\infty_{z_d}\frac{c_s(z')}{H(z')}dz'\,,
\end{equation}
which depends on the ratio of the sound speed and the expansion rate up to the redshift of baryon drag.
These solutions work because BAO measure the dimensionless quantity $H_0 \, r_s$. A larger value of $H_0$ requires decreasing $r_s$ by increasing $H$. 
Consistency between BAO and SNe (aka inverse distance ladder) introduces a relation between $H_0$ and $r_s$ which is largely insensitive to late-universe physics \cite{Heavens:2014rja,Bernal:2016gxb,Aylor:2018drw,Knox:2019rjx}. 
Combined inverse and direct distance ladder prefer a shorter acoustic scale than Planck+BAO in $\Lambda$CDM (figure \ref{fig:master_plot}).
This hints at an early-universe solution to the Hubble problem.

Early-solutions rely on new sources of energy contributing to the expansion rate before recombination, cf. Eq. (\ref{eq:acoustic_scale}).
Possible scenarios include additional radiation \cite{Eisenstein:2004an,Suyu:2020opl,Barker:2020gcp}, neutrinos with enhanced interactions \cite{DiValentino:2017oaw,Kreisch:2019yzn,Escudero:2019gvw,Blinov:2019gcj},  variation of fundamental constants \cite{Hart:2019dxi} and non-standard dark matter \cite{Raveri:2017jto,Archidiacono:2019wdp,Vattis:2019efj}.
Another idea is based on \textit{early dark energy}, an analog to time-dependent DE but active in the early universe.
Early DE can be studied via time-dependent parameterizations of the energy density in the Friedmann equation \cite{Karwal:2016vyq,Zhao:2017cud,Poulin:2018zxs} (see \cite{Doran:2006kp,Pettorino:2013ia} for earlier works) and/or the effective gravitational constants in the evolution of perturbations \cite{Lin:2018nxe}.

Dynamical early DE models introduce a quintessence scalar field to solve the Hubble problem \cite{Poulin:2018cxd,Agrawal:2019lmo,Smith:2019ihp}.
A potential $V(\phi)$ with a minimum is required to combine the phenomenology of thawing quintessence \cite{Caldwell:2005tm} and damped oscillations \cite{Turner:1983he}: 
the scalar field is initially subdominant and frozen by Hubble friction. It thaws as the energy density of matter becomes comparable to the potential. Then it begins rolling down the potential and oscillating around the minimum, losing energy in the process until it becomes subdominant again.
Data requires that the scalar starts evolving around the era of equality, setting the initial condition $\phi_i$ so $V(\phi_i)\sim \text{eV}^4$.
The scalar's energy density is constant before equality, with  $V(\phi_i)/\rho_{r}(z_{\rm BBN}) \propto  \left(T_{\rm eq}/T_{\rm BBN}\right)^4 \sim 10^{-24}$ when compared around Big-bang nucleosynthesis (BBN).
Without a mechanism to adjust $\phi_i$ \cite{Sakstein:2019fmf}, quintessence fields require very fine-tuned initial conditions to solve the Hubble problem.

Studies of dynamical early DE models have been restricted to quintessence scalars with different potentials and simple extensions \cite{Lin:2019qug,Niedermann:2019olb}. 
This covers but a narrow sliver in the space of known gravitational theories \cite{Ezquiaga:2018btd}. 
It is plausible that novel signatures and interesting features (e.g. reduced fine-tuning) can be found among extensions of early DE.
The goal of this work is to explore novel solutions to the Hubble problem in viable theories beyond GR, focusing on novel \textit{early Modified Gravity} mechanisms and their phenomenology.

\subsection{Summary and guide for the busy reader}

\begin{figure}
 \includegraphics[width=\columnwidth]{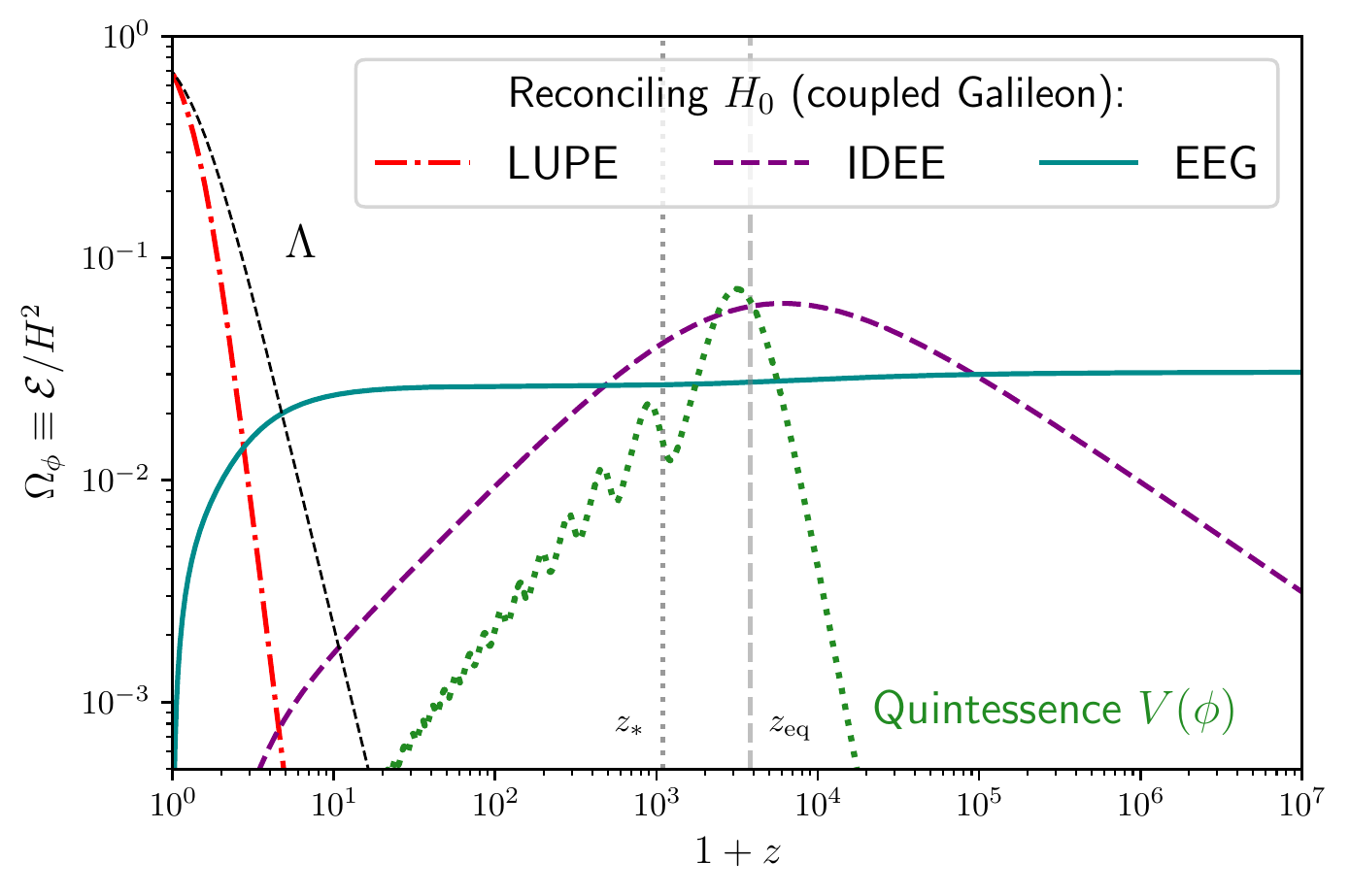}
 \caption{Scalar field energy density in scenarios reconciling early and late-universe values of $H_0$. LUPE acts at low $z$, while IDEE and EEG  reduce $r_s$. An early quintessence model is shown for comparison (Agrawal \textit{et al.}, Ref. \cite{Agrawal:2019lmo}).
 }\label{fig:dynamics_summary}
\end{figure}

This work considers \textit{three} potential solutions to the Hubble problem in scalar-tensor theories of gravity:
\begin{enumerate}
 \item \textit{Imperfect Dark Energy at Equality} (IDEE): the scalar kinetic energy dilutes faster than matter but more slowly than radiation, naturally peaking in the era of equality (\ref{sec:dynamics_idee}). 
 Minimal IDEE can not reconcile Planck+BAO and distance ladder (\ref{sec:constraints_uncoupled}).
 
 \item \textit{Enhanced Early Gravity}\, (EEG): the scalar field value modulates the strength of gravity via the Ricci coupling and can increase the expansion rate at early times (\ref{sec:dynamics_coupling}). 
 Planck+BAO allow EEG  to accommodate higher values of $H_0$, closer to the distance ladder measurement (\ref{sec:constraints_coupled}). Local tests of gravity strongly constrain EEG  (\ref{sec:challenges_local}).
  
 \item \textit{Late-Universe Phantom Expansion} (LUPE): the scalar energy density increases with time  at low redshift, $w_\phi<-1$ (\ref{sec:dynamics_late}). LUPE models with $\Lambda=0$ are ruled out, but coupled LUPE with $\Lambda\neq0$ can ease the $H_0$ tension (\ref{sec:constraints_late}).

\end{enumerate}
The constraints and evolution in each scenario are summarized in figures \ref{fig:master_plot} and \ref{fig:dynamics_summary}, respectively.

While IDEE, EEG  and LUPE are general mechanisms, the results refer to a coupled cubic Galileon scalar-tensor theory of gravity (\ref{sec:theory_cccg}). 
In this theory IDEE relies on the initial field velocity $\dot\phi_i$ and 
EEG  on the initial field value $\phi_i$ modulating the effective Planck mass (i.e. gravitational constant) via a coupling to curvature. 
LUPE requires a negative sign of the quadratic kinetic term (accelerating), causing the scalar energy density to grow in time.
The three mechanisms can operate together or independently.
Readers interested in either of the above mechanisms are directed to visit the sections cited above, in whatever order they consider appropriate. 

The main findings are summarized in the conclusions \ref{sec:conclusions}, along with ideas for further model building and observational tests. Section \ref{sec:theory} introduces the class of viable Galileon theories. Their cosmological dynamics are presented in Section \ref{sec:dynamics}. Section \ref{sec:constraints} presents the cosmological constraints (Planck, BAO, distance ladder) and Section \ref{sec:challenges} discusses the challenges faced by coupled models, including BBN, local gravity tests and GWs. The appendices contain additional discussions.

\section{Galileon Gravity after GW170817}\label{sec:theory}

This section presents the gravity theories studied as potential solutions to the Hubble problem. 
I will describe Galileon gravity theories and their status, focusing on constraints from cosmology and GW observations. In section \ref{sec:theory_cccg} I will narrow down to theories compatible with the GW speed and present (exponentially) coupled cubic Galileons, the class of models used here to investigate IDEE, EEG  and LUPE.

Most Galileon gravity theories are specific realizations of the Horndeski class \cite{Horndeski:1974wa,Deffayet:2011gz,Kobayashi:2011nu}, the most general action for a tensor and a scalar field, generally covariant, Lorentz invariant and leading to second order equations of motion in 4 space-time dimensions:
\begin{equation}
S[g_{\mu\nu},\phi]=\int\mathrm{d}^{4}x\,\sqrt{-g}\left[\sum_{i=2}^{5}{\cal L}_{i}[g_{\mu\nu},\phi]\,+\mathcal{L}_{\text{m}}\right]\,,\label{eq:action}
\end{equation}
where $\mathcal{L}_{\text{m}}=\mathcal{L}_{\text{m}}[g_{\mu\nu},\psi_{M}]$ is the matter Lagrangian density, minimally coupled to the Jordan frame metric $g_{\mu\nu}$ and the gravitational interaction is given by:
\begin{eqnarray}
{\cal L}_{2} & = & G_{2}(\phi,\,X)\,,\label{eq:L2}\\
{\cal L}_{3} & = & -G_{3}(\phi,\,X)\Box\phi\,,\label{eq:L3}\\
{\cal L}_{4} & = & G_{4}(\phi,\,X)R+G_{4X}(\phi,\,X)\left[\left(\Box\phi\right)^{2}-\phi_{;\mu\nu}\phi^{;\mu\nu}\right]\,,\label{eq:L4}\\
{\cal L}_{5} & = & G_{5}(\phi,\,X)G_{\mu\nu}\phi^{;\mu\nu}-\frac{1}{6}G_{5X}(\phi,\,X)\Big[\left(\Box\phi\right)^{3}
\nonumber \\ &&
+2{\phi_{;\mu}}^{\nu}{\phi_{;\nu}}^{\alpha}{\phi_{;\alpha}}^{\mu}-3\phi_{;\mu\nu}\phi^{;\mu\nu}\Box\phi\Big]\,.\label{eq:L5}
\end{eqnarray}
The four Lagrangians $\mathcal{L}_{i}$ encode the dynamics the scalar field $\phi$
of the Jordan-frame metric $g_{\mu\nu}$.
They contain four arbitrary functions $G_{i}(\phi,X)$ of the
scalar field and its canonical kinetic term, $2X\equiv-\partial_{\mu}\phi\partial^{\mu}\phi$. Subscripts $\phi,X$ to denote partial derivatives, e.g.~$G_{iX}=\frac{\partial G_{i}}{\partial X}$.
I will follow the conventions of the \texttt{hi\_class} code \cite{Zumalacarregui:2016pph,Bellini:2019syt}, including  natural units ($c=h=1$) and mostly-plus signature of the metric $(-,+,+,+)$.

The uncoupled covariant Galileon \cite{Nicolis:2008in,Deffayet:2009wt,DeFelice:2010pv} is is the most general Horndeski completion of a theory realizing the Galilean symmetry $\phi\to \phi+c+b_\mu x^\mu$ in flat space-time.
The theory is defined by the following Horndeski functions
\begin{eqnarray}
 && G_2=c_1\M^3\phi - c_2 X\,,\; G_3 = \frac{c_3}{\M^3}X\,,\; \\
 && G_4 = \frac{M_P^2}{2} - \frac{c_4}{\M^6}X^2\,,\; G_5 = \frac{3c_5}{\M_3^9}X^2\,, \label{eq:cov_gal_action}
\end{eqnarray}
where $M_P=1/\sqrt{8\pi G}$ is the reduced Planck mass and $\M_3 = (H_0^2 M_P)^{1/3}$. 
While the linear potential $\propto c_1\phi$ is compatible with Galilean symmetry, it does not lead to interesting phenomenology and it is common to set it to zero (see Ref. \cite{Bellini:2013hea} for an analysis including $c_1$).

Uncoupled covariant Galileons (\ref{eq:cov_gal_action}) have interesting cosmological solutions, including an unambiguous prediction of $H_0$ compatible with distance ladder.
If the quadratic kinetic term has negative sign ($c_2<0$), the theory predicts LUPE, accelerating solutions without the need of $\Lambda$.
These $\Lambda=0$ models require a sizeable neutrino mass $\sum m_\nu \sim 0.6$ eV to fit CMB+BAO observations \cite{Barreira:2014jha}, a value well within the range of laboratory experiments \cite{Aker:2019uuj} but that excluded by cosmological data if assuming $\Lambda$CDM. The high value of the neutrino mass is also necessary to solve the $H_0$ problem: the $\sum m_\nu \approx 0$ models do not only yield a worse fit, but also predict a value of $H_0$ above the distance ladder measurement \cite{Barreira:2013jma}. 

The evolution of the metric potentials constrains the parameter space of $\Lambda=0$ uncoupled Galileons (\ref{eq:cov_gal_action}). The minimal theory ($c_4,c_5=0$) always predicts growing potentials at low redshift, instead of decaying as in $\Lambda$CDM. The growth of the potentials leads to an anti-correlation between CMB temperature and the low redshift galaxies (CMB$\times$LSS) via the Integrated Sachs-Wolfe (ISW) effect, in stark disagreement with current measurements \cite{Ferraro:2014msa}.
General Galileons ($c_4,c_5\neq 0$) can accommodate decaying potentials in some regions of the parameter space \cite{Renk:2017rzu}. 

General uncoupled Galileons that agree with CMB$\times$LSS modify the GW speed \cite{Renk:2017rzu}.
The simultaneous detection of gravitational and electromagnetic radiation from the neutron star merger GW170817 \cite{Monitor:2017mdv} placed a tight bound on the GW speed $|c_g-1| \lesssim 10^{-15}$. 
This event severely constrains cosmologically viable $\Lambda=0$ uncoupled Galileons \cite{Ezquiaga:2017ekz,Creminelli:2017sry,Baker:2017hug,Sakstein:2017xjx} among many other theories beyond GR (see \cite{Lombriser:2015sxa,Brax:2015dma,Bettoni:2016mij} for earlier works). The limits are at the level $|c_4|, |c_5| \lesssim  5 \cdot 10^{-17}$ \cite{Ezquiaga:2017ekz}. Combining GW speed and CMB$\times$LSS seals the deal of all uncoupled Horndeski Galileons (\ref{eq:cov_gal_action}) without a cosmological constant.

Galileon theories beyond-Horndeski \cite{Zumalacarregui:2013pma,Gleyzes:2014dya,Langlois:2015cwa} can be made compatible with the GW speed but are ruled out by other GW observations \cite{Creminelli:2018xsv,Creminelli:2019nok} and cosmology \cite{Peirone:2019yjs}. 
Beyond-Horndeski theories in the Gleyzes-Langlois-Piazza-Vernizzi (GLPV) class can be constructed in which $c_g=1$ on any space-times \cite{Ezquiaga:2017ekz,Creminelli:2017sry,Babichev:2017lmw}. GLPV Galileons with $c_g=1$ have identical cosmological expansion than their Horndeski analogs, potentially providing a late-time solution to the Hubble problem compatible with the GW speed. 
However, GLPV theories predict a very rapid decay of GWs into scalar field excitations \cite{Creminelli:2018xsv,Creminelli:2019nok}, and the deviations from Horndeski need to be very suppressed for \textit{any} GW signal to be detected.
The remaining beyond-Horndeski term compatible with GW speed and decay does not have the Galileon form  \cite[Eq. 40]{Zumalacarregui:2013pma}

GW speed and decay bounds allow Horndeski theories with general $G_2, G_3$, but restrict $\mathcal{L}_4,\mathcal{L}_5$ to $G_4(\phi), G_5=0$. 
Galileon theories are equipped with the Vainshtein screening mechanism \cite{Vainshtein:1972sx} suppressing small-scale deviations from GR, including effects in the emission of GWs \cite{deRham:2012fw,Chu:2012kz,Dar:2018dra} (although see Ref. \cite{Brax:2020ujo} for a possible counter-example).
Still, these theories are still subject to GW constraints (instabilities induced by GWs \cite{Creminelli:2019kjy} and standard sirens) which I will discuss in section \ref{sec:challenges_GWs}.

\subsection{Coupled Cubic Galileon}\label{sec:theory_cccg}

I will explore the coupled Cubic Galileon \cite{Chow:2009fm,Silva:2009km,Appleby:2011aa,Appleby:2012ba,Neveu:2014vua,Bhattacharya:2015chc,Burrage:2015lla,Burrage:2019afs,Tsujikawa:2019pih}, a variant of the Galileons described above, restricted to be compatible with GW observations but extended through a non-minimal coupling between the scalar field and and the Ricci scalar.
The coupling is introduced via a $\phi$-dependence of $G_4$, the coefficient of the Ricci scalar in the Horndeski action
\begin{equation}
\mathcal{L}_{G3,C} = \frac{M_P^2}{2}C(\phi)R +2\frac{c_3}{M^3}X\Box\phi + c_2{X} - 2\Lambda + \mathcal{L}_{\rm m}\,,
 \label{eq:cccg_Gfuns} 
\end{equation}
or equivalently, $ G_4 = {C(\phi)}$,  $G_3 = -2\frac{c_3}{M^3}X$ and $G_2 = c_2{X}-2\Lambda$ in Eqs. (\ref{eq:L2}-\ref{eq:L4}).
The main effect of the coupling is to modify the strength of gravity, now depending on the value of the field.%
\footnote{The coupled theory (\ref{eq:cccg_Gfuns}) is a minimal extension of the uncoupled Galileons (\ref{eq:cov_gal_action}). One may also introduce the coupling directly into the matter action $\mathcal{L}_m(g_{\mu\nu},\psi_M)\to  \mathcal{L}_m(\tilde C(\phi) g_{\mu\nu},\psi_M)$ via the the so-called Einstein frame metric $\tilde C(\phi)g_{\mu\nu}$. 
This theory can be recasted into a minimally coupled Horndeski form via a field-dependent rescaling of the metric (into the so-called Jordan Frame).
The resulting action is not equivalent to Eq. (\ref{eq:cccg_Gfuns}) as $G_2$ is corrected by some terms depending on $\tilde C(\phi)$, see \cite{Ezquiaga:2017ner} for general expressions and Ref. \cite{Burrage:2019afs} for an explicit example.\label{foot:frames}}

Coupled Cubic Galileon theories admit a binary classification into \textit{Canonical} or \textit{Accelerating} models. 
An arbitrary redefinition of the scalar field by a constant factor
\begin{equation}\label{eq:field_rescaling}
 \phi\to\alpha\phi: \quad c_2\to \alpha^2 c_2,\, c_3\to \alpha^3 c_3,\, \beta\to\alpha\beta\,,
\end{equation}
in the Lagrangian (\ref{eq:cccg_Gfuns}) fixes one of the coefficients without any loss of generality \cite{Barreira:2013jma}. For a real-valued $\alpha$, the above transformation always preserves the sign of $c_2$, the quadratic kinetic term. Canonical and accelerating models models correspond to a positive and negative sign of $c_2$, respectively.
The differences between both will be explored in sections \ref{sec:dynamics_late} and \ref{sec:constraints_late}. 
\footnote{
The literature often refers to \textit{self-accelerating} models, in which the universe's acceleration is supported to the conformal coupling. 
This is defined as the acceleration condition being satisfied in the Jordan frame (used here), but not in the Einstein Frame (see footnote \ref{foot:frames}). I will not consider self-acceleration further.
}

For further simplicity, I will consider only an exponential form of the coupling
\begin{equation}\label{eq:cccg_G4_exp}
C(\phi) = e^{\beta\phi/M_p}\,.
\end{equation}
The exponential form is particularly simple to study. 
All couplings with $C_\phi\neq0$ break the shift-symmetry $\phi\to \phi + C$, but the exponential coupling introduces only a constant term in the scalar field equation. Thus, there is no dependence on $\phi$ in the scalar field equations. 
The dependence on the scalar field value is thus limited to the gravitational sector, as $\phi$ modulates the strength of gravity. 
Compared to other choices of the coupling function, the exponential form leads to convenient simplifications in the analysis of the cosmology described in the next section.

\section{Cosmological Dynamics}\label{sec:dynamics}

In this section I will discuss the cosmological dynamics of coupled cubic Galileons (\ref{eq:cccg_Gfuns}), specializing to the exponential form of the coupling (\ref{eq:cccg_G4_exp}).
Section \ref{sec:dynamics_general} introduces the dynamical equations and important concepts related to the theory.  
The following subsections detail how solutions of the coupled cubic Galileon lead to IDEE (\ref{sec:dynamics_idee}), EEG  (\ref{sec:dynamics_coupling}) and LUPE (\ref{sec:dynamics_late}).
The early time dynamics are further discussed in Appendix \ref{sec:dynamics_ic}.

\subsection{General Considerations}\label{sec:dynamics_general}

Let us start by presenting the general equations for the background metric and scalar field for coupled cubic Galileons. I will then review the classification of Galileons into canonical and accelerating and some properties of the exponential coupling.

\subsubsection{Equations \& Definitions}

The expansion history is governed by the modified Friedmann equation 
\begin{equation}\label{eq:M2_hubble_de}
M_*^2 H^2 = \rho_m + \hat{\mathcal{E}}\,,
\end{equation}
where $\rho_m$ is the total matter density in CLASS units [Mpc$^{-2}$] \cite{Blas:2011rf}.
The \textit{effective Planck mass}
\begin{equation}\label{eq:Mpl_def}
M_*^2 \equiv 2G_4 = C(\phi)\,,
\end{equation}
modulates the strength of gravity on the cosmological background and the \textit{kinetic energy density}
\begin{equation}\label{eq:rho_de_kin}
 \hat{\mathcal{E}} =  \frac{c_2}{6}\dot\phi^2-2\frac{c_3}{M^3} H \dot\phi^3 -H C^\prime \dot\phi\,,
\end{equation}
represents the remaining contributions of the scalar field to the expansion rate. Note that all the terms in $\hat{\mathcal{E}}$ are proportional to $\dot\phi$, while $M_{*}^2$ depends only on $\phi$.
The Galileon energy fraction today is then
\begin{eqnarray}
\Omega_{\phi,0} 
&=&\hat\Omega_{\phi,0} + (1-M_*^2) \label{eq:Omega_de_coupled}
\,,
\end{eqnarray}
where the kinetic contribution (\ref{eq:rho_de_kin}) reads
\begin{equation}
 \hat\Omega_{\phi,0} = \frac{c_{2}}{6}\xi^{2}-2c_{3}\xi^{3} -C^\prime\xi\,,
\end{equation}
and the \textit{dimensionless field velocity} \cite{Barreira:2014jha}
\begin{equation}\label{eq:xi_def}
\xi\equiv \frac{\dot{\phi}H}{M_{p} H_{0}^{2}}\,,
\end{equation}
provides a convenient variable.

The scalar field equation can be written in a current conservation form
\begin{equation} \label{eq:shift_current_eq}
 \dot {\cal J} + 3 H {\cal J} 
 = {\cal P}_\phi \,.
\end{equation}
Here the \textit{shift-charge density} (or \textit{shift-charge})
\begin{equation}\label{eq:shift_current_gal}
 {\cal J} = c_2\dot\phi - c_3\frac{6H}{H_0^2M_p}\dot\phi^2 = \frac{M_p H_0^2}{H }\left( c_2 \xi - 6c_3 \xi^2 \right)\,.
\end{equation}
is the time-component of a Noether current ${\cal J}^\mu$ associated to shift symmetry $\phi\to\phi+C$.
The \textit{kinetic term}
\begin{equation} \label{eq:kinetic_term}
 D\equiv \frac{\partial \mathcal{J}}{\partial\dot\phi} = c_2 - 12 c_3\xi \,,
\end{equation}
i.e. the coefficient of $\ddot\phi$ in Eq. (\ref{eq:shift_current_eq}) determines the stability of the theory. It needs to be positive for the stability of both the background and linear perturbations.
Finally, the source term is given by
\begin{equation}\label{eq:shift_violation_ricci}
 {\cal P}_\phi =  3 C_{,\phi}(\dot H + 2 H^2)\,,
\end{equation}
and is proportional both to the coupling strength $\beta= C_{,\phi}/C$ and the Ricci scalar evaluated on the cosmological background.

\subsubsection{Canonical vs Accelerating Galileon}

Let us now examine the kinetic structure and solutions of cubic Galileons, starting with the uncoupled case. 
Solutions to Eq. (\ref{eq:shift_current_eq}) with $\mathcal{P}_\phi=0$
\begin{equation}\label{eq:shift_symm_general_sol}
\mathcal{J}(a) = \frac{\mathcal{J}_0}{a^3}\,, \quad ({\cal P}_\phi=0)
\end{equation}
correspond to the shift-charge density diluting with the Universe's volume.
The scalar field is thus drawn towards ${\cal J} \propto c_{2}\xi-6c_{3}\xi^{2}\to 0$, corresponding to two solutions
\begin{equation}
 \begin{array}{l l l c}
  \xi= 0\,, & \hat\Omega_{\phi,0} = 0\,, & D=c_2 & \quad \text{(trivial)} \\
  \xi= \frac{c_2}{6c_3}\,, & \hat\Omega_{\phi,0} =  \frac{-c_2^3}{216 c_3^2} \,, & D= -c_2 & \quad \text{(tracker)}
 \end{array}\label{eq:gal_solutions}
\end{equation}
The sign of the kinetic term $c_2$ determines which solution is stable via the no-ghost condition $D>0$, Eq. (\ref{eq:kinetic_term}).

\begin{figure}[t!]
 \includegraphics[width=\columnwidth]{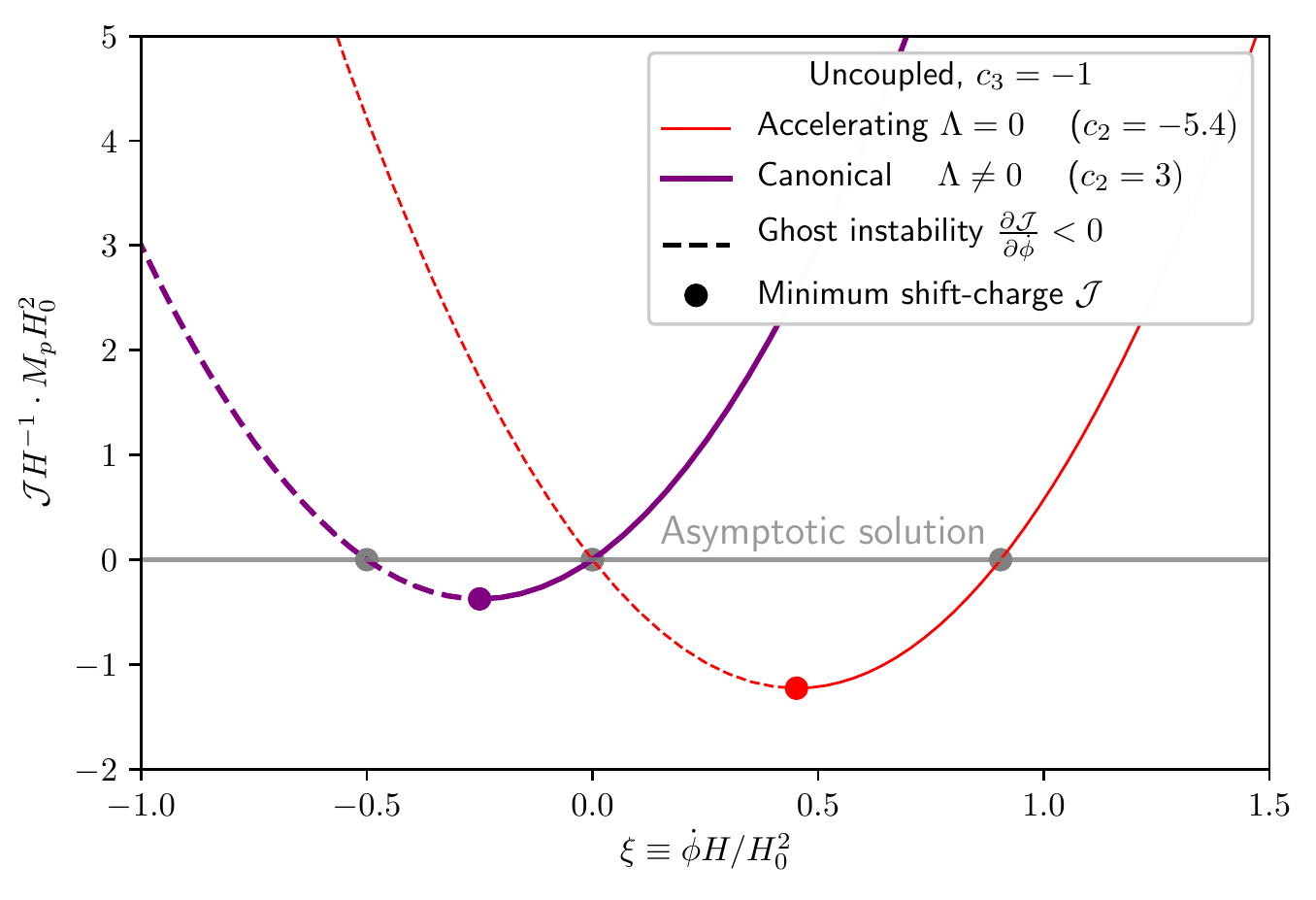}
 \caption{Kinetic structure of cubic Galileons. The relation between the shift-charge density (\ref{eq:shift_current_gal}) and the field derivative Eq. (\ref{eq:xi_def}) is shown for canonical ($c_2>0$, thick) and accelerating ($c_2<0$, thin) models. 
 Absence of ghosts requires a positive slope for the curve (\ref{eq:kinetic_term}), with the minimum of $\mathcal{J}$ corresponding to the transition to instability. 
 Stable accelerating/canonical models tend to $\xi\neq0$, $\xi=0$ respectively  (\ref{eq:gal_solutions})
 A positive coupling strength $\beta>0$ sources ${\cal J}$, delaying the approach to the asymptotic solution.
 Negative coupling strength $\beta<0$ drive the field towards the ghost region.  
 }\label{fig:kinetic_structure}
 \end{figure}

The above solutions reveal a binary classification of Galileons
\begin{itemize}
 \item \textit{Canonical Galileons} $c_2>0$ are driven towards the trivial solution. As $\Omega_{\phi,0}\to0$, a cosmological constant is necessary for these models to be viable.
 \item \textit{Accelerating Galileons} $c_2<0$ are driven towards the tracker solution. As $\Omega_{\phi,0}>0$, accelerating models produce LUPE and can accelerate the Universe without a cosmological constant.
\end{itemize}
The approach to these solutions is determined by the relationship between the shift charge and the scalar velocity, described in figure \ref{fig:kinetic_structure}.
The above classification is robust against rescalings of the scalar field, which preserve the sign of $c_2$. In contrast, the sign of either $\beta$, $c_3$ can be fixed by a field rescaling that preserves the form of the action, see Eq. (\ref{eq:field_rescaling}).

Canonical and accelerating Galileons are indistinguishable at early times, either because the cubic Galileon term dominates the dynamics ($|\dot\phi|\gg |c_2/6c_3|$) (e.g. IDEE) or because the kinetic energy is negligible ($\hat \Omega_\phi \sim 0$). 
The differences occur at late times, leading to the different values of $\hat \Omega_{\phi,0}$ in the asymptotic solutions (\ref{eq:gal_solutions}) and will be discussed in section \ref{sec:dynamics_late}. 
A non-zero coupling sources the shift-charge density, driving the solution away from $\mathcal{J}\to0$, Eq. (\ref{eq:gal_solutions}). This is shown in figure \ref{fig:kinetic_structure} and discussed below.

\subsubsection{Coupling \& Vainshtein Mechanism}

The coupling to curvature introduces a source to the shift-charge density (\ref{eq:shift_violation_ricci})
\begin{equation}\label{eq:shift_violation_rho}
 {\cal P}_\phi 
 = \frac{3}{2}\frac{C_{,\phi}}{C}\left(\rho_m+\hat{\mathcal{E}} - 3(p_m +\hat{\mathcal{P}})\right)\,,
\end{equation}
where the above expression uses the Friedman (\ref{eq:M2_hubble_de}) and acceleration equation, and $\hat{\mathcal{P}}$ is the scalar field pressure removing the effect of the strength of gravity (analog to $\hat{\mathcal{E}}$, cf. Ref. \cite[Eq. 3.5]{Bellini:2014fua}). 

The contribution of radiation and ultra-relativistic matter to the coupling is negligible since $\rho_{\rm rad} = 3p_{\rm rad}$. This follows from the coupling involving the Ricci scalar, which is sourced by the trace of the energy momentum tensor $T\propto 1-3w_m$. 
Sources to the coupling in the matter era will be discussed in section \ref{sec:dynamics_coupling}.  
Early-universe processes in the radiation era are presented in appendix \ref{sec:dynamics_ic}.

Analytic expressions exist for exponential coupling when the Galileon kinetic energy is negligible. 
For an exponential coupling $\beta\equiv C_{,\phi}/C$ is constant and the source term (\ref{eq:shift_violation_rho}) is independent of the field value. 
Then the field equation (\ref{eq:shift_current_eq}) can be integrated directly
\begin{equation}\label{eq:charge_sol_explicit}
 {\cal J} \approx  \frac{1}{a^3}\frac{3}{2}\beta \int_0^a da' a'^2H(a')\Sigma(a')\,.
\end{equation}
where the \textit{kick function} reads
\begin{equation}
 \Sigma \equiv \frac{\rho_m+\hat{\mathcal{E}} - 3(p_m +\hat{\mathcal{P}})}{H^2} \approx \frac{\rho_m-3p_m}{\rho}=1-3w_m\,.
\end{equation} 
This solution accounts for the effects of $M_*^2\neq 1$ on the expansion (\ref{eq:M2_hubble_de}) but neglects $\hat{\mathcal{E}},\hat{\mathcal{P}}\sim 0$, a very good approximation at early times. 
Note that the kick function also affects the integrand via 
\begin{equation}\label{eq:H_sigma}
 H(a) \propto a^{-3(1+w)/2} = a^{\Sigma/2-2}\,.
\end{equation}

It is possible to decompose the solution for the shift-charge \ref{eq:charge_sol_explicit} as
\begin{equation}\label{eq:sigma_decomposition}
 {\cal J} = \frac{{\cal J}_0}{a^3} + {\cal J}_M + {\cal J}_\Sigma
\end{equation}
where ${\cal J}_0$ describes a general initial condition, ${\cal J}_M$ is the contribution from the fraction of non-relativistic matter and ${\cal J}_\Sigma$ represents the contribution from deviations from radiation domination in the early universe.
The contribution from non-relativistic matter $\Sigma_M\approx \rho_{\rm mat}/\rho_{\rm rad} = a/a_{\rm eq}$ leads to a shift-charge
\begin{equation}
 {\cal J}_M = \frac{3}{4}H_0\sqrt{\Omega_r}\beta \frac{a^2}{a_{\rm eq}}
=\frac{3}{2}\beta\frac{\Omega_m H_0^2}{a^3}\cdot t\,.
\end{equation}
Appendix \ref{sec:dynamics_ic} describes additional sources ${\cal J}_\Sigma$ in the early Universe. 
No realistic source is able to contribute significantly to the scalar field initial conditions due to the non-linear derivative interactions.

The non-canonical nature of the cubic Galileon leads to the \textit{cosmological Vainshtein screening} \cite{Chow:2009fm,Burrage:2019afs}, an efficient suppression of the coupling at early times.
If the cubic Galileon term dominates, the scalar energy fraction is related to the shift-charge as
\begin{equation}\label{eq:shift_charge_to_omega3}
 \hat\Omega_{\phi,3} \approx \frac{1}{\sqrt{27 |c_3|}}\frac{H_0}{H}\left(\frac{\mathcal{J}}{H}\right)^{3/2}, 
 \quad (c_3\xi \gg c_2)\,,
\end{equation}
where Eqs. (\ref{eq:rho_de_kin}, \ref{eq:shift_current_gal}) have been used.
In contrast, if the quadratic term dominates, the equivalent expression reads
\begin{equation}\label{eq:shift_charge_to_omega2}
 \hat\Omega_{\phi,2} \approx \frac{\mathcal{J}^2}{6 H^2}\,,  \qquad (c_3\xi \ll c_2)\,.
\end{equation}
The ratio between the energy scales associated to the cubic Galileon and canonical kinetic term is
\begin{equation}
 \frac{\hat\Omega_{\phi,3}}{\hat\Omega_{\phi,2}}
= 12 \frac{c_3}{c_2}\frac{\dot\phi H}{H_0^2}
=12\frac{c_3}{c_2}\xi\,,
\end{equation}
so the cubic term dominates for large dimensionless field velocities. 

The cosmological Vainshtein screening stems from the $H_0/(\sqrt{|c_3|}H)$ factor in Eq. (\ref{eq:shift_charge_to_omega3}), suppressing the effects of the coupling on the shift-charge $\mathcal{J}$ at early times. 
While the cosmological Vainshtein screening may be circumvented by reducing the value of $c_3$, such a reduction will incur in constraints from local gravity tests in the late universe, unless the coupling is reduced accordingly (see section \ref{sec:challenges_local}).
The effects of the screening will be shown explicitly in section \ref{sec:dynamics_coupling} and appendix \ref{sec:dynamics_ic}.

\subsection{Imperfect Dark Energy at Equality} \label{sec:dynamics_idee}

\begin{figure*}
 \includegraphics[height=4.8cm]{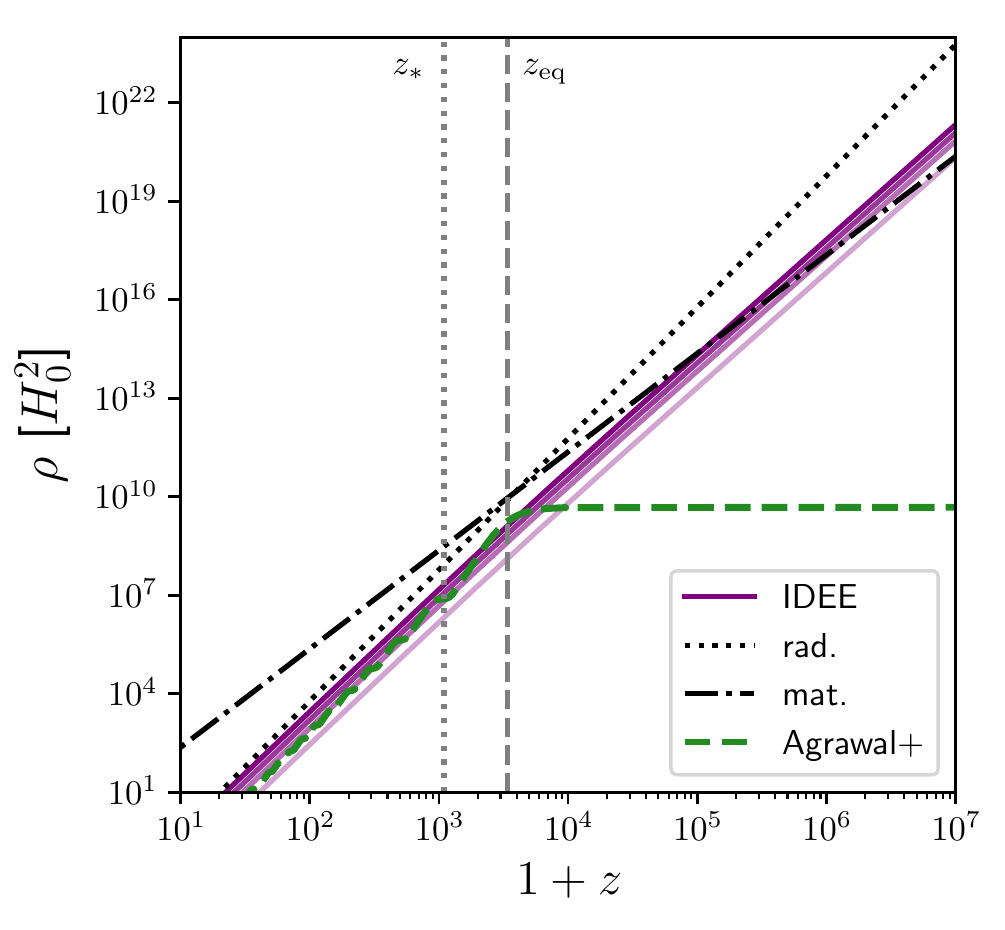}
 \includegraphics[height=4.8cm]{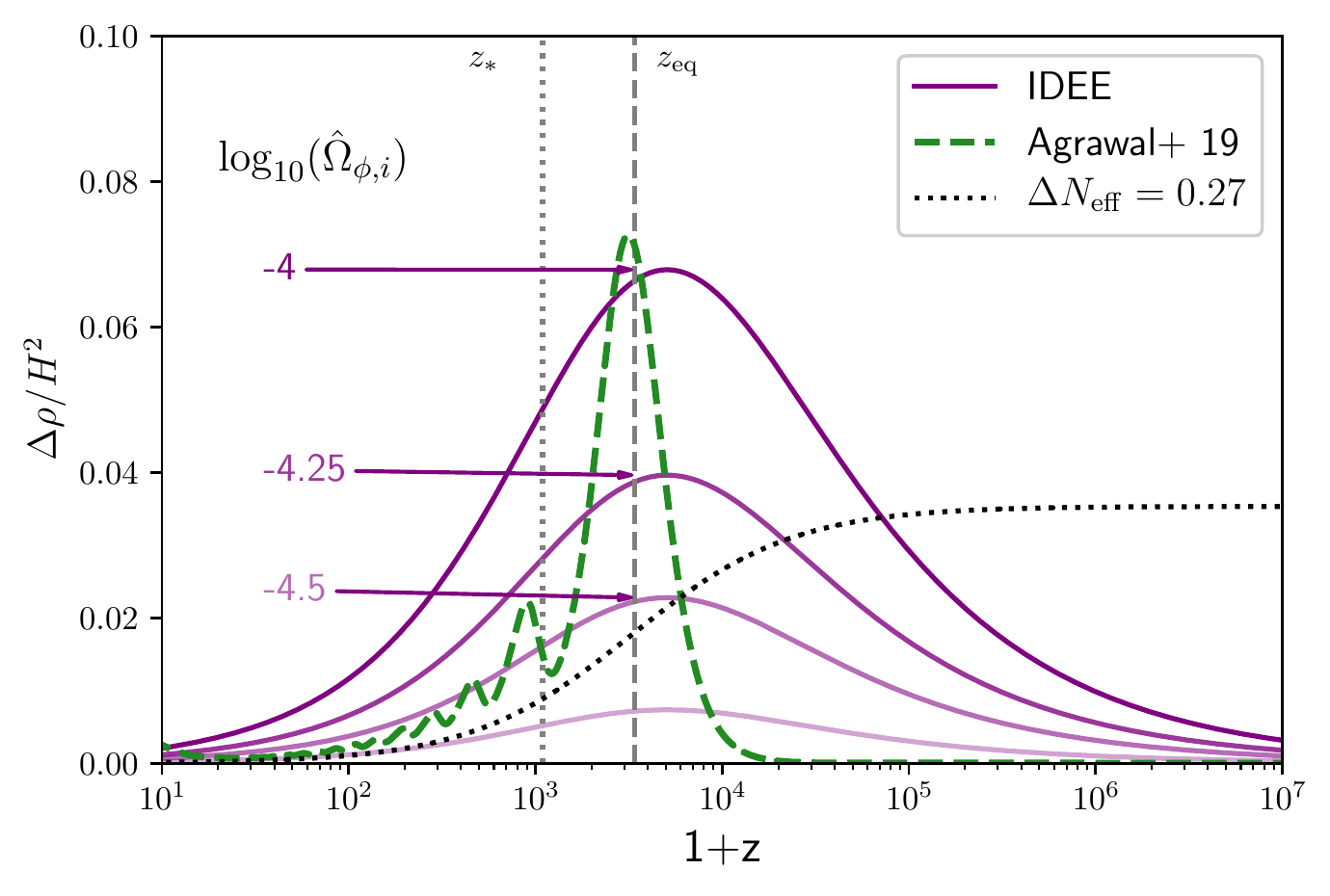}
 \includegraphics[height=4.8cm]{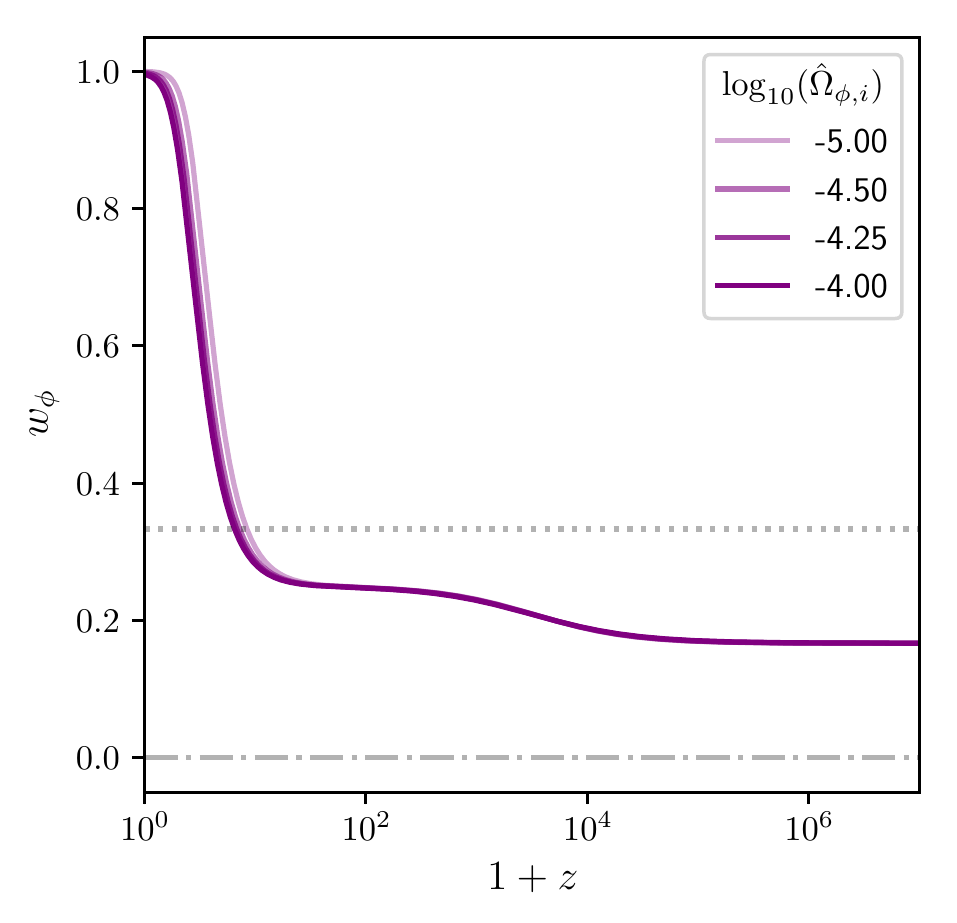}
 \caption{Imperfect Dark Energy at Equality (IDEE) in canonical uncoupled models. The initial energy density of the scalar field dilutes faster than radiation but more slowly than matter (left panel). By virtue of this scaling, the relative scalar field abundance peaks around the era of matter-radiation equality (middle panel), lowering $r_s$ and increasing $H_0$ for fixed $\theta_\star$.  Energy contributions of additional relativistic particles and an early quintessence model \cite{Agrawal:2019lmo} are shown for comparison.
The equation of state of the scalar remains in the range $w_\phi\in(0,1/3)$ until the kination phase at low $z$ (right panel).
 }\label{fig:idee_dynamics}
\end{figure*}

\textit{Imperfect dark energy at equality} (IDEE) is a distinct form of early dark energy beyond GR characterized by a contribution to the expansion history that peaks around matter-radiation equality.
IDEE is sourced by the cubic Galileon term, which effectively modifies gravity and changes the evolution of the perturbations (e.g. CMB).
In order to affect the acoustic scale IDEE requires a significant kinetic energy of the Galileon: an initial scalar field kinetic energy $\hat \Omega_{\phi,i}\sim 10^{-4}$ around the nucleosynthesis era evolves into a $\sim \%$ level contribution at equality, sufficiently to reconcile early and late measurements of $H_0$.

To understand the dynamics of IDEE I assume that the contribution from $\mathcal{L}_3\to c_3\dot\phi^3 H$ dominates the energy budget (\ref{eq:rho_de_kin})
\begin{equation}\label{eq:idee_energy_density}
\hat{\mathcal{E}} \approx -2 \frac{c_3}{H_0^2}\dot\phi^3 H \,, \quad (c_3\xi \gg c_2)\,,
\end{equation}
so 
  \begin{equation}\label{eq:idee_energy_fraction}
\hat\Omega_\phi 
\approx  -2 \frac{c_3}{H_0^2}\frac{\dot\phi^3}{H} = -2 c_3 \xi^3 \left(\frac{H_0}{H}\right)^4\,,
\end{equation}
where the final expression uses the dimensionless field velocity (\ref{eq:xi_def}).
These equations can be used to set the initial condition for the field derivative $\dot\phi$.
Note that the energy density scaling of IDEE relies only on the domination of the cubic term. It is otherwise independent of the Galileon energy scale $|c_3|$, provided that the initial field velocity is sufficiently high, as prescribed by Eq. (\ref{eq:idee_energy_density}).

The characteristic scaling of IDEE
\begin{equation}
 \hat{\mathcal{E}} \propto a^{-3/4(w_m-5)}
 = \left\{
 \begin{array}{l l l}
  a^{-3.5}\,, & w_\phi=\frac{1}{6} & \text{(rad.)} \\
  a^{-3.75}\,, & w_\phi=\frac{1}{4} & \text{(mat.)} \\
 \end{array}
 \right.\,.
\end{equation}
follows from substituting the solution for $\xi$ from the off-tracker evolution (\ref{eq:shift_current_eq}) and neglecting the coupling $\mathcal{P}_\phi\sim 0, n\propto a^{-3}$ determines the scaling of the energy density. 
This particular evolution, diluting faster than matter but more slowly than radiation, allows IDEE to emerge around matter-radiation equality.
Figure \ref{fig:idee_dynamics} shows the scaling of IDEE for different initial conditions, along with its effects on the acoustic scale (\ref{eq:acoustic_scale}).
Values of the initial field derivative such that $\hat\Omega_{\phi,i}\sim 10^{-4}$ at $z=10^{10}$ (around the BBN epoch) grow into sizeable early dark energy contributions $\sim 5\%$ at the epoch of equality, sufficient to lower the acoustic scale at the level needed to reconcile CMB+BAO and distance ladder inferences of the Hubble parameter.

IDEE models also induce deviations from general relativity. These are best parameterized by the dimensionless \textit{braiding} function \cite{Bellini:2014fua}
\begin{equation}\label{eq:idee_alpha_b3}
 \alpha_{B,3} \equiv -2\frac{c_3}{M_*^2}\frac{\dot\phi^3}{H_0^2 H}
 \approx \hat\Omega_{\phi}\,,
\end{equation}
where the second equality applies to the limit in which the cubic Galileon dominates the energy density, Eq. (\ref{eq:idee_energy_density}). 
$\alpha_B$ describes the kinetic mixing between the scalar field perturbations and the gravitational potentials on the cosmological background (see \cite{Bettoni:2015wta} for a covariant description). 
The function $\alpha_B$ also parameterizes the deviation from the uncoupled cubic Galileon from behaving as a perfect fluid \cite{Pujolas:2011he,Sawicki:2012re}.
The last equality shows that this deviation from GR is as important as the contribution to the expansion history. The deviations from GR induced by IDEE turn out to be very restrictive for IDEE models when compared with Planck data, as I will show in section \ref{sec:constraints_uncoupled}.

Non-cubic covariant Galileon theories (\ref{eq:cov_gal_action}) dilute more slowly with the expansion, restricting their early-universe dynamics. 
If the quartic Galileon $G_4\propto X^2$ term dominates, its energy density scales as $\mathcal{E}_4\propto a^{w_m-3}$, diluting faster than matter in the radiation era and tracking the matter density afterwards.
The quintic Galileon $G_5\propto X^2$ always dilutes more slowly than matter, as $\mathcal{E}_4\propto a^{\frac{3}{8}(w_m-7)}$, corresponding to $w_\phi = -\frac{1}{4}, -\frac{1}{8}$ in the radiation \& matter eras respectively.
Note that $c_4,c_5\neq 0$ may contribute to the early-universe dynamics,  provided that their effect on the speed of GWs is suppressed at late times. This could happen if the field velocity kinates away, Eq. (\ref{eq:canonical_kination}), soon after matter-radiation equality. 
While possible, this type of early modified gravity requires much more fine-tuning than the cubic Galileon implementation of IDEE.

The properties of IDEE in the simple cubic Galileon model were first discussed in Ref. \cite{Deffayet:2010qz}, where it was also pointed out that the initial kinetic energy of the field would grow until the epoch of equality and could lower the acoustic scale.
Previous works analyzing Galileons with general initial conditions focused on the general model \cite{Barreira:2012kk,Barreira:2013jma, Neveu:2013mfa,Neveu:2014vua,Neveu:2014kba,Neveu:2016gxp}, in which the cubic and quintic terms scale faster than matter, leading to tight constraints on $\hat\Omega_{\phi,i}$. A more recent analysis considered the cubic Galileon separately \cite{Leloup:2019fas}, but used the same priors as in previous models and did not explicitly discuss the relevant region in which early dark energy modifies the acoustic scale.

\subsection{Enhanced Early Gravity}\label{sec:dynamics_coupling}

\begin{figure*}
 \includegraphics[height=4.8cm]{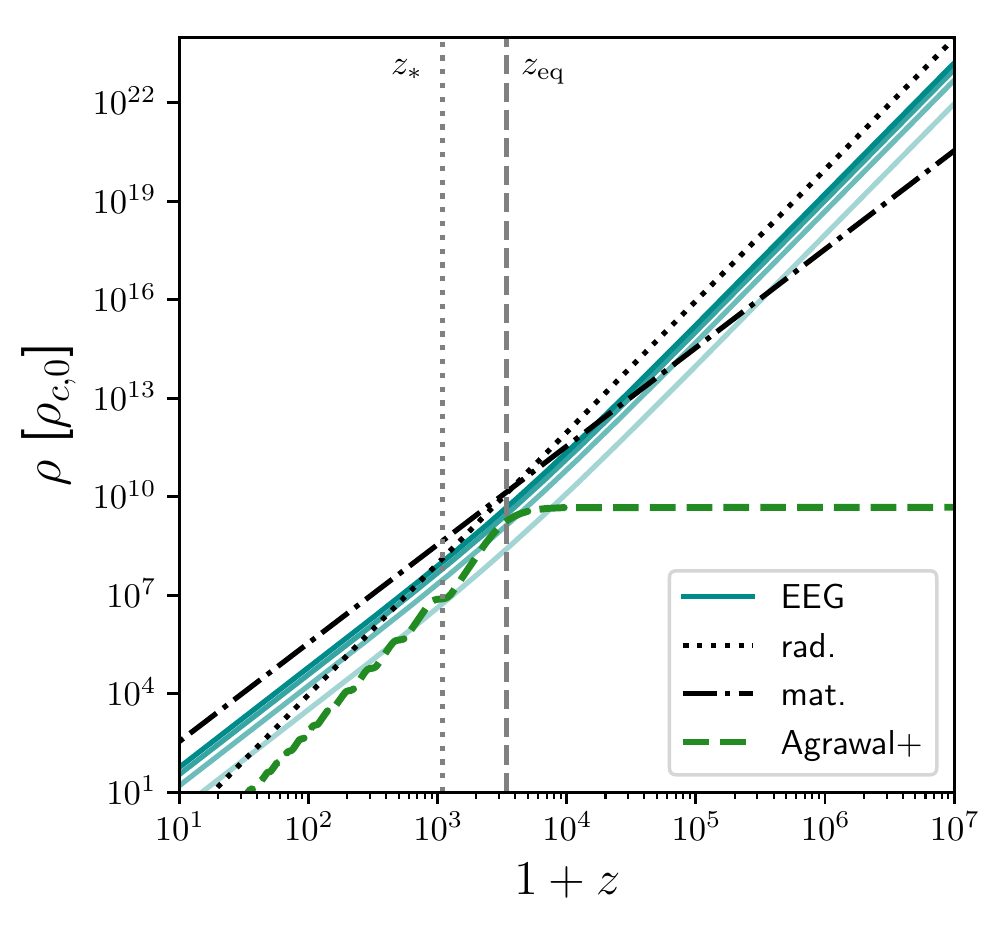}
 \includegraphics[height=4.8cm]{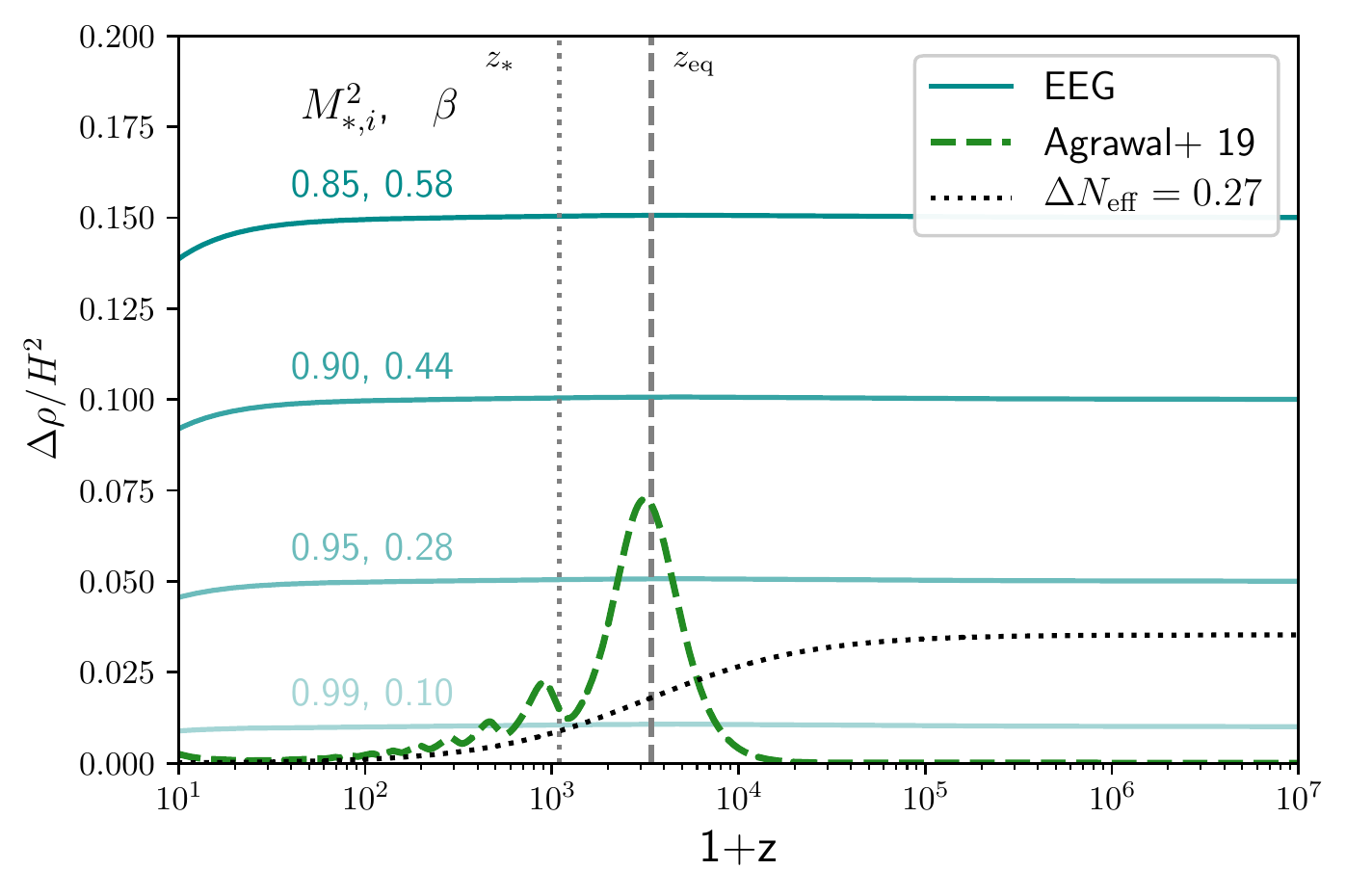}
 \includegraphics[height=4.8cm]{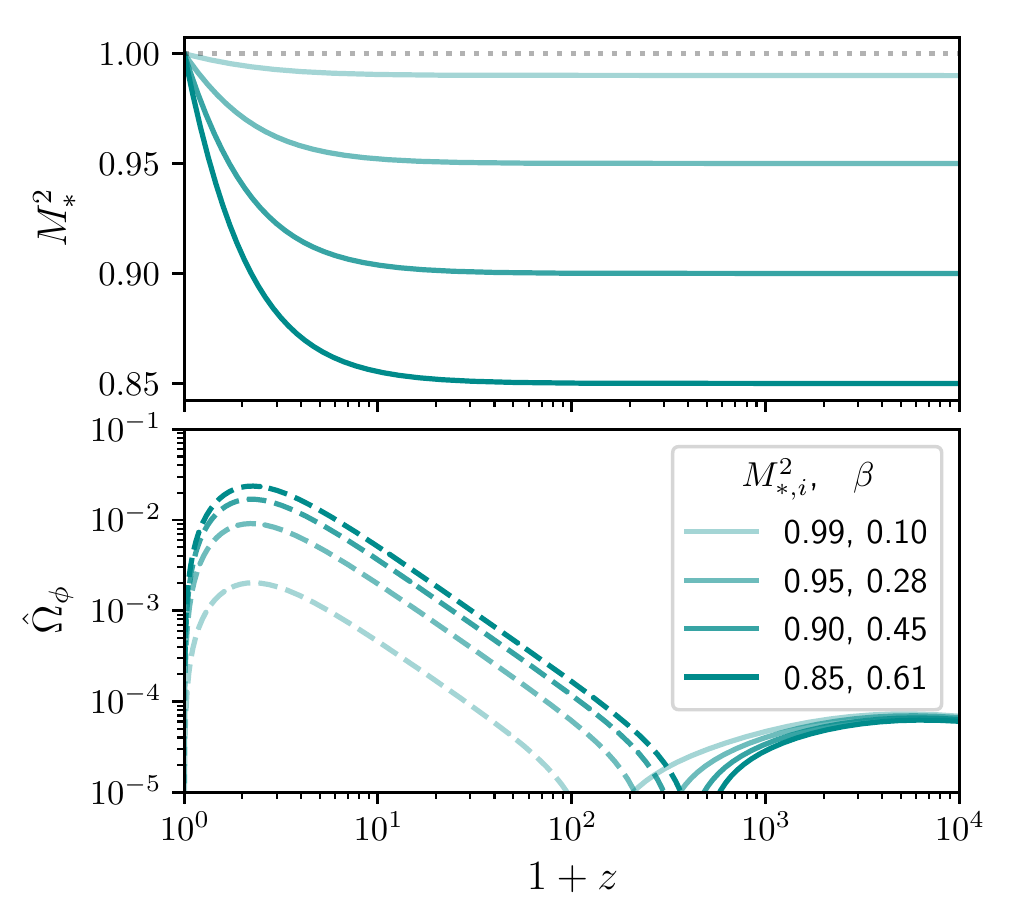}
 \caption{Enhanced Early Gravity (EEG) in canonical coupled models.
 The effective contribution to the expansion history Eq. (\ref{eq:M2_hubble_de}), including effect of $M_*^2$ on cosmic expansion, follows the dominant component at early times (left panel). 
 If $M_{*,i}^2<1$ the strengthening of gravity $\Delta\rho/\rho$ increases the expansion rate before recombination, lowering the acoustic scale and increasing $H_0$ for fixed $\theta_\star$.  Energy contributions of additional relativistic particles and the quintessence early dark energy model \cite{Agrawal:2019lmo} are shown for comparison.
 Right panel: effective Planck mass evolution (top), the coupling $\beta$ is chosen to fix the effective Planck mass today $M_{*,0}^2 = 1$ from the initial value $M_{*,i}^2$.
 Reduced scalar field density (bottom), $\hat\Omega_\phi$ excludes the contributions of $\Lambda$ and the effect of $M_*^2$ on the expansion.
 $H_0$ values are for fixed $\theta_\star$.
 }\label{fig:EGO_dynamcis}
\end{figure*}

Enhanced Early Gravity (EEG) consists of a time modulation of the effective Planck mass due to the scalar field dynamics and its coupling to curvature.  
At early times the strength of gravity is enhanced by a constant factor, as the cosmological Vainshtein mechanism prevents any significant evolution of $\phi$. 
At late times, the scalar's time variation weakens the strength of gravity, with potentially detectable signatures in local gravity and the large-scale structure of the Universe.

In EEG  models, the initial effective Planck mass affects the expansion rate at early times. This effect changes the expansion rate by
\begin{equation}\label{eq:deltaH_M2i}
 \frac{\Delta H}{H} \approx M_*^{-1} - 1\,,
\end{equation}
where $M_*=C(\phi)$ and other contributions to the Galileon energy density, including IDEE, have been neglected $\hat{\Omega}_\phi\sim 0$.
At a fixed matter content, reducing the Planck mass $M_*^2<1$ increases the expansion rate, in turn reducing the acoustic scale $r_s$, Eq. (\ref{eq:acoustic_scale}).

A successful EEG  model requires the strength of gravity to decrease between the early and the late universe. The effective Planck mass affects all scales in the homogeneous universe, including cosmological distances, e.g. the comoving angular diameter distance
\begin{equation}
 D_M(z) = \int_0^{z}\frac{dz'}{H(z')} 
 = \int_0^{z'}\frac{dz'}{M_*\sqrt{\rho + \hat{\mathcal{E}}}} \,.
\end{equation}
Thus, if $M_*^2$ were constant throughout, the angular diameter distance would be modified by the same multiplicative factor as the acoustic scale. This constant factor would cancel on the angular scale 
\begin{equation}\label{eq:theta_star_def}
\theta_* \equiv \frac{r_s(z_*)}{D_M(z_*)}\,,
\end{equation}
leaving the value of $H_0$ obtained from the CMB unchanged.
Decreasing $r_s(z_*)$ relative to $D_M(z_*)$ requires a positive coupling $\beta>0$. 
Ultimately, EEG  works because the same sign of the coupling strength $\beta$ required to increase $H_0$ drives the field away from the ghost region, cf. figure \ref{fig:kinetic_structure}.

The cosmological Vainshtein mechanism prevents $M_*^2$ from evolving at early times. Assuming matter domination the shift-charge density solution(\ref{eq:charge_sol_explicit}) is
\begin{equation}
 \mathcal{J} = \beta H(a)\,.
\end{equation}
where I have neglected any initial shift-charge (or equivalently $\hat\Omega_\phi\sim 0$). 
Cosmological Vainshtein screening occurs when the cubic Galileon term dominates, in which case the above shift-charge density translates to
\begin{equation}\label{eq:phi_dot_matter_screened}
 {\dot\phi} \approx \sqrt{\frac{\beta}{-6 c_3}}H_0 \,,\qquad (c_3\xi\gg c_2)\,,
\end{equation}
The scalar evolution is very suppressed compared to the characteristic evolution scale of other species, set by $H\gg H_0$. For this reason the coupling is extremely ineffective in giving the scalar field an initial velocity, as discussed in appendix \ref{sec:dynamics_ic}.
In contrast, the unscreened regime for canonical kinetic term corresponds to
\begin{equation}\label{eq:phi_dot_matter_unscreened}
 {\dot\phi} = \frac{\beta}{c_2} {H}\,, \qquad (c_3\xi\ll c_2)\,.
\end{equation}
In that case the scalar evolves at a rate $\propto H$ set by cosmic expansion. 
Note that the above expression applies to canonical models: in accelerating models the derivative of the field is set by the non-trivial tracker solution (\ref{eq:gal_solutions}).

The evolution of the scalar field leads to a \textit{running of the effective Planck mass}
\begin{equation}\label{eq:alpha_M}
 \alpha_M \equiv \frac{d\log(M_*^2)}{d\log(a)} = \beta\frac{\dot\phi}{H}\,,
\end{equation}
where the second equality corresponds to the exponential coupling.
Matter-domination solution in the screened regime (\ref{eq:phi_dot_matter_screened}) 
\begin{equation}\label{eq:alpha_M_c3}
 \alpha_M = \sqrt{\frac{\beta^3}{-6c_3}}\frac{H_0}{H} \,,\qquad (c_3\xi\gg c_2)  \,,
\end{equation}
leads to a negligible running at early times, as expected.
The unscreened regime (\ref{eq:phi_dot_matter_screened}) for canonical models
\begin{equation}\label{eq:alpha_M_c2}
 \alpha_M = \frac{\beta^2}{c_2} \,,\qquad (c_3\xi\ll c_2)  \,,
\end{equation}
leads to a constant running of $M_*^2$ in the matter era.

$\alpha_M$ is a standard parameterization of the impact of deviations from GR on cosmic structure formation. Just as a constant $M_*^2$ has no effect on background observables (cf. Eq. \ref{eq:theta_star_def}), a constant $M_*^2$ can be compensated by rescaling the abundances of all matter species so that $\Omega_i/M_*^2$ is constant, leading to no net effect on the perturbations \cite{Bellini:2014fua}. 
A running of the Planck mass produces deviations from GR in structure formation, potentially observable on the LSS of matter and the CMB.

Unscreened evolution (\ref{eq:alpha_M_c2}) is expected at intermediate and low redshifts, leading to effects in LSS and secondary CMB anisotropies.
Allowing $\alpha_M$ to affect early evolution and primary CMB requires very low values of $c_3\lesssim 10^{-9}$ for $|c_2|=1$ (cf. figure \ref{fig:GNdot}). 
Since this work is focused mainly on the CMB, I will set $c_3 = -1$ in the canonical models with $\Lambda \neq 0$ (in accelerating models it is set by $\Omega_{\phi,0}$).
Note that Brans-Dicke theories without Vainshtein screening also produce EEG, leading to a degeneracy between the coupling strength and $H_0$ \cite[Fig. 7]{Umilta:2015cta} (see also \cite{Ballardini:2016cvy,Rossi:2019lgt,Ballesteros:2020sik,Braglia:2020iik}). The field begins evolving at matter radiation equality in those models, when the coupling to curvature overcomes the Hubble friction.

The value of the scalar field is also related to the strength of gravity measured on small scales, including the Solar System. The potential to test EEG  using precision tests of GR as well as the difficulties in modeling the connection between cosmological and small scales will be discussed in Section \ref{sec:challenges_local}. 
While a full investigation of these issues is beyond the scope of this work, I remind the reader that all expressions in this section refer to the \textit{cosmological} evolution of the effective Planck mass.

\subsection{Late-Universe Dynamics} \label{sec:dynamics_late}

The late-time dynamics of Galileons are determined mostly by the sign of the quadratic kinetic term $c_2$. In accelerating models $c_2<0$ the stable solution (\ref{eq:gal_solutions}) corresponds to a growing $\hat\Omega_\phi$ and leads to Late-Universe Phantom Expansion (LUPE). 
In canonical models $c_2>0$ the stable solution (\ref{eq:gal_solutions}) corresponds to a trivial configuration $\hat\Omega_\phi\to 0$.

\begin{figure}[t!]
 \includegraphics[width=\columnwidth]{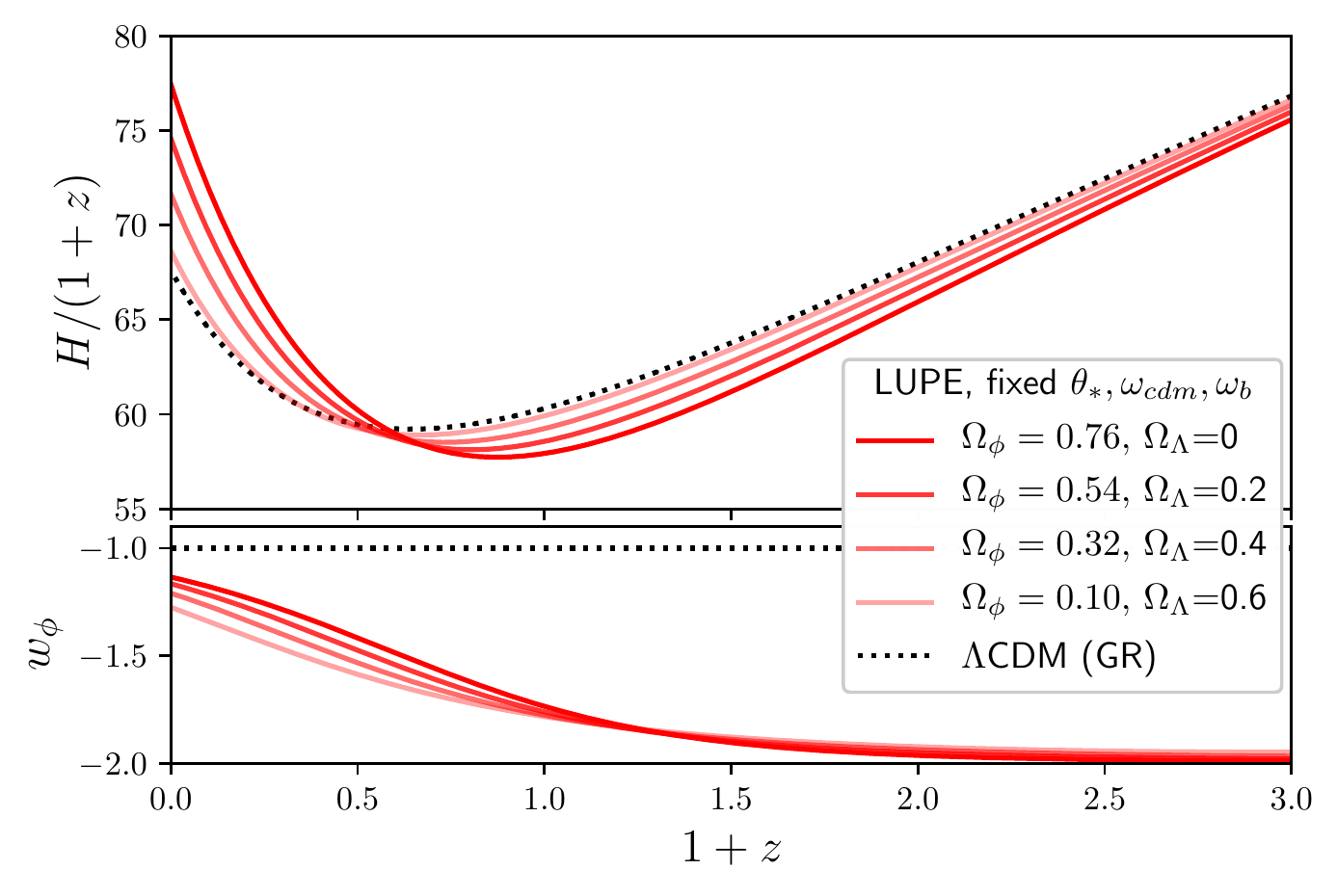}
 \caption{Late-Universe Phantom Expansion (LUPE). The energy density of accelerating Galileons ($c_2<0$) grows at low $z$, increasing the Hubble rate for fixed $\theta_*$. The curves assumed minimal neutrino mass, $\Sigma m_\nu\sim 0.6$eV is needed to reproduce the SH0ES result ($\Lambda=0$). The LUPE contribution to the equation of state (lower panel) to the dark energy is weighted by $\Omega_{\phi}$ in models with $\Lambda\neq 0$.
 }
 \label{fig:canonical_vs_accelerating}
\end{figure}

Accelerating models are very efficient at producing dark energy. The non-trivial \textit{tracker} solution with $\hat\Omega_{\phi,0}> 0$ in Eq. (\ref{eq:gal_solutions}) is stable. 
This solution is characterized by a growing scalar field velocity
\begin{equation}
 \xi\approx \text{const}\,,\quad \dot\phi\propto \frac{1}{H}\,,
\end{equation}
where the coupling has been neglected ($\beta\sim 0$). 
With this solution the scalar kinetic energy 
\begin{equation}
 \hat\Omega_\phi \approx \xi^2\left(\frac{H_0}{H}\right)^4\left(\frac{c_2}{6}-2c_3\xi\right) 
 - C^\prime \xi\left(\frac{H_0}{H}\right)^2
\end{equation}
rapidly dominates the energy budget (see figure \ref{fig:canonical_vs_accelerating}).

No cosmological constant is needed in accelerating models. Instead, the dark energy fraction today can be obtained by choosing the ratio of $c_2,c_3$ corresponding to the tracker solution in Eq. (\ref{eq:gal_solutions}), corrected by contributions due to the coupling, cf. Eq. (\ref{eq:Omega_de_coupled}).
Because the dark energy density grows (instead of being constant) $w_\phi < -1$, a larger value of $H_0$ can be obtained for fixed distance to the last-scattering surface.
This is the reason why Galileon models with $\Lambda=0$ predict a Hubble constant well above typical $\Lambda$CDM values, requiring sizeable neutrino masses $\sum m_\nu\sim 0.6$eV to both give a good fit and avoid a too-high value of $H_0$ (see appendix \ref{app:uncoupled_lupe_Lambda}).
Because of their interest as DE models, the late-time dynamics of accelerating Galileons have been studied extensively in previous works, e.g. Refs. \cite{DeFelice:2010pv,Barreira:2012kk,Barreira:2014jha,Leloup:2019fas}.

In canonical models the energy density of the scalar field decreases very fast once the quadratic term dominates the dynamics. This dynamical regime, known as \textit{kination}, is characterized by a rapid loss of kinetic energy of the field 
\begin{equation} \label{eq:canonical_kination}
\dot\phi \propto \mathcal{J}\propto  a^{-3}\,, \quad  \rho_{\phi}\sim  \dot\phi^2/2\sim a^{-6}\,,
\end{equation}
where the coupling has been neglected ($\beta\sim 0$). 
This loss of energy will continue until the coupling term becomes dominant. In uncoupled models it will evolve towards the trivial vacuum $\dot\phi=0$, $\Omega_{\phi,0}=0$, as anticipated in the solution Eqs. (\ref{eq:gal_solutions}). 
Uncoupled canonical models can thus provide only negligible amounts of dark energy in the late universe, requiring an additional cosmological constant to produce acceleration.
If the coupling is nonzero, the field will stabilize at a non-zero value of the shift-charge as described in section \ref{sec:dynamics_coupling}.

Canonical models with $\Lambda\neq0$ retain the freedom to set $c_2/c_3$ even after using up the scalar field rescaling (\ref{eq:field_rescaling}). 
This ratio determines onset of the kination phase, which begins when
\begin{equation}
 \xi \sim \frac{c_2}{6c_3}\,,
\end{equation}
Lowering $c_3$ allows for a conformal coupling ($\beta\neq0$) to play a role at earlier times, by weakening the cosmological Vainshtein screening (cf. section \ref{sec:dynamics_coupling}). Values $c_3\gtrsim 10^{-9}$ ensure kination occurs after recombination, and thus that the primary CMB is only affected by IDEE and EEG, as described in the above sections.
For these reasons, I will set $c_3=-1$ in this analysis. Some of the consequences of varying $c_3$ are shown in figure \ref{fig:GNdot}, but a more detailed study of the role of $c_3$ is left for future work.

\section{Cosmological Constraints}\label{sec:constraints}

This section presents tests of different solutions to the $H_0$ problem, as implemented in coupled cubic Galileon theories. Section \ref{sec:constraints_overview} contains an overview of the models, data and methods used. 
Section \ref{sec:constraints_uncoupled} presents the limits on IDEE and uncoupled models.
Section \ref{sec:constraints_coupled}) discusses EEG  in canonical coupled models.
Section \ref{sec:constraints_late} addresses the status LUPE in accelerating models, the role of the coupling at late-times and prospects to reduce the tension between Planck and weak lensing surveys.
Appendix \ref{app:uncoupled_lupe_Lambda} discusses uncoupled LUPE models.

\subsection{Overview of Models, Datasets and Analysis}\label{sec:constraints_overview}

The models under study can be classified along two separate properties:
\begin{itemize}
 \item By coupling, into uncoupled $\beta=0$ and coupled $\beta\neq0$.
 Uncoupled models can impact $r_s$ only via IDEE, coupled models produce EEG  (cf. sections \ref{sec:dynamics_idee}, \ref{sec:dynamics_coupling}).

 \item By kinetic term sign, into canonical ($c_2>0$) and accelerating ($c_2<0$). 
 Canonical models require $\Lambda\neq 0$. 
 Accelerating models produce LUPE and need either $\Lambda\neq0$ or $\sum m_\nu\gg 0.06$ eV if $\Lambda=0$ (cf. section \ref{sec:dynamics_late}).
\end{itemize}
I will consider several combinations of models and datasets, as shown in Table \ref{tab:models_vs_data}. Uncoupled, LUPE, $\Lambda\neq0$ models are discussed in appendix \ref{app:uncoupled_lupe_Lambda}.

\begin{table}
 \begin{tabular}{l  @{\quad} c  @{\quad} c @{\quad} c}
 & Uncoupled  &  Coupled  &  \\
  & (IDEE)  &  (EEG)  &  \\
  & $\Omega_{\phi,i}$ & $M_{*,i}^2,\, \beta$ & Section\\
  \hline
\multirow{2}{*}{Canonical + $\Lambda$}
& PB, PBS &  & \ref{sec:constraints_uncoupled} \\
   &  & PB, PBS & \ref{sec:constraints_coupled} \\
 \hline
Accelerating 
(LUPE) & & &  \\ 
\quad  + $\sum m_\nu$\quad ($\Lambda=0$) & PB & PB &  \multirow{2}{*}{\ref{sec:constraints_late}}\\
 \quad + $\Lambda$\qquad ($\sum m_\nu$ fixed) & - & PB &  \\
 \hline
 \end{tabular}
\caption{Overview of coupled cubic Galileon, Eq. (\ref{eq:cccg_Gfuns}) model and datasets combinations. The PB, PBS parameter constraints are shown in tables \ref{tab:model_parameters_PB}, \ref{tab:model_parameters_PB}, respectively, including reference $\Lambda$CDM results. Uncoupled accelerating $\Lambda\neq0$ models are discussed in appendix \ref{app:uncoupled_lupe_Lambda}.}
\label{tab:models_vs_data}
\end{table}

\begin{table*}
 \begin{tabular}{c|c|cc|c|cc}  & $\Lambda$CDM & Unc. can. + $\Lambda$ & Coup. can. + $\Lambda$ & Coup acc. + $\Lambda$ & Unc. acc + $m_\nu$ & Coup. acc + $m_\nu$\\  & GR & IDEE only & + EEG & + EEG, LUPE & + LUPE & + EEG, LUPE\\ \hline$H_0$ & $67.72 \pm 0.46$\, & $67.84 \pm 0.49$\, & $68.7 \pm 1.5$\, & $69.1 \pm 1.4$\, & $71.97 \pm 0.71$\, & $71.5 \pm 1.3$\,\\$100\,\omega_b$ & $2.242 \pm 0.014$\, & $2.242 \pm 0.014$\, & $2.244 \pm 0.021$\, & $2.23 \pm 0.02$\, & $2.228 \pm 0.014$\, & $2.224 \pm 0.017$\,\\$\omega_{cdm}$ & $0.119 \pm 0.001$\, & $0.1195 \pm 0.0011$\, & $0.1200 \pm 0.0012$\, & $0.1205 \pm 0.0013$\, & $0.1203 \pm 0.0012$\, & $0.1204 \pm 0.0013$\,\\$\tau_{\rm reio}$ & $0.0554 \pm 0.0078$\, & $0.0552 \pm 0.0079$\, & $0.0553 \pm 0.0084$\, & $0.0532 \pm 0.0074$\, & $0.0507 \pm 0.0078$\, & $0.0506 \pm 0.0079$\,\\$n_s$ & $0.9664 \pm 0.0038$\, & $0.9668 \pm 0.0039$\, & $0.9683 \pm 0.0074$\, & $0.9614 \pm 0.0068$\, & $0.964 \pm 0.004$\, & $0.9626 \pm 0.0057$\,\\$\sigma_8$ & $0.8086 \pm 0.0074$\, & $0.8092 \pm 0.0073$\, & $0.82 \pm 0.01$\, & $0.836 \pm 0.016$\, & $0.792 \pm 0.022$\, & $0.788 \pm 0.025$\,\\$\sum m_\nu$& $0.06$& $0.06$& $0.06$& $0.06$ & $0.576 \pm 0.082$\, & $0.590 \pm 0.097$\,\\\hline$\Omega_{\phi,0}$& $-$ & $<3 \cdot 10^{-7}$ (95\%) & $-0.023 \pm 0.056$\, & $0.110 \pm 0.082$\, & $0.7128 \pm 0.0072$\, & $0.708 \pm 0.012$\,\\{\small $\log_{10}(\hat\Omega_{\phi,i})$}& $-$ & $<-5.22$ (95\%) & $<-5.26$ (95\%) & $<-5.33$ (95\%) & $<-5.44$ (95\%) & $<-5.03$ (95\%)\\$\beta$& $-$& $0$ & $0.135 \pm 0.099$\, & $0.08 \pm 0.07$\,& $0$ & $-0.002 \pm 0.024$\,\\$M_{*,i}^2$& $-$& $1$ & $0.988 \pm 0.035$\, & $1.015 \pm 0.031$\,& $1$ & $1.012 \pm 0.025$\,\\\hline 
$-\log(\mathcal{L})$ & $1388.08$ & $-0.81$ & $-0.47$ & $-0.57$ & $+9.80$ & $+10.03$\\\hline \end{tabular}
\caption{Planck + BAO marginalized constraints on cosmological and Galileon parameters. 
Quantities show mean and 68\% confidence level, upper limits correspond to 95\% confidence.
$\Lambda$CDM and uncoupled canonical are practically indistinguishable due to the stringent bounds on IDEE. 
In EEG  models the coupling increases the uncertainty on $H_0$ by a factor $\sim 3$ in coupled models with $\Lambda\neq 0$ (canonical and accelerating) slightly increasing the central value as well.
$\Lambda=0$ accelerating models (coupled and uncoupled) predict a high value of $H_0$, but have a bad fit and are ruled out by other observations \cite{Renk:2017rzu}. 
Note that the central value of the initial effective Planck mass is $M_{*,i}^2\sim 1$: CMB+BAO data has no preference in the absence of late-universe information.
The last line shows the best fit log-likelihood for the reference $\Lambda$CDM and differences for each Galileon: all models with $\Lambda\neq0$ have a slightly better fit, while accelerating models with $\Lambda=0$ are disfavoured.
\label{tab:model_parameters_PB}
} 
\end{table*}

\begin{table}
\begin{tabular}{c|c|cc}  & \, $\Lambda$CDM\, & \, Unc. can. + $\Lambda$\, & \, Coup. can. + $\Lambda$\,\\  & GR & IDEE & + EEG\\ \hline$H_0$ & $68.42$ & $68.75$ & $70.37$\\$100\,\omega_b$ & $2.265$ & $2.264$ & $2.259$\\$\omega_{cdm}$ & $0.1180$ & $0.1183$ & $0.1183$\\$\tau_{\rm reio}$ & $0.0550$ & $0.0533$ & $0.0585$\\$n_s$ & $0.9670$ & $0.9691$ & $0.9778$\\$\sigma_8$ & $0.8016$ & $0.8023$ & $0.814$\\\hline$\Omega_{\phi,0}$& $-$ & $10^{-7}$ & $0.032$\\{\small $\log_{10}(\hat\Omega_{\phi,i})$}& $-$ & $-5.53$ & $-6.67$\\$\beta$& $-$& $0$ & $0.09$\\$M_{*,i}^2$& $-$& $1$ & $0.955$\\\hline 
$-\log(\mathcal{L})$ & $1397.53$ & $-2.04$ & $-4.07$\\\hline \end{tabular} 
\caption{
Best fit Planck + BAO + SH0ES cosmological and Galileon parameters.
These are shown instead of the marginalized constraints because the datasets are in tension.
The last line shows the best fit log-likelihood for the reference $\Lambda$CDM and differences for each Galileon model, for the PBS datasets.
\label{tab:model_parameters_PBS}
} 
\end{table}

IDEE is produced by the initial field velocity $\dot\phi_i$. This is specified via a flat prior on the initial dark energy abundance $\log_{10}(\hat\Omega_{\phi,i})\in [-8,0]$, cf. Eq. (\ref{eq:idee_energy_fraction}) evaluated at $z_i = 10^{10}$ (around the BBN era). 
The lower limit in the logarithmic prior of $\Omega_{\phi,i}$ is indistinguishable from $\Lambda$CDM, while the upper limit corresponds to the scalar field dominating the energy budget in the radiation era.
The initial field velocity will be varied freely for all models presented below.

EEG  relies on the initial value of the scalar $\phi_i$, which is approximately constant at early times, cf. section \ref{sec:dynamics_ic}. I will set $\phi_i$ through a flat prior on the initial Planck mass $M_{*,i}^2 = e^{\beta\phi_i} \in (0,\infty)$, where $M_{*,i}^2>0$ is necessary for the stability of tensor perturbations. 
Since $M_{*,i}^2 = 1+\beta\phi_i+\mathcal{O}(\phi_i^2)$, a prior on $M_{*,i}^2$ is equivalent to a prior on the initial condition for small deviations in the strength of gravity.
In uncoupled models I set $\phi_i=0$, as the initial value is irrelevant due to shift symmetry.
$M_{*,i}^2$ and the coupling strength $\beta$ will be varied freely for all coupled models.

The coupling strength is varied in the range $\beta\in[-0.5,\infty)$.%
\footnote{The Planck+BAO analysis of canonical models included an upper limit $\beta\in[-0.5,0.5]$. This was removed in the Planck+BAO+$H_0$ analysis for which $\beta\in[-0.5,\infty)$. Both analyses yield very similar bounds on $\beta$, suggesting that the more restrictive prior was broad enough, cf. section \ref{sec:constraints_late}.}
Ghost instabilities can occur for negative coupling $\beta <0$ (figure \ref{fig:kinetic_structure}), the prior allows the data to explore that region as well.
Note that $\beta$ could be set instead by fixing the final effective Planck mass $M_{*,0}^2$. In this analysis I will not be concerned about $M_{*,0}^2$, deferring the issue to the discussion of local gravity tests and GW-induced instabilities in sections \ref{sec:challenges_local}, \ref{sec:challenges_GWs}.

Galileon coefficients govern the low redshift Galileon dynamics, including LUPE.
In canonical models the scalar field is normalized to $c_2=+1$ and the cubic coupling is fixed to $c_3=-1$ to simplify the pre-recombination dynamics, cf. sections \ref{sec:dynamics_coupling} and \ref{sec:dynamics_late}.
In accelerating models the values of the Galileon coefficients $c_2,c_3$ are fully fixed by normalization of the field and fixing the scalar field abundance today $\Omega_{\phi,0}$.

Accelerating $\Lambda = 0$ LUPE models require sizable neutrino masses \cite{Barreira:2014jha}. In those cases I will vary $m_\nu \in (0,\infty)$ assuming a degenerate hierarchy.
Neglecting the neutrino mass splittings has negligible differences in cosmological predictions, note that the total mass required in LUPE $\Lambda=0$ models, $\sum m_\nu \approx 0.6$eV \cite{Renk:2017rzu} is significantly larger than both the minimal mass and the mass allowed assuming $\Lambda$CDM \cite{Giusarma:2016phn,Vagnozzi:2017ovm} (see Ref. \cite{Peirone:2017vcq} for analysis of uncoupled Galileons using different hierarchies).
All other cases will assume a single massive neutrino with minimal mass $m_\nu = 0.06$eV.

Other cosmological parameters were chosen following the Planck analyses \cite{Aghanim:2018eyx}. 
I will assume the universe to have zero spatial curvature, with the fraction of scalar field energy density $\Omega_\phi$ given by the closure relation $\sum_i \Omega_i = 1$.
The standard cosmological parameters $100\theta_*$ (or $H_0$), $\omega_{cdm}$, $\omega_{b}$, $\ln(10^{10}A_s)$, $n_s$ and $\tau_{\rm reio}\in [0.04,\infty)$ are varied with flat priors unless explicitly stated.  
By default I will consider the Helium fraction $Y_{\rm He}$ to be set by BBN given $\omega_b$ and the expansion rate at early times. I will discuss constraints from light element abundances in section \ref{sec:challenges_bbn}.

To test solutions to the Hubble problem, I will consider CMB data from Planck (P), distances from Baryon Acoustic Oscillations (B) and a prior on the $H_0$ from the SH0ES collaboration (S) in the following combinations: 
\begin{itemize}
 \item Planck + BAO (PB), as the default combination. This determines the model-dependent early-universe inference of $H_0$ and the room to accommodate late-universe measurements. PB results are summarized in table \ref{tab:model_parameters_PB}.
 \item Planck + BAO + SH0ES (PBS), including a distance ladder prior on $H_0$. This analysis will serve to find the global best fit. I will consider this combination in few selected cases. PBS results are summarized in table \ref{tab:model_parameters_PBS}.
\end{itemize}

The CMB data choice follows the Planck 2018 baseline analyses \cite{Aghanim:2018eyx}. It includes high-$\ell$ temperature TT, E-mode polarization EE, their cross correlation (TE) as well as low-$\ell$ TT and EE spectra \cite{Aghanim:2019ame}.
I will not consider the Planck lensing likelihood to focus on testing primary anisotropy effects, as much as possible.
Omitting CMB lensing will not significantly impact uncoupled IDEE results, which are strongly constrained by temperature and polarization alone.
Coupled models can be further constrained by CMB lensing, as late-time dynamics of the scalar field will modify the lensing potential via non-zero $\alpha_M$, Eq. (\ref{eq:alpha_M}). This analysis will be left for future work.

BAO data is necessary for a precise inference of $H_0$, anchoring $r_s$ as determined by the CMB to the late-universe expansion. For BAO data I will use the measurements from galaxy samples from the Baryon Oscillation Spectroscopic Survey (BOSS) data release 12 \cite{Alam:2016hwk} and the low-$z$ sample combining the 6dF survey \cite{Beutler:2011hx} and the main galaxy sample from SDSS data release 7 \cite{Ross:2014qpa}. 
I will use the galaxy BAO data as given, including density field reconstruction. This methodology is conservative for canonical uncoupled models where the late-time dynamics is indistinguishable from $\Lambda$+GR. However, the use of reconstructed data for coupled or accelerating models assumes the validity of reconstruction. This has been tested in simple extensions of $\Lambda$CDM which assume GR \cite{Sherwin:2018wbu,Carter:2019ulk}. However, modified gravity can enhance non-linear effects, including the shift of the BAO scale \cite{Bellini:2015oua}.

As direct $H_0$ measurement I will use the SH0ES project 2019 measurement $H_0 = 74.03 \pm 1.42\, [{\rm km}\,s^{-1}\,{\rm Mpc}^{-1}]$ \cite{Riess:2019cxk}, resulting in a 4.4$\sigma$ tension with Planck+BAO and $\Lambda$CDM. This value relies on a distance-ladder measurement of the expansion rate with improved Cepheid variable star measurements from the Large Magellanic Cloud. The methodology has been shown to be robust by other analyses \cite{Cardona:2016ems,Zhang:2017aqn,Follin:2017ljs,Feeney:2017sgx,Dhawan:2020xmp}. 
Other late-universe measurements of the Hubble parameter exist tend to produce larger values of $H_0$ than Planck+BAO within $\Lambda$CDM that are either in tension (lensing time delays \cite{Wong:2019kwg}) or compatible (standard sirens \cite{Abbott:2017xzu}, tip of the red-giant branch \cite{Freedman:2019jwv}), see Ref. \cite{Verde:2019ivm} and \cite[section 5.4]{Aghanim:2018eyx} for recent overviews.
Adding a prior on $H_0$ serves to find the best-case scenario and its goodness-of-fit in light of all available (although possibly discrepant) datasets.

Type Ia SNe data will not be included in this analysis, but left for future work. In coupled Galileons the interpretation of SNe data requires modeling the variable strength of gravity on small scales and its effect on the intrinsic SNe luminosity, as discussed in section \ref{sec:challenges_local_GN}. Note that a time-variation of SNe luminosity invalidates the inverse standard ladder method (BAO+SNe) of inferring the acoustic scale (figure \ref{fig:master_plot}).
SNe modelling issues are absent in uncoupled Galileons, but SNe will not qualitatively change the conclusions of this analysis.
In canonical uncoupled models ($\Lambda\neq0$) the expansion history is indistinguishable from $\Lambda$CDM at low redshift.
Accelerating uncoupled models (LUPE) are disfavoured by SNe, but the tension can be read directly by comparing the contours with the inverse-distance ladder (BAO+SNe) in figure \ref{fig:master_plot}.

To obtain the theoretical predictions I used the \texttt{hi\_class} code%
\footnote{\url{www.hiclass-code.net}}
\cite{Blas:2011rf,Zumalacarregui:2016pph,Bellini:2019syt}, where the exponentially coupled cubic Galileon model (section \ref{sec:theory_cccg}) was implemented using the covariant Lagrangian approach developed in version 2.0 (see Ref. \cite{Bellini:2019syt} for details).
The parameter space of each model and dataset combination was sampled using a Markov chain Monte Carlo (MCMC) analysis with a Metropolis-Hastings proposal distribution. The sampling relied MontePython (version 3) \cite{Audren:2012wb,Brinckmann:2018cvx}, modified to record errors whenever model predictions can not be computed, such as unstable regions of the parameter space.
To ensure convergence the MCMC runs until the variance across chains over in-chain variance (Gellman-Rubin convergence ratio) is smaller than $0.05$.
The resulting chains were analyzed with MontePython, Getdist \cite{Lewis:2019xzd} and CosmoSlik  \cite{2017ascl.soft01004M}.

\subsection{Uncoupled Models: Autopsy of IDEE}\label{sec:constraints_uncoupled}

\begin{figure}
 \includegraphics[width=0.49\textwidth]{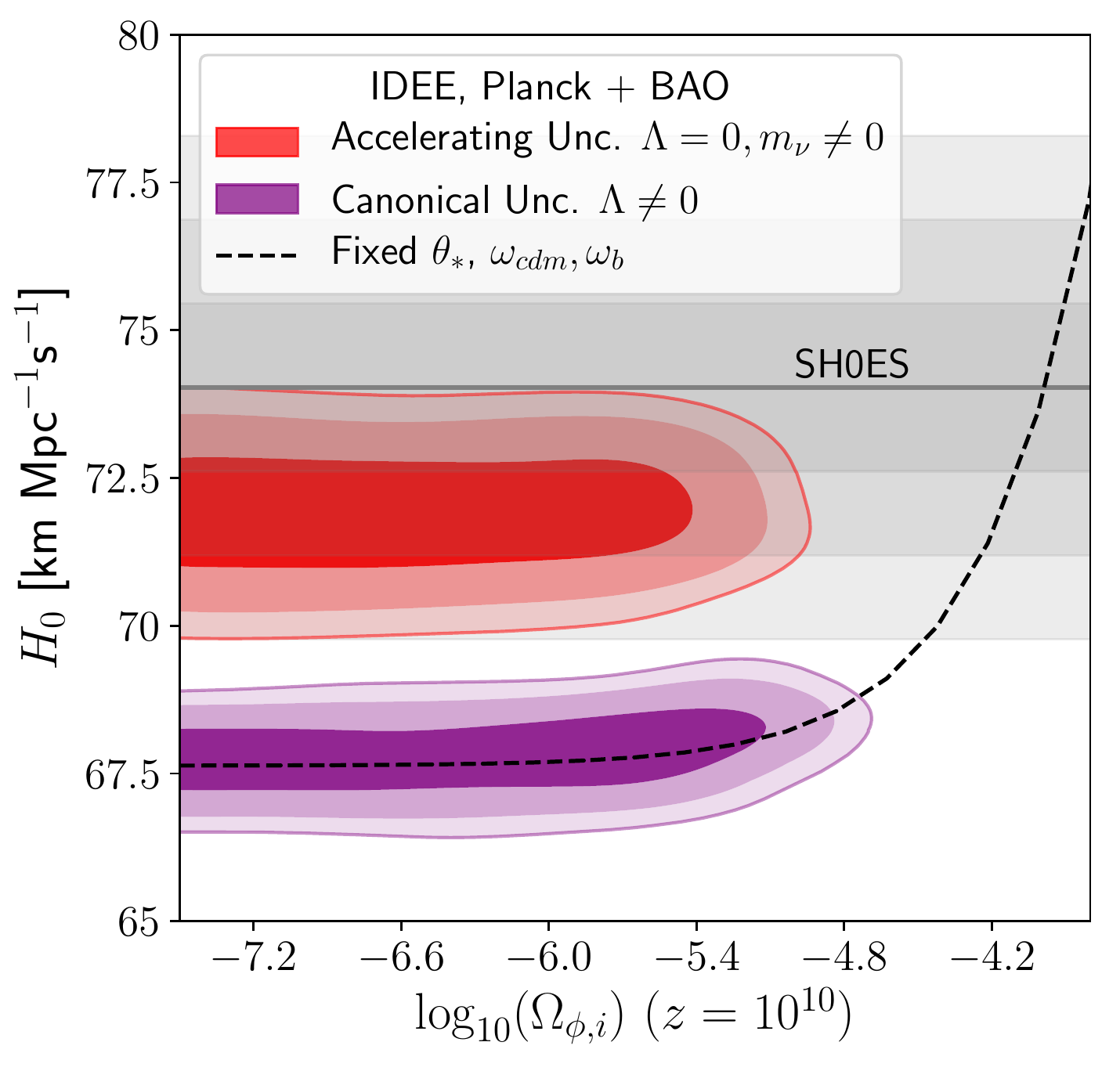}
 \caption{Planck + BAO constraints on the initial IDEE abundance and the Hubble parameter. 
 Contours show 68, 95 and 99\% c.l. posteriors for uncoupled canonical $\Lambda\neq0$ (purple) and accelerating $\Lambda=0,m_\nu>0$ (red) models. 
 The black dashed line shows the effect of IDEE on $H_0$ via the acoustic scale, for fixed cosmological parameters. Gray bands correspond to the distance ladder measurement.
 }\label{fig:idee_bounds}
\end{figure}

IDEE provides a source of early dark energy that peaks in the era of equality. The cosmological limits are too stringent for IDEE to play any role in solving the Hubble problem. One reason is that scalar field perturbations have a period of fast growth that affects the CMB spectrum in a characteristic scale-dependent manner.

\begin{figure*}
\centering
 \includegraphics[width=0.49\textwidth]{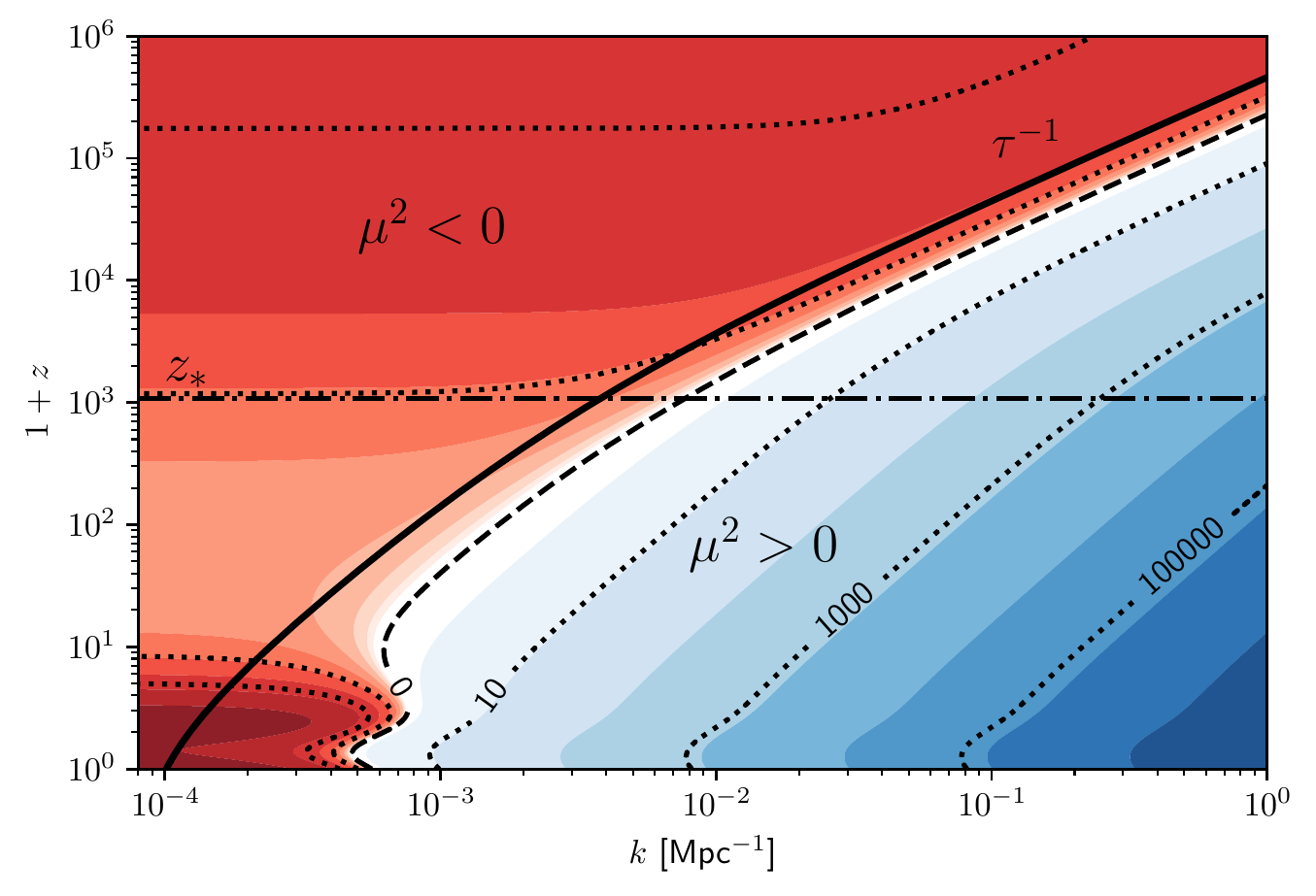}
 \includegraphics[width=0.49\textwidth]{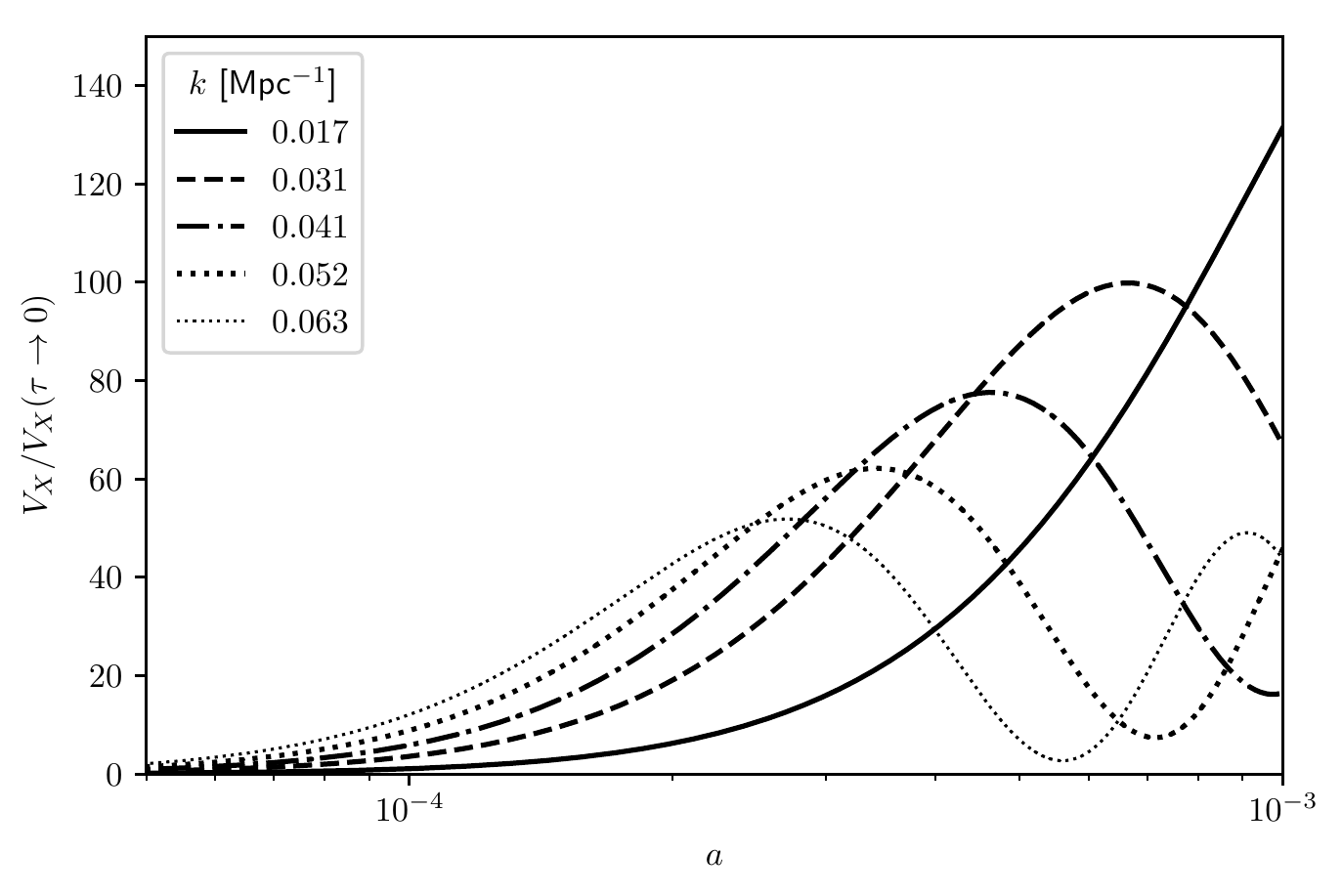}
 \caption{Autopsy of IDEE in uncoupled canonical Galileons with $\Omega_{\phi,i}=10^{-4}$. 
 Left panel: effective squared mass of scalar perturbations (\ref{eq:eff_mass}). Negative values (red region) induce growth of perturbations after horizon crossing (solid line) and before the scalar-field pressure stabilizes the growth (dashed line), leading to oscillatory behaviour (blue region).
 Right panel: evolution of the field perturbation $V_X\equiv \delta\phi/\dot\phi$ for scales corresponding to the first peaks and troughs of the CMB. For the modes dominating the first acoustic peaks, recombination occurs when the modes are still growing due to the tachyon or have barely began oscillating.
}
 \label{fig:idee_autopsy}
\end{figure*}

\begin{figure*}
\includegraphics[width=0.49\textwidth]{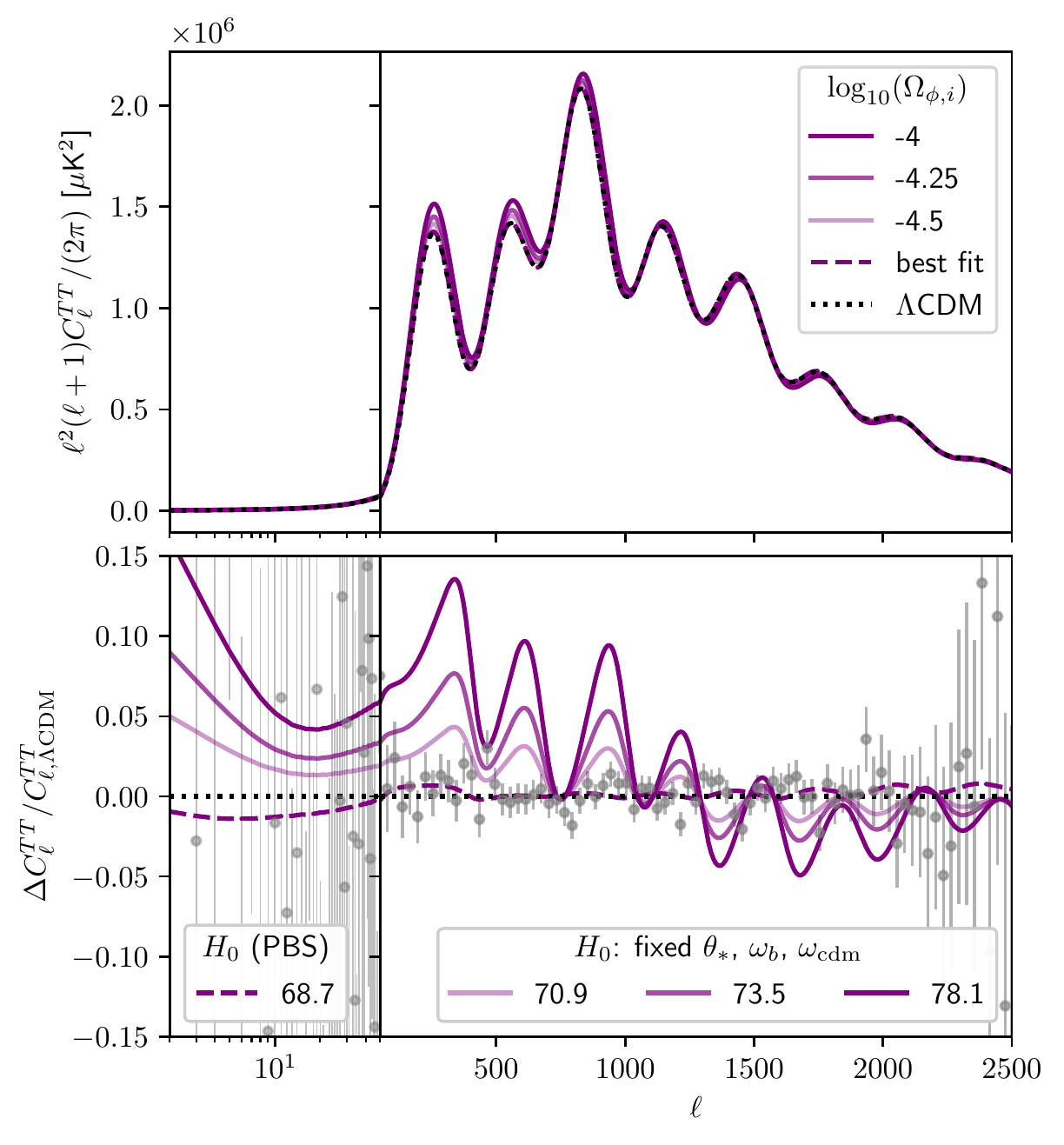}
\includegraphics[width=0.49\textwidth]{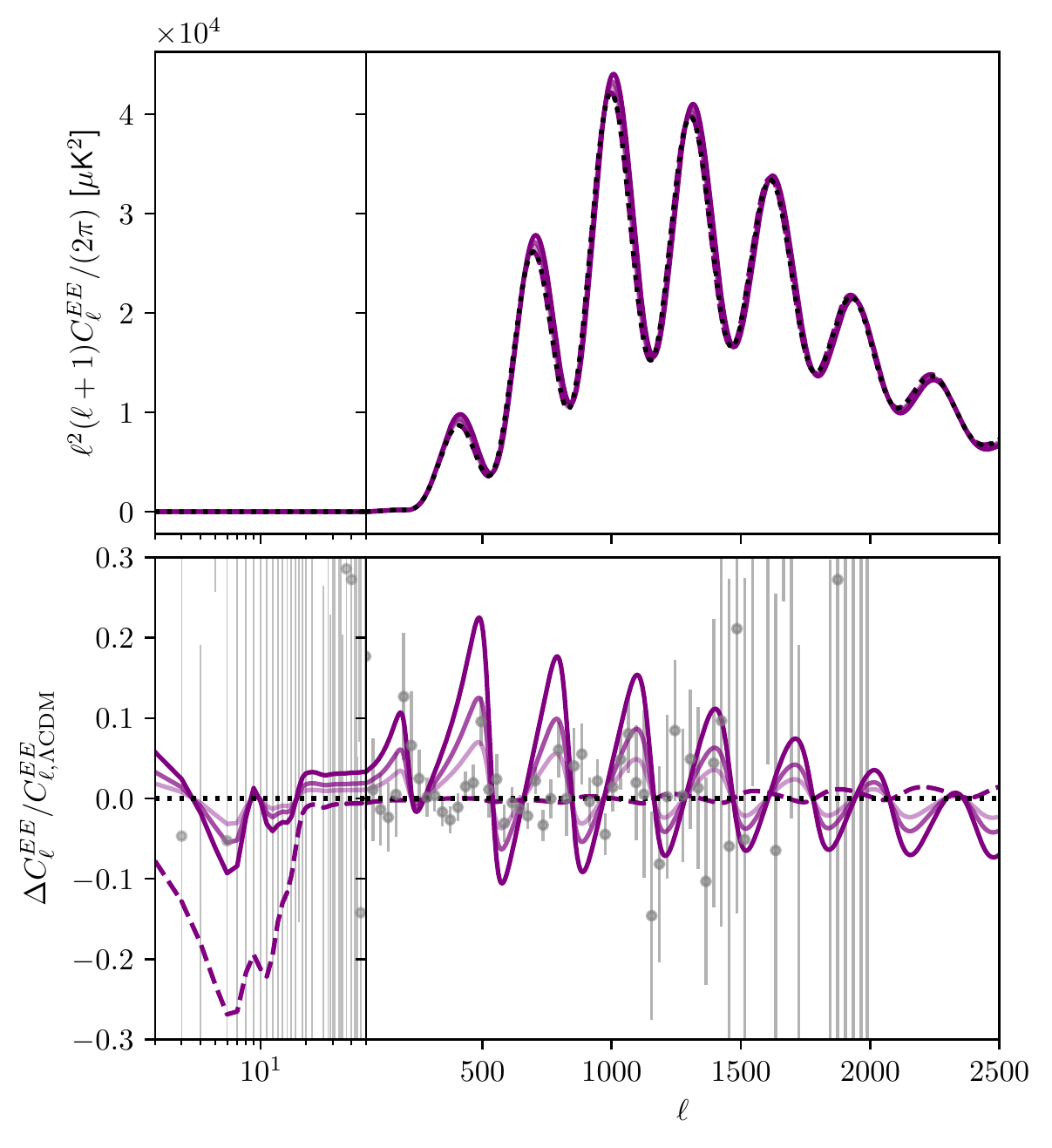}
 \caption{Impact of IDEE on the CMB. Solid lines show the predictions for canonical uncoupled for different values of $\Omega_{\phi,i}$, along with their predictions for the Hubble parameter. $\theta_*$ and other cosmological parameters are fixed. The dashed line shows the best fit to PBS data. Left and right panel show TT and EE spectra, respectively. Residuals (lower panels) are compared to binned Planck data. Low multipoles ($\ell<50$) are shown in logarithmic scale and compared with unbinned Planck data.
 }\label{fig:idee_cmb}
\end{figure*}

The initial fraction of dark energy $\Omega_{\phi,i}$ is constrained by Planck+BAO to the point where its effect the acoustic scale and $H_0$ is negligible. 
Figure \ref{fig:idee_bounds} shows the posteriors marginalized over the initial energy density and the Hubble parameter for uncoupled models. 
The relationship $\Omega_{\phi,i}-H_0$ in the absence of constraints is shown for fixed $\theta_*$ and other cosmological parameters.
Using IDEE to solve the Hubble problem would require $\Omega_{\phi,i}\gtrsim 10^{-4.2}$ in the canonical model (cf. section \ref{sec:dynamics_idee}), while CMB+BAO bounds are at the level of $\Omega_{\phi,i}\lesssim10^{-5.2}$ at 95\% c.l. (see table \ref{tab:models_vs_data}).

The bounds on IDEE make canonical uncoupled models indistinguishable from $\Lambda$CDM. The impact of IDEE in the late universe is bound to be smaller than on the primary CMB, due to the IDEE scaling in the matter era and the late kination phase (cf. figure \ref{fig:canonical_vs_accelerating}). As a consequence, the parameter bounds are almost identical to those of the reference $\Lambda$CDM analysis.
The role of IDEE is also negligible on accelerating uncoupled models. Those cases are indistinguishable from setting $\dot\phi_i$ to the tracker value, Eq. (\ref{eq:gal_solutions}). The fit best fit likelihood to Planck+BAO is significantly worse than the canonical case. Moreover, other analyses rule out the model including CMB$\times$LSS cross-correlations \cite{Kimura:2011td,Renk:2017rzu} and a combination of late-universe datasets \cite{Peirone:2017vcq}.

Including a distance ladder prior on $H_0$ in does not alter these conclusions significantly. The bound on IDEE becomes slightly higher $\log_{10}(\Omega_{\phi,i})<-4.94$ 95\% c.l. for the canonical model. Trying to fit datasets in tension leads to larger shifts on the remaining cosmological models, with a change $\Delta H_0/\sigma_{H_0} = 1.51$ driven mainly by $\Delta n_s/\sigma_{n_s} = 1.05$, $\Delta \omega_{\rm b}/\sigma_{\omega_{\rm b}} = 0.80$, $\Delta \omega_{\rm cdm}/\sigma_{\omega_{\rm cdm}} = -0.82$, as can be seen comparing tables \ref{tab:model_parameters_PB} and \ref{tab:model_parameters_PBS}.

An autopsy of IDEE shows that the strong limits on $\Omega_{\phi,i}$ originate from the growth of the scalar field perturbations around horizon crossing. 
This can be understood by examining the mass-squared for the field fluctuations
\begin{equation}\label{eq:eff_mass}
 \mu^2 = \frac{c^2_s k^2}{a^2H^2} + m_\phi^2 \,,
\end{equation}
where $c_s$ is the scalar sound speed and the time and scale-dependence of the mass is shown in the left panel of figure \ref{fig:idee_autopsy}.
A consequence of cubic Galileon domination is that the scale-independent contribution is negative $m_\phi^2<0$, a feature known as \textit{tachyon instability}. 
Tachyons are associated with growing scalar field perturbation $V_X \equiv \frac{\delta\phi}{\dot\phi} \sim e^{\pm \mu t}$ (i.e. imaginary frequency) on scales larger than the scalar field sound horizon $k < c_s/(aH)$.
For perturbations at a scale $k$, the growth begins around horizon crossing. 
The rate of tachyonic growth is modulated by $|m_\phi^2|$. This is proportional to $\Omega_{\phi}$ and thus enhances the scalar field perturbations significantly before recombination in models able to affect $r_s$.

The growth of scalar perturbations is tamed by the scalar-field pressure, i.e. the scale-dependent term in the effective mass (\ref{eq:eff_mass}). Stability on small scales requires $c_s^2>0$, or equivalently, that small-scale perturbations in the field have oscillatory solutions. This oscillatory regime begins once the sound speed dominates the effective mass, with scalar field perturbations decaying by virtue of the Hubble friction. This transition corresponds to the red/blue border in the left panel of figure \ref{fig:idee_autopsy}.
Different physical scales undergo growth and oscillations at different times, leading to different amplitudes at recombination (figure \ref{fig:idee_autopsy}, right panel). The scale dependence of the field perturbations is transferred to the gravitational potentials and other species via the modified Einstein's equations.

The interplay between growth and oscillations leave a characteristic imprint on the CMB spectra.
The tachyonic growth is largest for modes that enter the horizon soon before recombination and correspond to the first peaks and troughs of the CMB spectra (figure \ref{fig:idee_autopsy}). 
These modes have no time for the oscillatory phase to stabilize their growth, leading to a larger impact on relatively low multipoles in the CMB, as shown in figure \ref{fig:idee_cmb}.
The differences are strongest for the TT spectrum on the larger angular scales ($\ell \lesssim 1000$), particularly on the first peak and through. 
Overall there is an enhancement of the odd peaks (1, 3 and 5) and a suppression of the even peaks (2 and 4). However, this effect can not be compensated adjusting the value $\omega_b$. 
This odd/even pattern is overlaid with an overall suppression of intermediate angular scales ($1000\lesssim \ell \lesssim 2000$), and an enhancement of small angular scales $\ell\gtrsim 2000$. 
The EE polarization spectrum (right panel in figure \ref{fig:idee_cmb}) shows a similar trend, with deviations becoming smaller in higher multipoles.

\subsection{Coupled Models: Viability of EEG}\label{sec:constraints_coupled}

Now I will discuss EEG  models, focusing on the initial effective Planck mass $M_{*,i}^2$, its effect before recombination and its impact on the Hubble rate and other cosmological parameters. 
I will also describe the main features of EEG  and the differences to IDEE and other early dark energy scenarios.
Due to the Vainshtein mechanism both canonical and accelerating models have the same early-time behavior. For this reason, I will focus on canonical models and leave the discussion of both accelerating models and constraints on the coupling $\beta$ for section \ref{sec:constraints_late} below.

\begin{figure}
\centering
\includegraphics[width=0.46\textwidth]{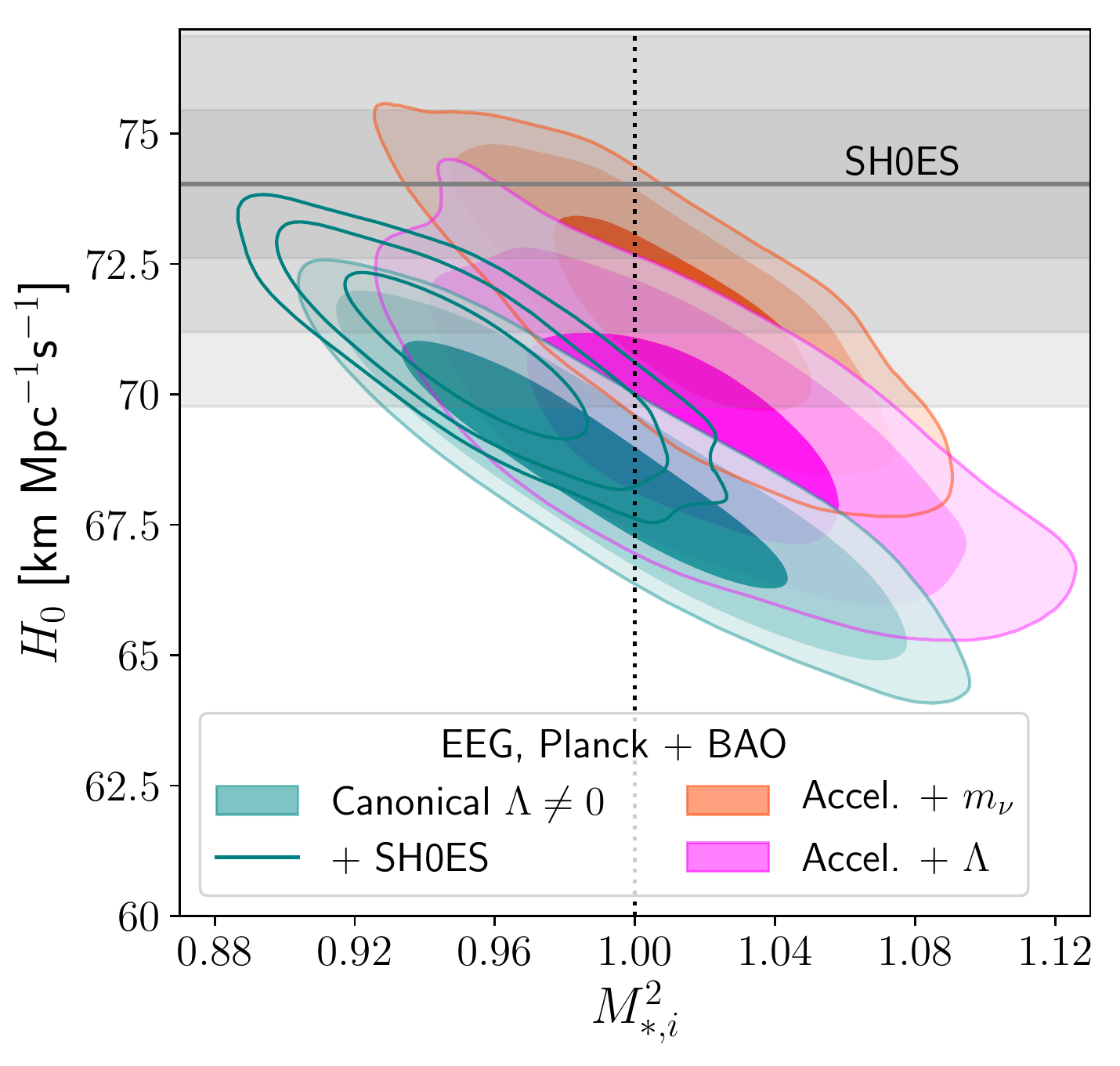}
 \caption{Degeneracy between the initial effective Planck mass $M_{*,i}^2$ and the Hubble parameter for coupled cubic Galileons.  Regions correspond to 68, 95 and 99\% c.l. marginalized posteriors for Planck+BAO (filled), Planck+BAO+$H_0$ (unfilled) and distance ladder measurement of $H_0$ (filled gray). 
 The anti-correlation between $H_0$ and $M_{*,i}^2$ is due mainly to the impact of the early expansion rate on the acoustic scale (cf. section \ref{sec:dynamics_coupling}) and other parameter degeneracies (figure \ref{fig:coupled_triangle}). 
}
 \label{fig:coupled_Mi_H0}
\end{figure}

A non-zero coupling introduces a significant degeneracy between the initial effective Planck mass and the Hubble parameter with the potential to accommodate high values compatible with late-universe constraints. 
Figure \ref{fig:coupled_Mi_H0} shows the marginalized posteriors on the $M_{*,i}^2$-$H_0$ plane, exhibiting the anti-correlation between both quantities, as anticipated in figure \ref{fig:master_plot} and described in section \ref{sec:dynamics_coupling}. 
This relation can be understood as follows: a weakening of gravity at early times $M_{*,i}^2<1$ increases the early expansion rate before recombination and reduces the acoustic scale. Then the same projected CMB scales correspond to a larger value of $H_0$, as long as the late time effective Planck mass is larger than the pre-recombination value (e.g. today $M_{*,0}^2<M_{*,i}^2$), which requires a positive coupling constant $\beta>0$. While the effect of the initial effective Planck mass is the basis of EEG  other parameter degeneracies also play a role in relieving the Hubble tension.

\begin{figure*}
\centering
\begin{tikzpicture}
\node at (0,0){\includegraphics[width=0.68\textwidth]{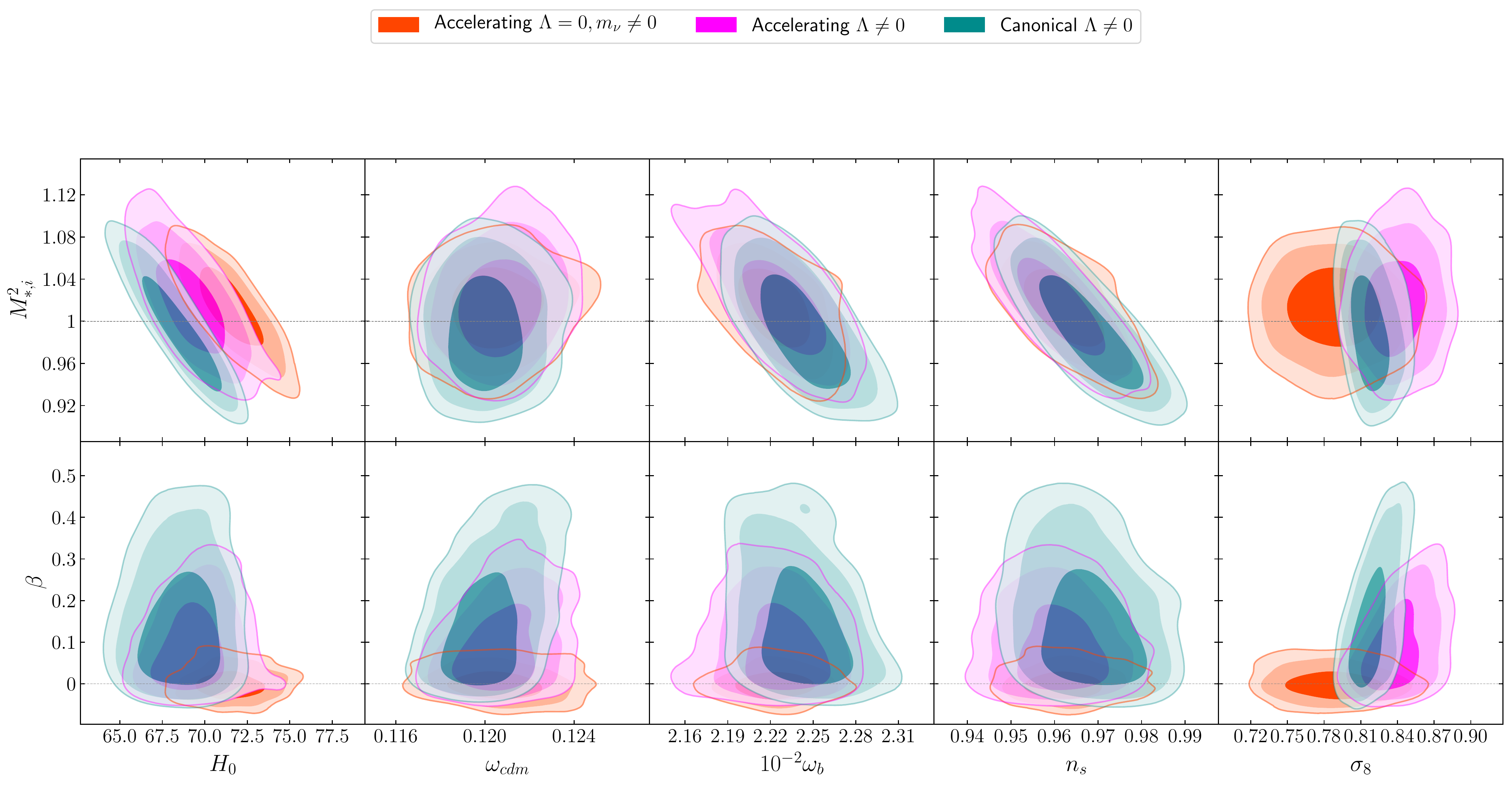}};
 \node at (.15,8.7) {\includegraphics[width=0.71\textwidth]{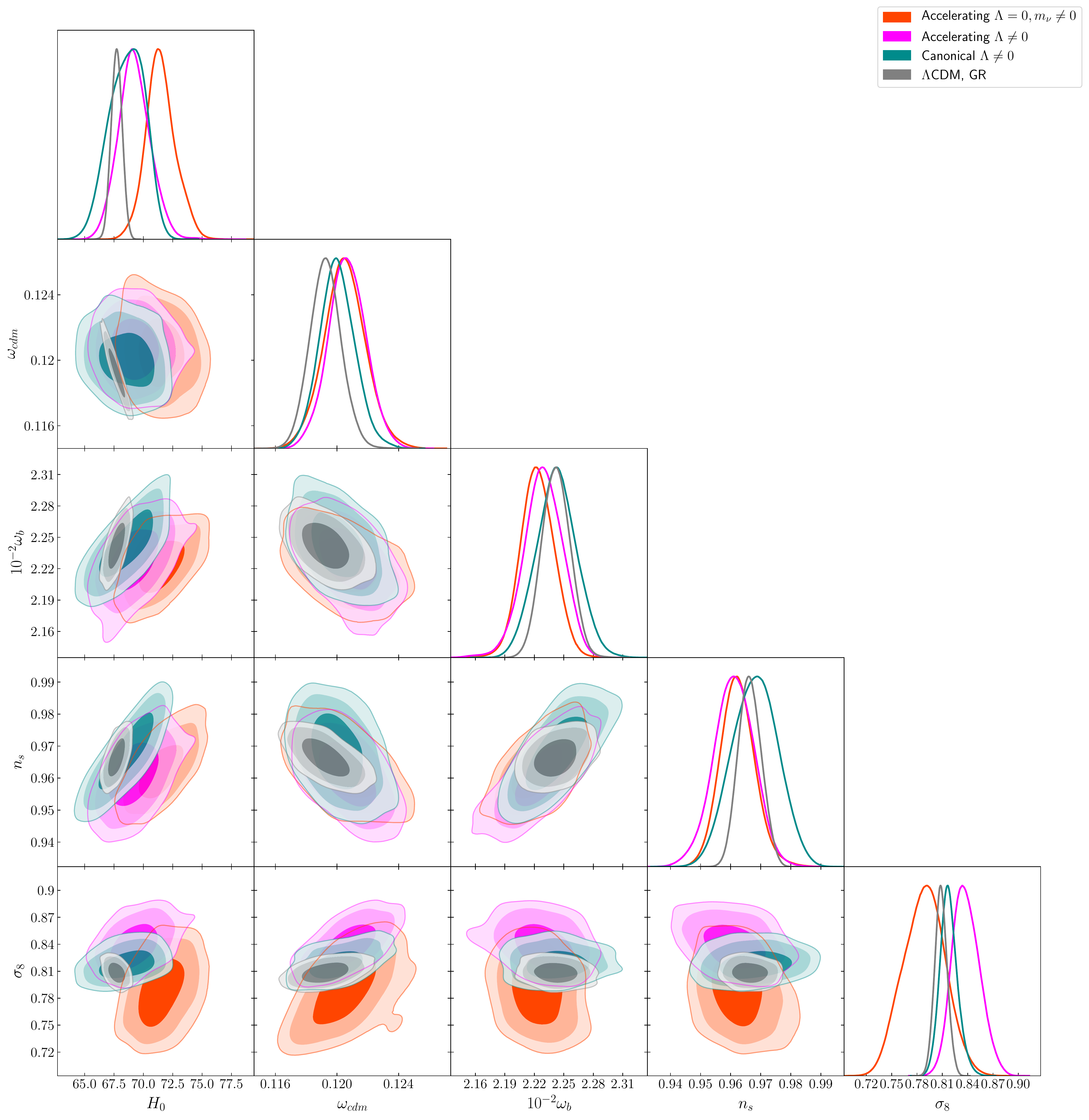}};
 \node at (5,10.8) [draw=black!50!white,thin] {\includegraphics[width=0.35\textwidth]{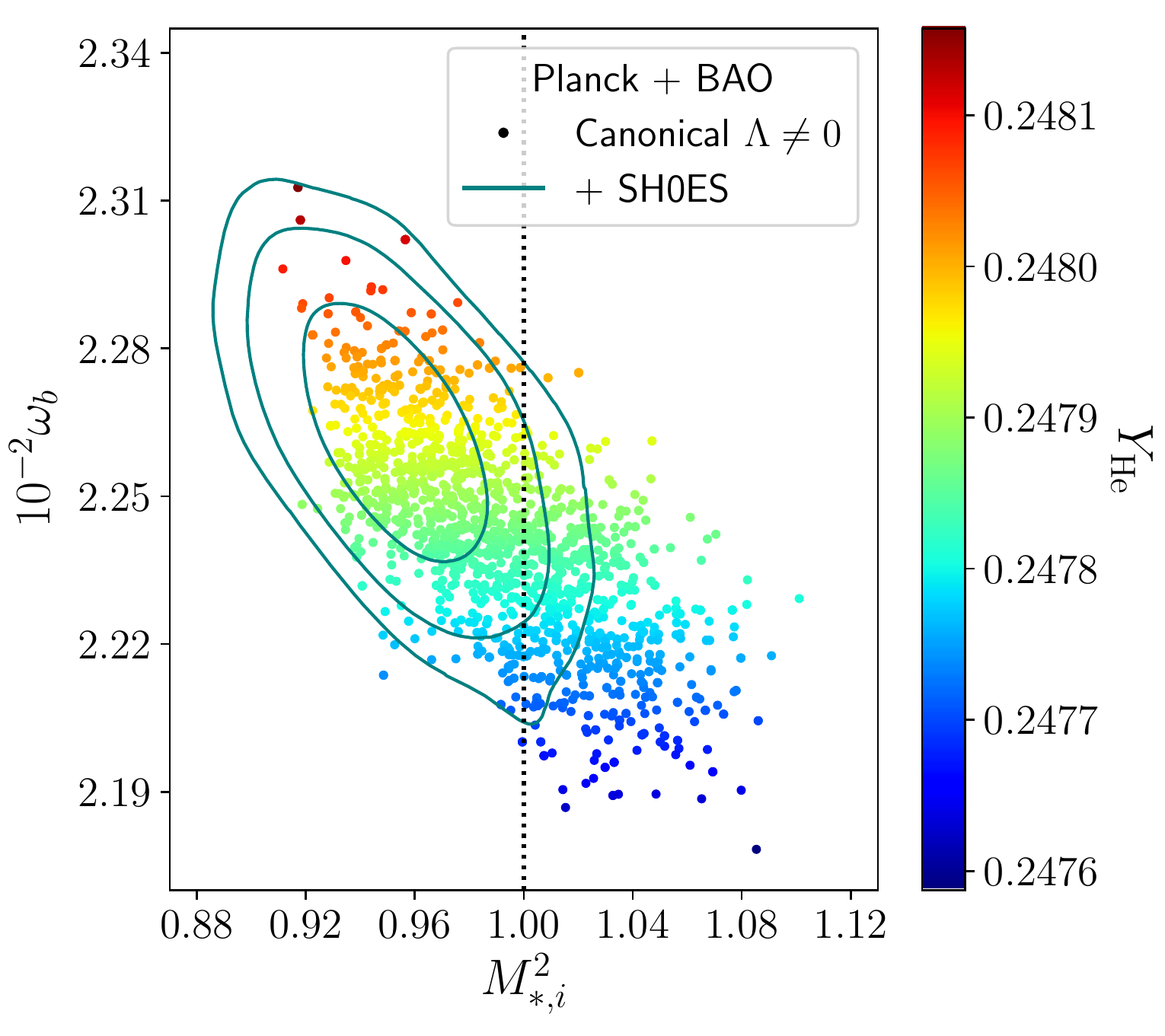}};
\end{tikzpicture}
 \caption{Planck + BAO constraints on cosmological parameters for coupled Galileon models.  Regions correspond to 68, 95 and 99\% c.l. 
 The $\beta$-$M_{*,i}^2$ constraints are shown in figure \ref{fig:coupling_constraints} and constraints on $\log_{10}(\Omega_{\phi,i})$ can be found in table \ref{tab:model_parameters_PB}
 The top right panel shows the effect of $M_{*,i}^2$ on the baryon-Helium abundance induced by BBN.
}
 \label{fig:coupled_triangle}
\end{figure*}

\begin{figure*}
\includegraphics[width=0.49\textwidth]{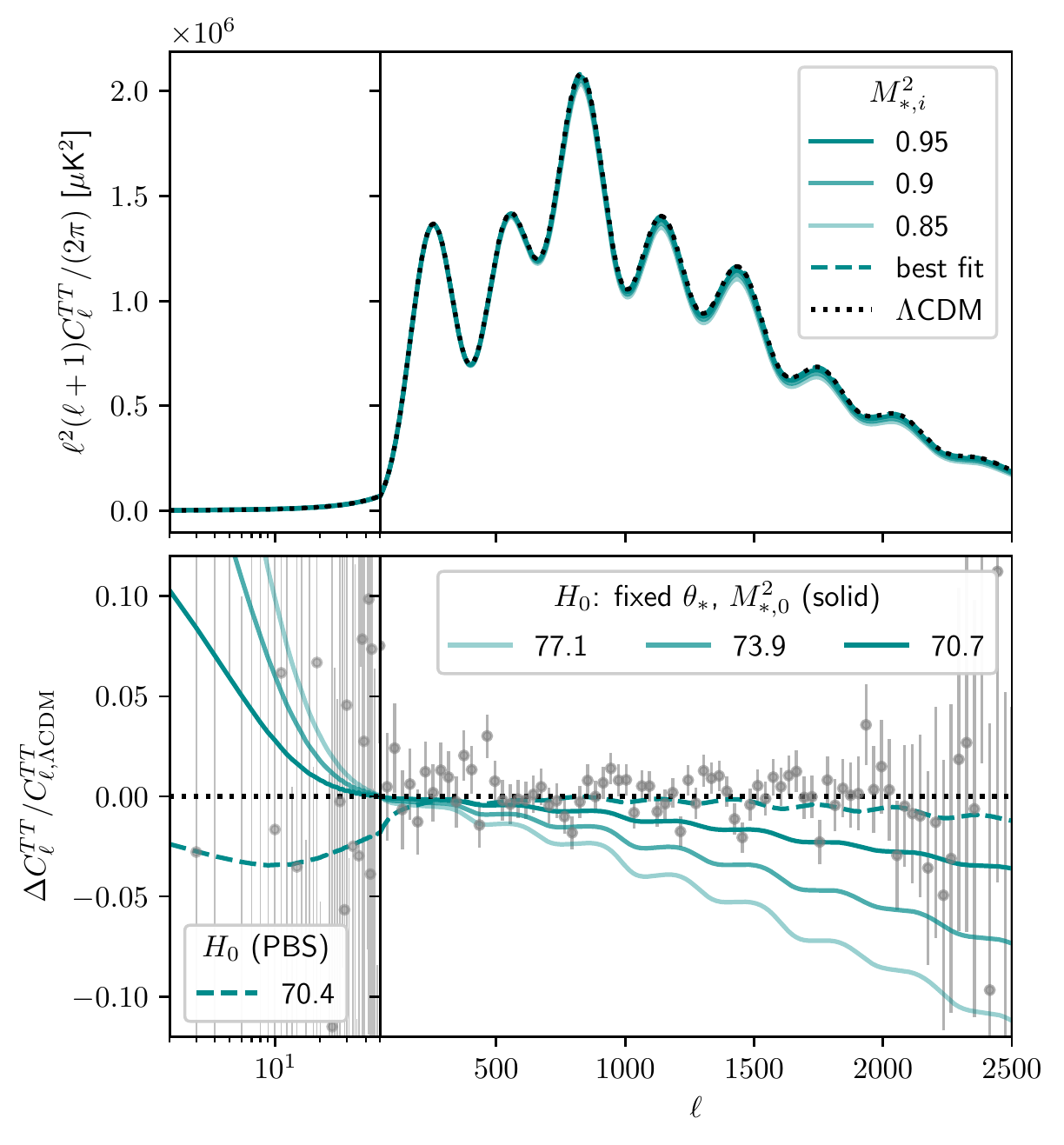}
\includegraphics[width=0.49\textwidth]{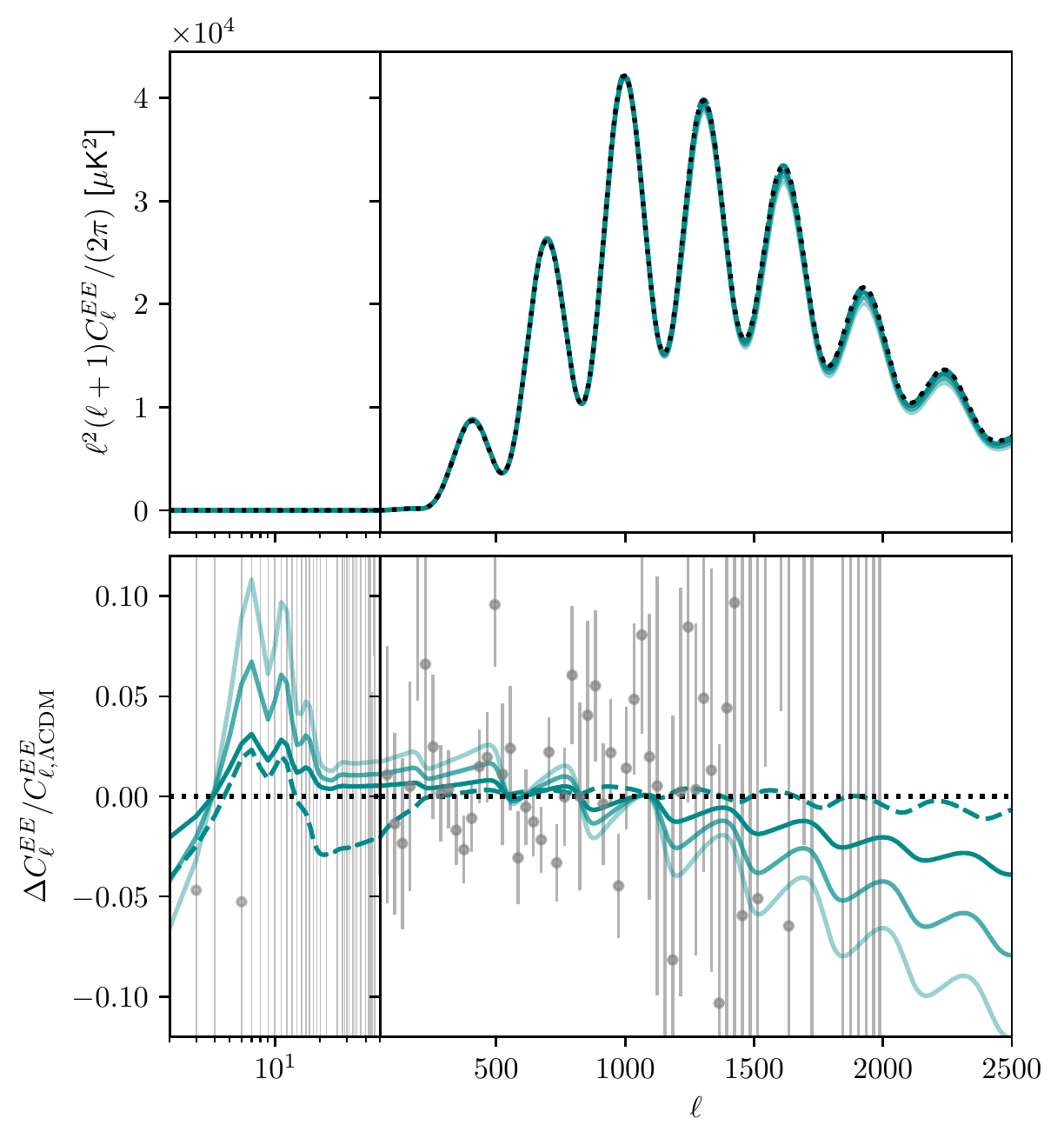}
 \caption{CMB effects of canonical EEG  models. Solid lines represent different  values of the initial effective Planck mass $M_{*,i}^2$ for fixed $\theta_*$ and cosmological parameters and fixing the coupling $\beta$ to set the effective Planck mass to $M_{*,0}^2=1$ today ($\beta = 0.28,0.45,0.61$ for decreasing $M_{*,i}^2$). Degeneracies between $M_{*,i}^2$ and cosmological parameters allow for a better fit for large values of $H_0$ relative to the $\Lambda$CDM fit to the same data (cf. table \ref{tab:model_parameters_PBS}). The dashed line shows the best-fit EEG  to Planck+BAO+$H_0$.
 Left and right panel show TT and EE spectra, respectively.
 Residuals (lower panels) are compared to binned Planck data. Low multipoles ($\ell<50$) are shown in logarithmic scale and compared with unbinned Planck data.
 }\label{fig:M2i_cmb}
\end{figure*}

EEG  introduces new degeneracies with cosmological parameters, contributing to accommodate larger values of $H_0$ in EEG  with respect to $\Lambda$CDM and IDEE. 
These degeneracies and the resulting enlarged posteriors are apparent from a triangle plot, figure \ref{fig:coupled_triangle}. 
For the purpose of the Hubble problem, the most important is the anti-correlation between $M_{*,i}^2$ and both the baryon density $\omega_b$ and spectral index $n_s$. As both $\omega_b,n_s$ are themselves correlated with $H_0$, decreasing $M_{*,i}^2$ leads to a higher Hubble rate by virtue of these degeneracies. These effects are on top of the direct reduction in the acoustic scale caused by EEG.
There is an additional, mild anti-correlation between $M_{*,i}^2$ and the amplitude of perturbations $\sigma_8$ (or equivalent $A_s$).
Interestingly, the dark matter abundance $\omega_{\rm cdm}$ has no apparent correlation with $M_{*,i}^2$, although it correlates weakly with the coupling $\beta$.
The introduction of a coupling increases the limits on IDEE $\Omega_{\phi,i}$ only slightly and does not allow it to play any role on the Hubble tension (cf. table \ref{tab:model_parameters_PB}).

The anti-correlation between $\omega_b$ and $M_{*,i}^2$ is driven by the BBN relation between the baryon and Helium abundances assumed in the analysis (cf. figure \ref{fig:coupled_triangle}, upper right panel). 
This relation limits the damping scale by linking the helium fraction to $\omega_b$ and $M_{*,i}^2$. 
The degeneracy is analog to bounds on additional relativistic species (cf. \cite[figure 39]{Aghanim:2018eyx}), as both increasing $N_{\rm eff}$ or decreasing $M_{*,i}^2$ lead to a faster expansion rate in the BBN era, although the CMB is independently sensitive to relativistic species via perturbations (see section \ref{sec:challenges_bbn}).
Lifting the standard BBN assumption will thus increase the range of allowed values for $M_{*,i}^2$ and $H_0$, but considering the measured primordial helium abundances will limit this range, cf. \cite[figure 41]{Aghanim:2018eyx}.

Note that for a constant $M_*^2$ the equations for the cosmological expansion and linear perturbations depend only on the ratio $\tilde\omega_i \equiv \omega_i/M_*^2$ of the different matter components \cite{Bellini:2014fua}. Therefore, the predictions remain unchanged if all physical densities $\omega_i$ are rescaled, leaving $\tilde\omega_i$ invariant. 
This might suggest a correlation $\omega_b\propto M_*^2$, very different from the anti-correlation observed in the data, roughly $\omega_b\propto M_{*,i}^{-1/2}$. 
This apparent contradiction can be explained by noting that 
1) $M_*^2$ is not constant, with $M_{*,0}>M_{*,i}$ due to the coupling and 
2) unlike $\omega_b,\omega_{\rm cdm}$, the radiation density is fixed by the CMB temperature today and can not be rescaled.
Moreover, as discussed above the baryon fraction degeneracy is set by standard BBN and the effect of helium on the CMB damping.

While Planck+BAO constraints on EEG  models allow a $\sim 3\times$ wider range of values for $H_0$, they show no significant preference towards either lower or higher values relative to $\Lambda$CDM.
Including a prior from distance ladder shifts the posteriors towards high values of $H_0$. 
This analysis results in a $\sim 2\sigma$ preference towards stronger gravity at early times $M_{*,i}^2<1$, with $\Delta{M_{*,i}^2}/\sigma_{M_{*,i}^2}\equiv ({M_{*,i}^2(\text{PBS})-M_{*,i}^2(\text{PB})})/\sigma_{M_{*,i}^2}(\text{PB}) =-1.01$ relative to the Planck+BAO case. 
The shift of other cosmological parameters when including the $H_0$ prior follows the same trends as for uncoupled IDEE models, with the dominant shifts being
$\Delta \omega_b/\sigma_{\omega_b}=0.93$, 
$\Delta \omega_{cdm}/\sigma_{\omega_{cdm}}=-0.43$,
$\Delta n_s/\sigma_{n_s}=0.97$,
 $\Delta \sigma_8/\sigma_{\sigma_8}=0.50$ and
$\Delta \beta/\sigma_{\beta}=0.51$,
leading to change in the Hubble rate by $\Delta H_0/\sigma_{H_0}=1.37$.
While the relative shifts caused by the $H_0$ prior on cosmological parameters are similar as in uncoupled IDEE models, the larger uncertainties of coupled models lead to a stronger net shift.

The main effect of enhanced early gravity is to lower the amplitude of the CMB spectra when $M_{*,i}^2<1$. 
Figure \ref{fig:M2i_cmb} shows the impact of $M_{*,i}^2$ on the temperature and polarization power for fixed cosmological parameters and choosing the coupling $\beta$ so the effective Planck mass today is $M_{*,0}=1$ (this is not assumed in the MCMC analysis).
The lower temperature power is caused by an enhanced Sachs-Wolfe effect: stronger gravity deepens the gravitational potentials, increasing the redshift of photons emitted from over-dense regions. The dependence of this effect on the angular scale is mild, allowing small shifts in cosmological parameters to partially compensate for the differences.
These degeneracies are not accounted for in the solid lines of figure \ref{fig:M2i_cmb}, leading to a seemingly worse fit than if other cosmological parameters had been varied.

EEG  models present important differences relative to IDEE and canonical models for early dark energy. In EEG  models the scalar field modulates the strength of gravity, but because of the Vainshtein mechanism the value of the field remains approximately constant in the early universe, cf. section \ref{sec:dynamics_coupling}. This is equivalent to a constant energy density contribution, which does not affect the ratio of energy densities of all matter and radiation species before recombination.
This is in sharp contrast with both IDEE and quintessence models of early dark energy in which scalar energy density evolves before recombination (cf. figure \ref{fig:idee_dynamics}).
The constancy of the scalar field before recombination also prevents deviations from GR to affect the perturbations (i.e. $\alpha_M\sim \alpha_B\sim 0$), whose dynamics is the same as in standard cosmology but with abundances rescaled by $M_*^2$ \cite{Bellini:2014fua}. 
In contrast, IDEE models induce deviations from GR proportional to $\Omega_\phi$, including the tachyonic growth described in section \ref{sec:constraints_uncoupled}.

\subsection{Late-time dynamics of coupled models} \label{sec:constraints_late}

\begin{figure}
\includegraphics[width=0.95\columnwidth]{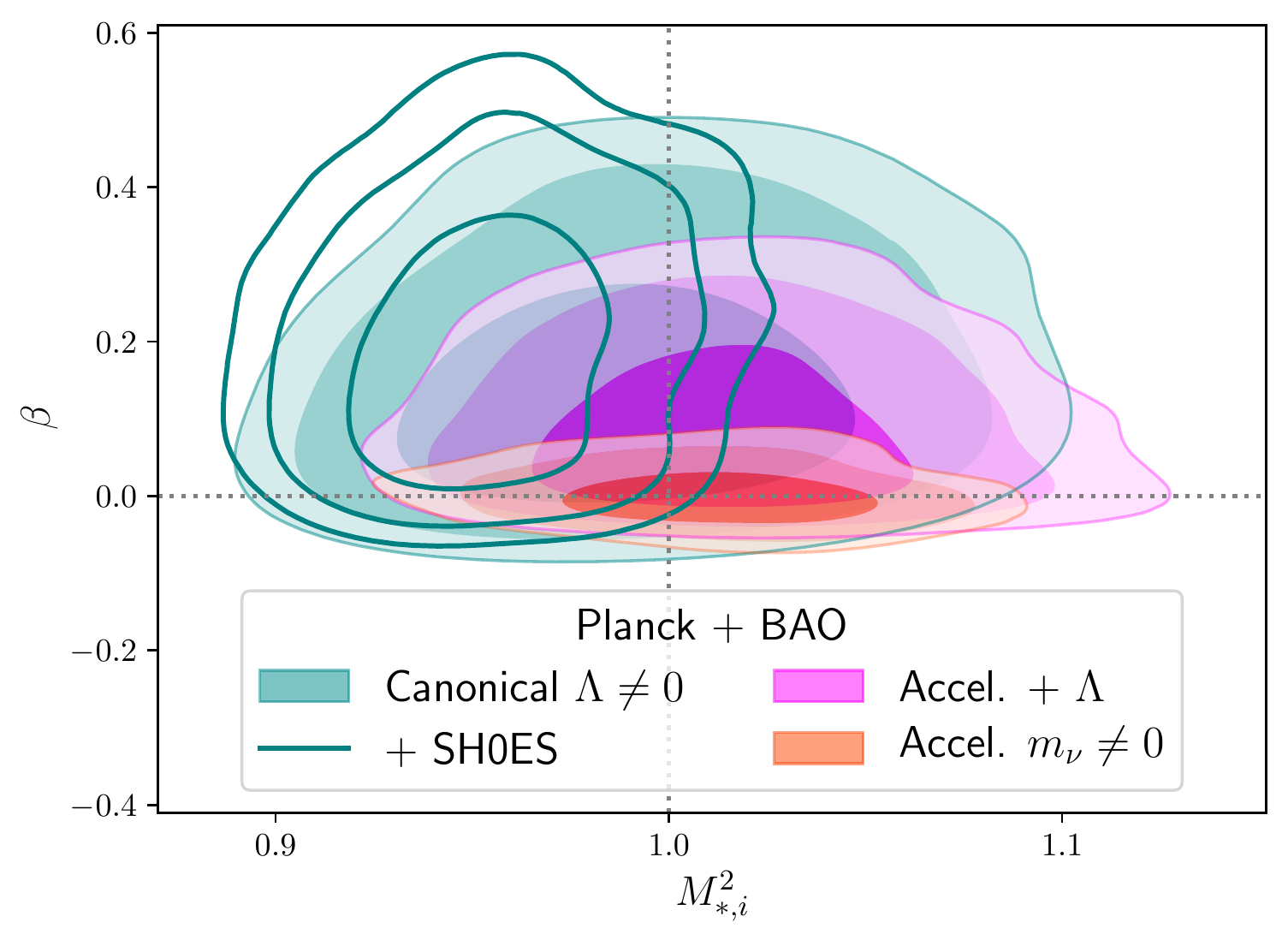}
 \caption{Constraints on the initial effective Planck mass $M_{*,i}^2$ and the coupling strength $\beta$ for coupled luminal Galileons.  Regions correspond to 68, 95 and 99\% c.l. marginalized posteriors for Planck+BAO (filled) and Planck+BAO+$H_0$ (unfilled).}
 \label{fig:coupling_constraints}
\end{figure}

Let us now examine the late-time dynamics of coupled cubic Galileons. I will discuss the constraints on the coupling and the status of canonical (EEG) and accelerating (LUPE+EEG) models. Uncoupled, LUPE-only models are discussed in appendix \ref{app:uncoupled_lupe_Lambda}, including the role of $\Lambda$ and $\sum m_\nu$.
The constraints on the coupling strength $\beta$ and the initial effective Planck mass $M_{*,i}^2$ are shown in figure \ref{fig:coupling_constraints}.

The effect of $M_{*,i}^2$, described in the previous subsection, is very similar across all coupled models. 
The preferred values of $M_{*,i}^2$ depend only mildly on the model, although including a distance-ladder prior on $H_0$ shows a preference for $M_{*,i}^2<1$ corresponding to EEG  cf. figure \ref{fig:coupled_Mi_H0}.
There is a significant widening of the $H_0$ posteriors due to EEG  and the parameter degeneracies already discussed in the previous subsection. The main difference is the central value of the Hubble parameter, which is sensitive to the late-universe expansion and differs in accelerating models via LUPE. 
In the case of $\Lambda=0$ accelerating Galileons, that central value is much closer to the distance ladder measurement than in canonical models with $\Lambda$. 
The accelerating model with $\Lambda \neq 0$ is an intermediate case between the two.

The coupling is constrained by stability criteria and the late-time evolution of the model.
Negative values are mostly excluded as they drive the field evolution towards a ghost instability (cf. section \ref{sec:dynamics_general}). Very small negative values may be supported by the initial field velocity, but this is related to the initial energy density $\Omega_{\phi,i}$ and very limited by the analysis of IDEE models (section \ref{sec:constraints_uncoupled}). 
The Vainshtein mechanism prevents $\beta$ from playing any role before recombination (cf. section \ref{sec:dynamics_coupling}). Therefore, the coupling is constrained by late-universe physics, including low redshift expansion history and secondary CMB anisotropies (ISW effect,  CMB lensing).
As late-time dynamics depends greatly on the presence of $\Lambda$ and the accelerating or canonical nature of the model, each sub-class has different limits on $\beta$, as evident from figure \ref{fig:coupling_constraints}.

The strongest constraints on $\beta$ occur in accelerating models with $\Lambda = 0$, where the field time derivative is largest. 
The absence of a cosmological constant requires a large $\dot\phi$ at late times to support $\Omega_{\phi,0}\approx 0.7$, Eq. (\ref{eq:gal_solutions}). This variation translates on a sizeable running of the effective Planck mass $\alpha_M \propto \beta \dot\phi $, Eq. (\ref{eq:alpha_M}), which is severely constrained by the ISW effect's impact on the CMB's large angular scales.  
Thus, a coupling is very constrained and tends to exacerbate the problems of accelerating Galileons.
Coupled models fare no better than the uncoupled version. They are disfavoured by Planck+BAO  (the best-fit likelihood is even worse for the coupled model, despite being an extension cf. table \ref{tab:models_vs_data}).
Being so close to the uncoupled version they are also strongly ruled out by other observations, such as LSS$\times$CMB cross-correlations \cite{Renk:2017rzu}.
The exponential coupling can thus not save accelerating $\Lambda=0$ Galileons, offering no solution to the $H_0$ problem.

Accelerating models with a cosmological constant have milder bounds on $\beta$. Allowing for $\Lambda\neq0$ eliminates the burden of cosmic acceleration from the Galileon field, which becomes a sub-dominant contribution to the energy density. The velocity of the field is not tied anymore to the requirement of cosmic acceleration and can be lowered significantly. Note that Table \ref{tab:model_parameters_PB} shows $\Omega_{\phi,0}\approx 0.11$, but this includes the contribution from the effective Planck mass today $M_{*,0}^2$, which does not contribute to either $\dot\phi_{0}$ or $\alpha_M$.
Coupled accelerating models with nonzero $\Lambda$ give a reasonably good fit to CMB+BAO, while increasing the allowed value of $H_0$ both due to EEG  and sub-dominant LUPE contribution.
Other cosmological data (such as Ia SNe and LSS$\times$CMB) may place further limits on this scenario.

Canonical models have the loosest constraints on $\beta$.
In this case, cosmic acceleration is entirely supported by the cosmological constant, and thus the contribution from the scalar field to the energy density can be arbitrarily small. Planck+BAO prefer a very subdominant contribution $\Omega_{\phi,0}\sim 0.02$, which is further split into the strength of gravity $\propto M_{*,0}^2-1$ and the kinetic terms $\hat{\mathcal{E}}\propto \dot\phi$, cf. Eq. (\ref{eq:rho_de_kin}).
The kinetic terms $\hat{\mathcal{E}}$ are typically small, as the field derivative is sourced by the coupling $\propto \beta\rho_{\rm mat}\propto a^{-3}$, which reduces the value of $\alpha_M$ cf. Eq. (\ref{eq:alpha_M_c2}). In contrast, accelerating models are driven by the non-trivial solution, Eq. (\ref{eq:gal_solutions}), associated with larger field derivatives.

\begin{figure}
\includegraphics[width=0.95\columnwidth]{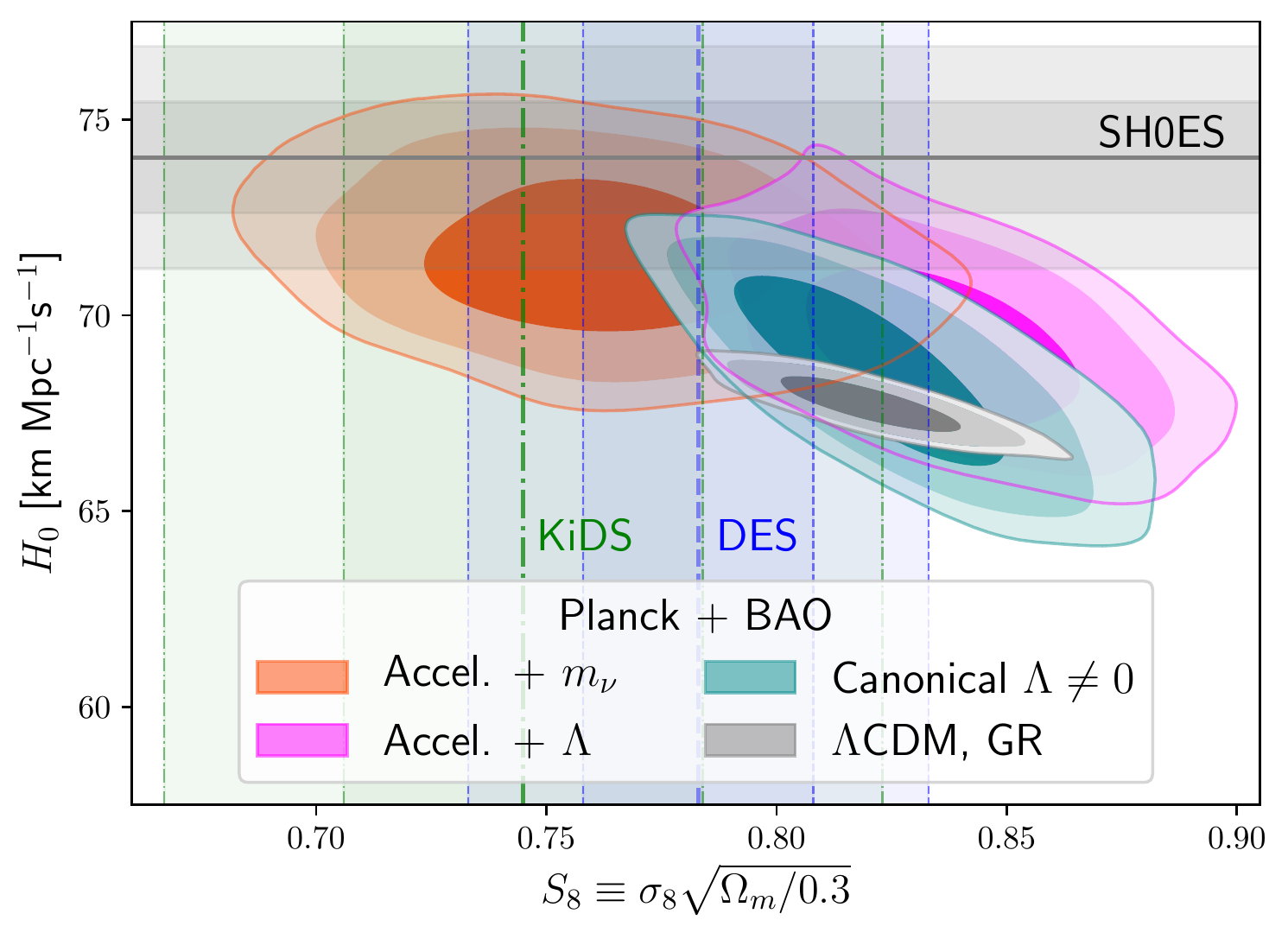}
 \caption{Summary of cosmological tensions in $H_0$ and $S_8 \equiv \sigma_8 \sqrt{\Omega_m/0.3}$. Vertical bands correspond to 68 and 95\% c.l. $\Lambda$CDM inferred values from weak lensing surveys \cite{Abbott:2017wau,Hildebrandt:2016iqg}.
 These bands are model-dependent and do not include deviations from GR in gravitational lensing. 
 Regions correspond to 68, 95 and 99\% c.l. marginalized posteriors for Planck+BAO in EEG models.}
 \label{fig:eeg_tensions}
\end{figure} 

Finally, let us examine the status of the tension between Planck and weak gravitational lensing of galaxies. 
The quantity $S_8\equiv \sigma_8\sqrt{\Omega_m/0.3}$ captures this tension, with weak lensing surveys \cite{Hildebrandt:2016iqg,Abbott:2017wau} finding lower values than Planck for $\Lambda$CDM and simple extensions.
Figure \ref{fig:eeg_tensions} shows an anti-correlation between $H_0$ and $S_8$, leading to $\Lambda\neq0$ EEG models increasing $H_0$ to produce lower values of $S_8$. 
The trend is due to the impact of $H_0$ on $\Omega_m=(\omega_{m}+\omega_b)/h^2$.
The accelerating $\Lambda=0$ scenario lowers $S_8$ via lower values of $\sigma_8$ (see figure \ref{fig:coupled_triangle}).
While these results suggest that a common solution to the Hubble and weak lensing tensions might be possible in EEG models, it is important to emphasize that the $S_8$ values in figure \ref{fig:eeg_tensions} were derived for $\Lambda$CDM. Weak lensing observations depend on the cosmological model and are very sensitive to the properties of gravity: addressing the weak lensing tension requires a comparison of EEG models with weak lensing data.

While this analysis has focused on Planck, BAO and SH0ES, all coupled models can be further constrained by additional cosmological probes and tests of gravity.
In the next section I will outline some remaining challenges for coupled models, including big-bang nucleosynthesis, precision tests of gravity and gravitational waves.

\section{Challenges for Coupled Galileons}\label{sec:challenges}

In this section I describe further observational constraints that may challenge coupled models implementing Enhanced Early Gravity  and/or Late Universe Phantom Expansion. I will first discuss the effect of the effective Planck mass on primordial nucleosynthesis (section \ref{sec:challenges_bbn}). Then I will address the issue of local tests of gravity, including scalar fifth forces, the value of the Planck mass and its time variation (section \ref{sec:challenges_local}).
Finally, I will discuss how GWs may induce instabilities in the scalar perturbations, pushing the theory beyond its regime of validity (section \ref{sec:challenges_GWs}).

\subsection{Primordial Element Abundances} \label{sec:challenges_bbn}

The primordial abundance of light elements is sensitive to the expansion rate in the era of Big-bang nucleosynthesis (BBN). It can be used to place constraints on the initial effective Planck mass $M_{*,i}^2$ independent of the CMB. I will explain how to translate known BBN limits on the expansion history to EEG  models and discuss their implications.

BBN limits are often quoted in terms of additional light particles, such as the number of neutrino species $N_\nu$. By comparing the Hubble law in coupled models (\ref{eq:M2_hubble_de}) to the effects of additional radiation in the expansion history, one can derive a relation between the initial effective Planck mass and extra radiation
\begin{equation}\label{eq:Dnu_to_M2i}
 \frac{1}{M_{*,i}^2}=1+\frac{\Delta\rho}{\rho} = 1+\frac{7}{43}\Delta N_\nu  \,,
\end{equation}
where the second equality uses $\rho = \frac{\pi^2}{30}\left(2+\frac{7}{2}+\frac{7}{4}N_\nu\right)$ accounting for photons, electrons and neutrinos active in the BBN era \cite{Cyburt:2015mya}. Note that $\Delta N_\nu$ is defined relative to a fiducial value $N_\nu=3$, neglecting the small correction from the energy injected by positron annihilation that lead to the difference between $N_\nu$ and the more widely used $N_{\rm eff}$ \cite{deSalas:2016ztq}.
Note that the equivalence between $N_\nu$ and $M_{*,i}^2$ can be applied to nucleosynthesis constraints because BBN is sensitive to additional components only through the expansion history. 
In contrast, CMB anisotropies are sensitive to perturbations in the additional species, including a phase shift due to the super-sonic propagation of neutrinos \cite{Bashinsky:2003tk,Baumann:2015rya,Follin:2015hya}.

Limits on the initial effective Planck mass from primordial abundances of deuterium and helium can be translated using Eq. (\ref{eq:Dnu_to_M2i}) using no CMB data. 
Because the CMB responds differently to relativistic particles and modified gravity, I will only quote values not involving any input from Planck (see Ref. \cite[section 7.6]{Aghanim:2018eyx} for the Planck implications on BBN and $N_{\rm eff}$).
From more to less conservative, several 95\% c.l. limits with no CMB information are
\begin{itemize}
 \item $M^2_{*,i} > 0.860$ ($\Delta N_\nu<1$) for helium only \cite{Cyburt:2015mya,Aver:2015iza}, 
 \item $M^2_{*,i} > 0.911$ ($\Delta N_\nu<0.6$) including the degeneracy with $\omega_b$,  \cite[Fig. 10]{Cyburt:2015mya} and 
 \item $M^2_{*,i} > 0.939$ ($\Delta N_\nu<0.4$) marginalized over $\omega_b$, which is the value quoted in the most recent review of particle properties \cite{Tanabashi:2018oca}.
\end{itemize}
All the above values are less stringent than the Planck + BAO constraints (section \ref{sec:constraints_coupled}).
Other measurements of primordial abundances may produce stronger constraints or even a preference for EEG. For instance, an early work \cite{Bean:2001wt} reports BBN limits on early dark energy equivalent $M_{*,i}^2>0.957$ at 95\% c.l. or $\Delta N_\nu<0.2-0.3$.
In contrast, the helium abundance reported in Ref. \cite{Izotov:2014fga} 
translate to $0.862 <M_{*,i}^2< 0.972$ ($\Delta N_\nu = 0.58  \pm 0.40$) using BBN theory and $0.912 <M_{*,i}^2< 0.976$ ($\Delta N_\nu = 0.37  \pm 0.22 $) when also including Planck data \cite{Aghanim:2018eyx}.

Besides avoiding CMB data, the above bounds assume that the only effect on the expansion history is from a constant effective Planck mass. 
This is an excellent assumption in IDEE models, for which $\Omega_{\phi}(z_{\rm BBN})\lesssim 10^{-5}$ (Planck+BAO limits) and even $\Omega_{\phi}(z_{\rm BBN})\sim 10^{-4}$ is required to reconcile $H_0$ values. 
The assumption remains valid in the presence of a coupling thanks to the cosmological Vainshtein screening, which prevents the scalar field to vary significantly at early times (section \ref{sec:dynamics_ic}). 
BBN constraints can be more stringent for coupled theories without cosmological screening, as the effective Planck mass can vary during the BBN era \cite{Coc:2006rt,Bambi:2005fi,Erickcek:2013dea,Iocco:2008va}.

It is worth emphasizing that the BBN predictions have been included in the CMB+BAO constraints on EEG , and play an important role by relating the helium fraction, $\omega_b$ and $M_{*,i}^2$. As discussed in section \ref{sec:constraints_coupled} and emphasized in figure \ref{fig:coupled_triangle} (top right), lifting the assumption of standard BBN will lead to looser constraints. In that case, including bounds on helium and deuterium abundances will be particularly important to supplement Planck data (see Ref. \cite[figure 41]{Aghanim:2018eyx} for the case of additional relativistic particles).
While current CMB+BAO places more stringent limits than BBN on $M_{*,i}^2$ and EEG, any improvement on the measured primordial abundances can be used as a further test.

\subsection{Gravity on Small Scales}\label{sec:challenges_local}

Deviations from Einstein's general relativity are very well constrained by local gravity tests. There are at least three effects that may be used to constrain coupled Galileons and limit EEG  and LUPE solutions to the Hubble problem:
\begin{itemize}
 \item Scalar forces
 \item Local strength of gravity
 \item Time-variation of the local strength of gravity
\end{itemize}
Reliable constraints based on these effects require solutions of the coupled Galileon theory connecting the cosmological solution to very small-scales. 
In this section I will discuss the challenges to connect cosmological and local dynamics of coupled Galileons, limits on the above effects from lunar laser ranging (LLR) and other precision gravity tests and the challenges in interpreting type Ia supernovae (SNe) observations in coupled models.

\subsubsection{Scalar Forces} \label{sec:challenges_local_Fphi}

The Galileon scalar field mediates an additional, attractive interaction. While the scalar force is very suppressed within the \textit{Vainshtein radius} \cite{Vainshtein:1972sx}
\begin{equation}\label{eq:vainshtein_radius_def}
 R_V^3 = \frac{4GM}{H_0^2}\beta\frac{c_3}{c_2^2}\,,
\end{equation}
it leads to a small deviation form the $1/r^2$ dependence of the gravitational force that causes a small shift in the orbital phase of bound objects and can be probed by sensitive enough measurements (e.g. \cite{Dvali:2002vf}). 
Lunar laser ranging (LLR) measurements set the following bounds
\begin{equation}\label{eq:beta_llr_higher}
 \frac{|c_3|}{\beta^3} > 0.1\,,
\end{equation}
for the phase shift of the Moon to be within the observational limits \cite{Brax:2011sv}.
Note that screening implies that the coefficient of the cubic kinetic term $c_2$ does not enter this bound.

The upper bound (\ref{eq:beta_llr_higher}) leads to $\beta \lesssim 2.15$ for the canonical EEG  models studied here (fixed $c_2 = 1, c_3=-1$), a value well well above the cosmological bounds, cf. figure \ref{fig:coupling_constraints}.  
The above bounds implicitly assume 
\begin{equation}\label{eq:beta_llr_lower}
 \frac{|c_3|}{c_2^2}\beta > 3.4 \cdot 10^{-25}\,,
\end{equation} 
i.e. the Moon's orbit is confined within the Vainshtein radius of the Earth (\ref{eq:vainshtein_radius_def})
The lower limit on the coupling (\ref{eq:beta_llr_lower}) is only relevant only if $c_3\sim 0$. Even in the lack of screening $|c_3|/c_2^2\to 0$, a change in the gravitational strength is proportional to 
\begin{equation}\label{eq:beta_unscreened}
\phi' = \frac{2GM}{r}\frac{\beta^2}{c_2}\,, 
\end{equation}
and thus negligible if $\beta$ is sufficiently small.
An estimate for the bounds in the uncoupled regime (\ref{eq:beta_unscreened}) can be obtained by comparing the coupled free theory to the Brans-Dicke Lagrangian \cite[section 3.1]{Clifton:2011jh}, where one can identify $\beta\sim 1$, $c_2\sim \omega_{\rm BD}\gtrsim 4 \cdot 10^4$ and the lower limit is required for compliance with Solar System tests. Comparing the theory-dependent coefficient of Eq. (\ref{eq:beta_unscreened}) with the Brans-Dicke case suggests that $\beta\lesssim 2\cdot 10^{-2}\sqrt{c_2}$ is in agreement with observations even in the lack of screening.

The above limits on the scalar force follow from an expansion around the Minkowski solution. However, the time evolution of the scalar field modifies the Galileon terms. These corrections have been computed only on de-Sitter backgrounds in which the field evolution is stationary $\dot\phi=\text{constant}$, cf Ref. \cite[appendix B]{Nicolis:2008in}.
For cubic Galileon only the quadratic term is affected. This can be seen from expanding the action for the total field locally $\phi_{\rm loc}=\phi + \varphi$ as
\begin{equation}
\mathcal{L} \sim [c_2-4\frac{c_3}{H_0^2}\Box\phi](\partial\varphi)^2 -4\frac{c_3}{H_0^2}(\partial\varphi)^2(\hat\Box\varphi + \delta\Gamma\partial\bar\phi)\,,
\end{equation}
where $\phi=\phi(t)$ is the cosmological solution and $\varphi=\varphi(t,\vec x)$ is the local correction.
Since cosmological evolution is slow compared to dynamical time-scales of the Solar System, the term in brackets can be taken as constant and the difference between the connections $\delta\Gamma\sim H$ can be neglected.
Then, cosmological evolution amounts to a redefinition of $c_2$, which does not affect the constraints on the scalar force in the screened regime (\ref{eq:beta_llr_higher}).

The local time-variation will be most decisive on the coupling function, as it affects the strength of gravity measured in the Solar System.

\subsubsection{Local Strength of Gravity \& Supernovae} \label{sec:challenges_local_GN}

The scalar field coupling modulates the local value of the Newton's constant, which depends on the local value 
\begin{equation}\label{eq:local_vs_global}
\phi_{\rm loc}(t,\vec x) = \phi + \varphi 
\end{equation}
where $\phi=\phi(t)$ is the cosmological solution and $\varphi=\varphi(t,\vec x)$ is a local correction. For exponential couplings the measured value is recovered if $\beta \phi_{\rm loc}\approx 0$ in the Solar System today.
Because of the space-time dependence of the field, this condition does not necessarily reduce to fixing $M_{*,0}^2=1$ (or $\phi(t_0)=0$) on the cosmological solution.  
A detailed calculation of $\phi_{\rm loc}$ needs to account for vastly different scales, including how the cosmological solutions adapt to the local dark matter halo, how that solution adapts to the galaxy, and so forth, all the way to the Solar System. In addition, it is necessary to model the evolution of the scalar field over the timescales in which those structures form.
While such an analysis is well beyond the scope of this work, I will discuss possible outcomes for the local solution.%
\footnote{See Ref. \cite{Belgacem:2018wtb} for a detailed analysis of this issue in non-local gravity theories without the Vainshtein mechanism and Ref. \cite{Afshordi:2014qaa} for a study of the interplay between local solutions and cosmological time dependence in cubic Horndeski theories. The time-evolution and stability of spherically symmetric systems approaching the Vainshtein screened solution  was studied in Ref. \cite{Brito:2014ifa}. Ref. \cite{Lagos:2020mzy} finds a suppression of the local field velocity in Chameleon models.}

Shift-symmetry $\phi_{\rm loc}\to \phi_{\rm loc} + C$ guarantees the existence of solutions where the field evolves at the cosmological rate around a matter source. 
For a spherically symmetric configuration such a \textit{stationary solution} takes the form \cite{Babichev:2011iz}
\begin{equation}\label{eq:BDE_ansatz}
 \phi_{{\rm loc},s} = \dot\phi(t_0)\cdot t + \varphi_s(r)\,,
\end{equation}
where the only differences with Eq. (\ref{eq:local_vs_global}) is that the field velocity $\dot\phi$ is constant and the local correction is static $\dot\varphi_s=0$. 
Introducing the above ansatz in the dynamical equations explicitly neglects field accelerations $\ddot\phi_{\rm loc},\ddot\phi$, losing any information about how that solution is reached. 
Stationary solutions (\ref{eq:BDE_ansatz}) are a likely endpoint for the dynamical evolution near a matter source, once the local value values of the scalar field reaches equilibrium with the cosmological evolution. 
The main question is whether this occurs before the present time or in the cosmological future.

Two scenarios are possible depending on the relation between the local and cosmological evolution:
\begin{enumerate}
 \item \textit{Homogeneous evolution:} if $\dot\phi_{\rm loc} \approx \dot\phi$ in the Solar System, then $M_{*,0}^2\approx 1$ is a necessary condition. In this case $\alpha_M$ is constrained directly by the variation of Newton's constant (section \ref{sec:challenges_local_GNdot}).
 
 \item \textit{Inhomogeneous evolution:} if $\dot\phi_{\rm loc} \ll \dot\phi$ then  $\phi_{\rm loc}(t_0,0)\sim 0$ requires $M_{*,0}^2>1$. The latter condition is compatible with EEG, but requires a sizeable value of the coupling $\beta$, which enhances the scalar force (\ref{eq:beta_llr_higher}). In this case $\dot\phi_{\rm loc}(t_0,\vec x_0)$ could be small enough to satisfy bounds on the time variation of Newton's constant (section \ref{sec:challenges_local_GNdot}), but cosmological effects will be larger.%
\footnote{In Ref. \cite{Burrage:2020jkj} the authors argue that inhomogeneous evolution leads to either a violation of the equivalence principle, a scalar force larger than the gravitational force in some intermediate region (e.g. between the Solar system and cosmological scales) or a fine-tuned suppression of $\dot\phi_{\rm loc}$, occurring during a short epoch only (the reference appeared after this article's first version).}
\end{enumerate}
The case $\dot\phi_{\rm loc} \gg \dot\phi$ is both nonviable and unlikely. Nonviable because it would yield a large time variation of Newton's constant. And unlikely because the Vainshtein mechanism slows down the field evolution in screened regions.

The analysis of the time-dependent Galileon equation in a screened region suggest that inhomogeneous evolution could happen in small scales.
The starting point is a spherically symmetric field configuration in Minkowski space, where the field evolution is governed by \cite{Brito:2014ifa}
\begin{equation}
Z_{tt}\ddot{\hat\phi}
 = \beta T 
 + c_2 \frac{1}{r^2}\left(r^2\hat\phi^\prime\right)^\prime 
 + 4 c_3 \left((\dot{\hat\phi}')^2 + \frac{\hat\phi^\prime}{r^3}\left(r^2\hat\phi^\prime\right)^\prime 
 \right)\,,
\end{equation}
where primes denote radial derivatives, $t,r$ are in units of $M = (H_0^2 M_p)^{1/3}$ and $\hat\phi(r,t)\propto\phi_{\rm loc}$, $\rho$ have been made dimensionless. 
The \textit{kinetic coefficient} is given by
\begin{equation}\label{eq:vainshtein_kinetic_general}
 Z_{tt} = c_2 + 4c_3\left(\hat\phi'' +4\frac{\hat\phi'}{r}\right)\,,
\end{equation}
and screened region is characterized by $Z_{tt}\gg c_2$. 
While one does not expect the field to evolve in a strictly static space-time, this simple configuration was used to model the approach to the static, Vainshtein-screened solution. A similar analysis might shed light on the interplay between the small-scale and the cosmological solution, which would appear as a boundary condition in this situation.

The evolution timescale in a screened region can be estimated evaluating $Z_{tt}$ on the static screened solution. For $\hat\phi \approx \frac{1}{4}\sqrt{M(r)/(\hat\phi c_3 r)}$ valid deep in the Vainshtein radius (\ref{eq:vainshtein_radius_def}) it reads
\begin{equation} \label{eq:vainshtein_kinetic_screened}
 Z_{tt} \sim \frac{7}{2}\frac{c_3}{\pi}\left(\frac{r_V}{r}\right)^{3/2} \quad (r\ll r_V)\,.
\end{equation}
Thus the characteristic timescale for the field evolution is rescaled by a factor $\sim \sqrt{Z_{tt}}\propto \left({r}/{r_V}\right)^{3/4}$, slowing down evolution in screened regions. This is analog to the cosmological screening mechanism discussed in section \ref{sec:dynamics_coupling}.%

SNe observations need to be reinterpreted in coupled Galileons to reflect the variable strength of gravity. 
The intrinsic luminosity of SNe is expected to depend on the Chandrasekhar mass, the threshold for a white dwarf to be supported by electron degeneracy pressure. 
Its dependence on the strength of gravity
\begin{equation}
 M_{\rm ch} \propto G^{-3/2} \propto C(\phi_{\rm loc})^{3/2} \,,
\end{equation}
implies that the intrinsic luminosity of a SNe will vary with $\phi_{\rm loc}$.
Early works on the subject argued that stronger gravity (lower $M_{\rm ch}$) leads to dimmer SNe (lower ejected mass) \cite{Amendola:1999vu,GarciaBerro:1999bq,Riazuelo:2001mg}. However, a more recent study based on a semi-analytical model for SNe light curves concludes that the opposite is true, with stronger gravity producing brighter SNe after standardization \cite{Wright:2017rsu}.%
\footnote{This calculation suggests that SNe luminosity scales as $\propto G^{1.46}$ \cite{Sakstein:2019qgn}. See Ref. \cite{Hanimeli:2019wrt} for an analysis of a model with variable gravitational constant using different prescriptions for its effect of SNe luminosity.}

If the scalar field evolves homogeneously, the variation of $M_*^2$ produces a redshift-dependent correction to the luminosity distance observed by SNe (this can be tested even independently of the specific model \cite{Linden:2009vh}). 
If the evolution is inhomogeneous, the SNe luminosity will also depend on the properties of the host galaxy/halo, leading to an additional scatter in the Hubble diagram. If the scatter is significant, it can be probed by methods used to search for lensing signatures of compact dark matter \cite{Zumalacarregui:2017qqd}.

Any other observation that may rely on the strength of gravity needs to be reinterpreted in coupled models. One example is the measurement of the Hubble rate inferred from lensing time delays, for which the Hubble parameter inferred from a lens at redshift $z_L$ scales as $H_{0,z_L}=H_{0,\rm true}/M_*^2(z_L)$ \cite{Liao:2020zko}. 
GW standard sirens observations need to be reinterpreted along similar lines once they become available over a larger redshift range.
Unlike for SNe, in both gravitational lensing and GWs the relationship between the luminosity and the strength of gravity is well understood, and the only challenge is modelling the connection between the cosmological evolution and the relevant scales.

\subsubsection{Time Variation of Newton's Constant} \label{sec:challenges_local_GNdot}

The variation of the local scalar field value is equivalent to a time-varying Newton's constant, an effect that can be constrained via precision tests of gravity.
The most precise current bound is based on LLR are \cite{Hofmann:2018myc} 
$ \frac{\dot G_N}{G_N} = (7.1\pm 7.6)\cdot 10^{-14} {\rm yr}^{-1}$,
or equivalently \cite{Belgacem:2018wtb}
\begin{equation}\label{eq:bound_GN_variation}
 \frac{C^\prime}{C}\frac{\dot\phi_{\rm loc}}{H_0} = \beta\frac{\dot\phi_{\rm loc}}{H_0} = -(0.99\pm 1.06)\cdot 10^{-3} \left(\frac{0.7}{h}\right)\,,
\end{equation}
where it has been assumed that $\ddot G_N=0$ and the result is quoted in terms of the scalar field variation using $G_N\propto C(\phi_{\rm loc})^{-1}$. 

The time variation of Newton's constant is strongly correlated with the vector describing the rotation of the Moon's core \cite[section 4.1]{Hofmann:2018myc}. Since the core rotation is poorly constrained independently, the above limits assume the core rotation vector obtained from the standard with $\dot G_N=0$. More conservative assumptions about the Moon's inner structure might lead to weaker constraints.

Note that the central value of $\dot G_N/G_N$ corresponds to a growing strength of gravity. In contrast, coupled cubic Galileons predict a decrease $\dot G_N<0$. This is a theoretical and observational requirement. Theoretically, it follows from the need of positive coupling constant $\beta>0$, necessary required both to prevent ghosts (cf. \ref{sec:dynamics_general}). Observationally it is required for solving the Hubble problem via EEG  and the need to increase the strength of gravity.

The impact on the model parameters and the $H_0$ tension requires understanding the connection between the global and local dynamics of the scalar.
If the scalar field evolves homogeneously, the bound on the variation of the gravitational constant (\ref{eq:bound_GN_variation}) can be translated directly into a stringent limit on the effective Planck mass running today \cite{Tsujikawa:2019pih}
\begin{equation}\label{eq:Gdot_homog}
 \beta\frac{\dot\phi_{\rm loc}}{H_0} \approx \alpha_M(t_0)\,.
\end{equation}
Comparison with the approximate expressions in section \ref{sec:dynamics_coupling} indicate that only very small values of the coupling would be allowed, ruling out EEG.

\begin{figure}
\includegraphics[width=\columnwidth]{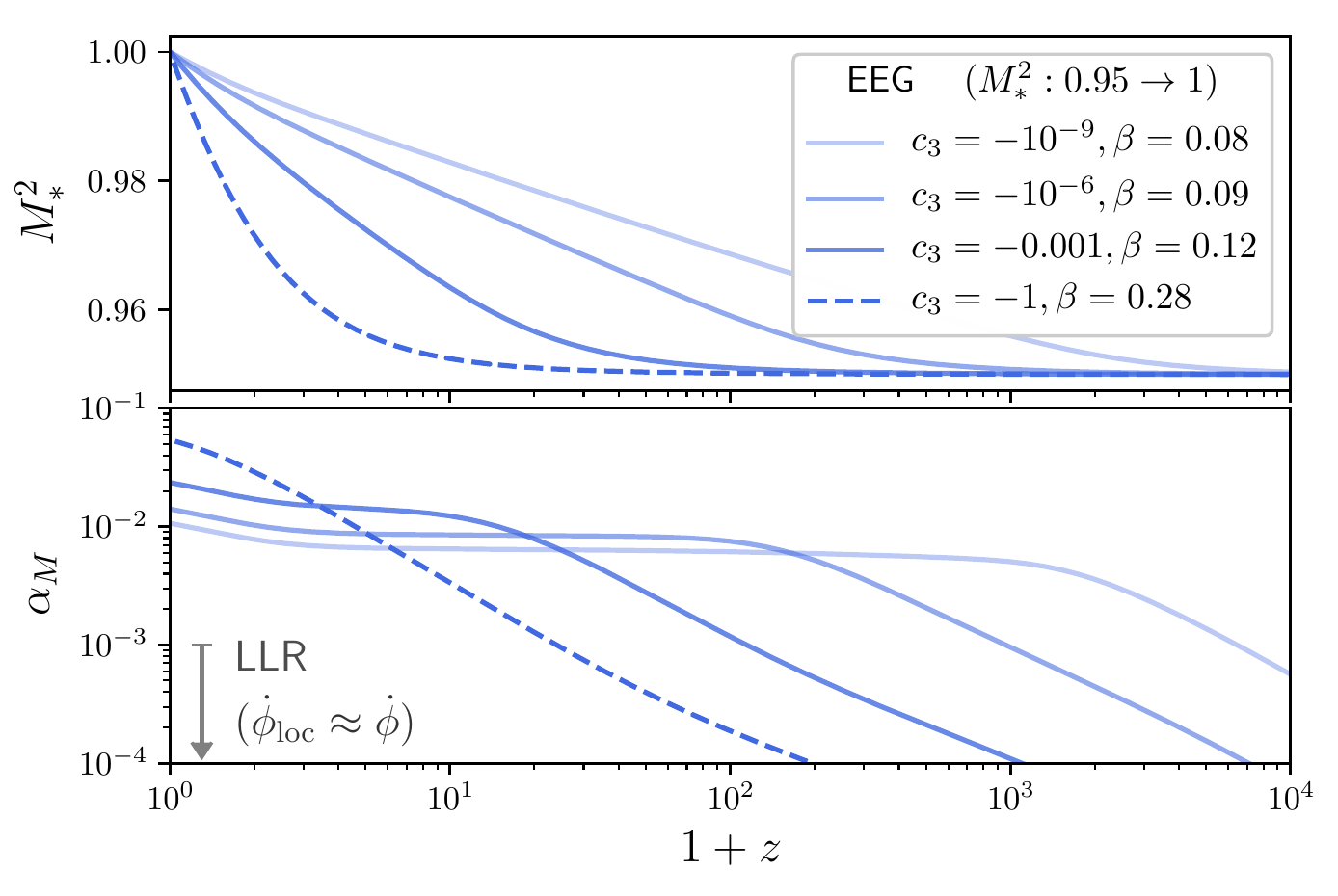}
\caption{Effective Planck mass $M_*^2$ (top) and its running $\alpha_M$ (bottom) for EEG. Lines show the dependence on the cubic Galileon coupling $c_3$, with $c_3=-1$ corresponding to the value analyzed in section \ref{sec:constraints_late}. Lowering $c_3$ reduces the running of the Planck mass.
All models have $M_{*,i}^2=0.95$, and coupling strength $\beta$ fixed so $M_{*,0}^2=1$. The constraint on Newton's constant variation assuming homogeneous evolution (\ref{eq:Gdot_homog}) is shown for comparison. 
\label{fig:GNdot}
}
\end{figure}

Reducing the cubic coupling $c_3$ reduces $\alpha_M$ for fixed value of today's effective Planck mass $M_{*,0}^2=1$. 
This slowdown works by weakening the cosmological Vainshtein mechanism, allowing the field to start evolving earlier, as shown in figure \ref{fig:GNdot} for models where $\beta$ is adjusted so $M_{*,0}^2=1$. 
Values $c_3\lesssim 10^{-9}$ correspond to evolution before recombination, potentially impact primary CMB anisotropies.
The slowdown achievable is not enough to prevent LLR constraints on Newton's constant variation, at least assuming a sizeable EEG  ($M_{*,i}^2\sim 0.95$, assuming homogeneous evolution (\ref{eq:Gdot_homog}) and assuming the latest result with standard value of the Moon's core rotation (\ref{eq:bound_GN_variation}).
In addition, reducing the cubic coupling makes Galileons vulnerable to 

If the local evolution is inhomogeneous the limit needs to be satisfied for the value of the scalar field in the Solar system. A tentative order of magnitude estimate of the Vainshtein suppression suggest applying the constraint (\ref{eq:bound_GN_variation}) to the Planck mass dressed by the kinetic term as discussed in section \ref{sec:challenges_local_GN} 
\begin{equation}\label{eq:Gdot_inhomog}
\beta\frac{\dot\phi_{\rm loc}}{H_0} \sim \frac{\alpha_M}{\sqrt{Z_{tt}^{\rm eff}}}\,. 
\end{equation}
If the above scaling holds, the dependence of the kinetic term with the radius in a screened region (\ref{eq:vainshtein_kinetic_screened}) indicates that the time variation of the Newton's constant could be very suppressed locally, allowing EEG  to remain compatible with time-variation of Newton's constant.

The back-of-the envelope slowdown in screened regions (\ref{eq:Gdot_inhomog}) is likely an overestimation. While better modeling is needed, the true solution is likely to lie within the two limits, Eqs. (\ref{eq:Gdot_homog},\ref{eq:Gdot_inhomog}).
The homogeneous case is well beyond the limit (\ref{eq:bound_GN_variation}) for the EEG , models in which the effective Planck mass evolves significantly $M_{*,i}^2\sim 0.95 \to M_{*,0}^2\approx 1$, since that evolution occurs mostly at low redshift.
A very efficient suppression of $\dot\phi_{\rm loc}$ implies that a large coupling $\beta$ is required to connect EEG  at early times to the correct local value $\phi_{\rm loc}\to 1$ today. Large $\beta$ would be problematic for both cosmology (figure \ref{fig:coupling_constraints}) and scalar force constraints Eq. (\ref{eq:beta_llr_higher}).

\subsection{Gravitational Waves}\label{sec:challenges_GWs}

Coupled cubic Galileon Gravity avoids constraints from the GW speed and decay by construction. In this section I will discuss other GW tests of coupled cubic Galileons, focusing on the scalar instabilities induced by passing GW.

Cubic Galileon interactions may induce instabilities in scalar sector: a background GW can flip the sign of the kinetic term for scalar-field perturbations triggering a ghost or gradient instability  \cite{Creminelli:2019kjy}. This requires a GW with sufficient amplitude propagating on a non-screened region, which is estimated to have happened in a significant fraction of the universe unless
\begin{equation}\label{eq:gw_cubic_braiding}
 \alpha_{B,3}\lesssim 10^{-2}\,, 
\end{equation}
where the cubic Galileon term contribution to the braiding is given by Eq. (\ref{eq:idee_alpha_b3}).
The relevant quantity above is the the contribution of the cubic term to the braiding $\alpha_B$ (the coupling also contributes to $\alpha_B$, but not to the instability).
Whether a given model triggers the instability depends on the time variation of the field at low redshift.

\begin{figure}
\includegraphics[width=\columnwidth]{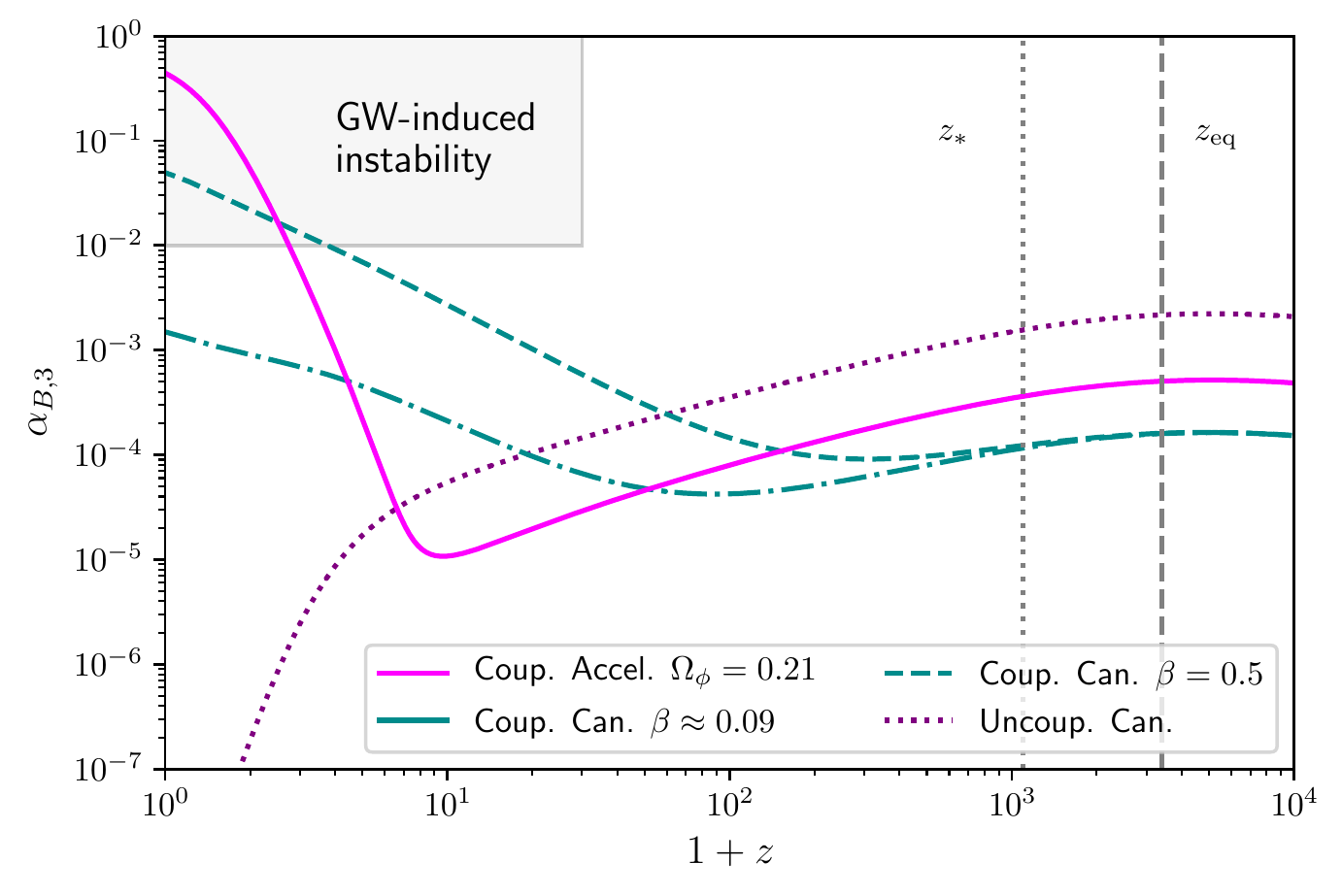}
\caption{GW-induced scalar instabilities for different models. Curves show the cubic braiding (\ref{eq:idee_alpha_b3}) for some best-fit models resulting from the analysis of section \ref{sec:constraints}). Accelerating (magenta) or canonical models with sizeable couplings (dark cyan dashed) are able to trigger the instability at late times (shaded region), Eq. \ref{eq:gw_cubic_braiding}, where it has been assumed that no GW sources exist for $z>30$.
The role of the coupling $\beta$ is shown for a canonical model (dark cyan) with the best fit-value (solid) and a value close to the excluded region (dashed).
Accelerating $\Lambda=0$ models produce even larger values $\alpha_{B,3}(t_0)\sim 1$ (not shown).
\label{fig:gw_instabilities}
}
\end{figure}

Models in which the field evolves rapidly are most susceptible to the instability. Figure \ref{fig:gw_instabilities} shows examples selected from the best-fit models in the cosmological analysis (cf. section \ref{sec:constraints}). The instability is triggered in all accelerating models unless a cosmological constant is allowed and the contribution of the scalar field to the energy density is very subdominant. In canonical $\Lambda\neq0$ models the field variation at late times is driven by the coupling. The instability is triggered only for values of $\beta$ that are sizeable, yet allowed by cosmology (cf. figure \ref{fig:coupling_constraints}). 
In canonical uncoupled models the field's kinetic energy dilutes very fast at late times and remains well below the unstable region.
Note that any feasible amount of IDEE can not trigger the instability. This is due both to the stringent bounds from CMB and the fact that GWs sources with enough amplitude exist only at relatively low redshift, after IDEE peaks.
While including limits from the instability improves over the CMB+BAO constraints, these improvements are milder than suggested by studies based on parameterizations of the alpha-functions \cite{Noller:2020afd}.

While taking the GW-induced instability as a hard constraint is complementary to the cosmological analysis, it is important to remember that they are conservative from a theoretical point of view. The fate of the theory after the instability is reached is uncertain. Specifically, it is not clear whether the instability is associated with any prediction which violates current experimental bounds, and simple models exist in which a similar instability is associated with no pathological behaviour \cite{Creminelli:2019kjy}. All that can be said for sure is that a high-energy completion of the theory is needed to address the consequences of entering the unstable region.

Coupled Galileons also predict a mismatch between distances measured from GWs (standard sirens) and electromagnetic or geometric observations (e.g. SNe, BAO). This difference is produced by the effect of the conformal coupling $G_{4,\phi}\neq 0$ on the GW propagation. 
Current bounds are very weak $|\frac{C^\prime}{C}\frac{\dot\phi}{H_0}|\lesssim \mathcal O(10)$ \cite{Lagos:2019kds}, well below the level of other probes discussed here. 
Upcoming GW observation campaigns and new detectors will improve these limits considerably \cite{Ezquiaga:2018btd,Belgacem:2019pkk,Maggiore:2019uih}. However, it has been argued that the interpretation of standard sirens needs to be reconsidered in theories with screening mechanisms \cite{Dalang:2019fma,Dalang:2019rke,Garoffolo:2019mna} (see also Ref. \cite{Wolf:2019hun}).
Because standard sirens are not yet a competitive test, I will not discuss them further.

\section{Conclusions}\label{sec:conclusions}

Discrepancies in the Hubble constant inferred by different methods could be an indication of physics beyond the simple $\Lambda$CDM model and its underlying assumptions. 
Here I have examined three different mechanisms by which gravity theories beyond Einstein's GR may alleviate the discrepancy on the $H_0$ values inferred via BAO+CMB and distance ladder observations.
Imperfect Dark Energy at Equality (IDEE) and Enhanced Early Gravity  (EEG) modify the pre-recombination expansion history to reduce the acoustic scale $r_s$ for fixed angular projection $\theta_*$. Late-Universe Phantom Acceleration (LUPE) is based on the dark energy density growing at low redshift $w_\phi<-1$.
Each mechanism can operate individually or in combination with the others. 

The three mechanisms exist in the coupled cubic Galileon, a simple scalar-tensor theory compatible with the speed of GWs, lack of GW decay and equipped with the Vainshtein screening mechanism.
This investigation focused on an exponential form of the coupling $G_4=C(\phi)\propto e^{\beta\phi}$, and considers the two possible signs of the quadratic kinetic term $\propto c_2(\partial\phi)^2$, dubbed canonical ($c_2>0$) and accelerating ($c_2<0$).
Different combinations of model properties (coupled/uncoupled $\times$ canonical/accelerating) were tested against Planck+BAO, including in some cases the SH0ES distance ladder measurement of $H_0$ to address the tension in extended models.

The main findings regarding the cosmology of these models can be summarized as follow:
\begin{enumerate}
 
 \item IDEE relies on the scaling of the scalar field energy density, which dilutes faster than matter but more slowly than radiation (figure \ref{fig:idee_dynamics}). It requires a large initial velocity for the field $\dot\phi_{i}$. Values corresponding to $\Omega_{\phi}(z_{\rm BBN})\sim 10^{-4}$ would lower $r_s$ enough to reconcile CMB+BAO with SH0ES for fixed $\theta_*$.
 
 \item Planck+BAO constrain IDEE to $\Omega_{\phi}(z_{\rm BBN}) \lesssim 10^{-5}$, below the level necessary to solve the $H_0$ problem (figure \ref{fig:idee_bounds}). The strong bounds on IDEE stem from modified gravity and expansion. A tachyon instability enhances the growth of scalar-field perturbations after Hubble crossing, impacting mainly the first CMB peaks (figures \ref{fig:idee_autopsy} and \ref{fig:idee_cmb}).

 \item EEG  relies on the coupling $C(\phi)R$, allowing the scalar field to modulate the strength of gravity via the effective Planck mass $M_*^2=G/G_{\rm eff} = C(\phi) = \exp(\beta\phi)$. EEG  requires for the field to roll in the late-universe to reduce $r_s$ at fixed $\theta_*$ (figure \ref{fig:EGO_dynamcis}).
 Initial conditions corresponding to $M_{*,i}^2\sim 0.95$ evolving to $M_{*,0}^2=1$ could solve the $H_0$ problem.

 \item Planck+BAO data is compatible with EEG  in the range $0.92< M_{*,i}^2 < 1.06$ (95\% c.l.). The degeneracy between $M_{*,i}^2-H_0$ and other parameters weakens CMB+BAO bounds to $H_0 = 68.7 \pm 1.5$ (68\% c.l.) with a $\approx 3$-fold increase in uncertainty, relative to $\Lambda$CDM (figure \ref{fig:coupled_Mi_H0}). This reduces the tension with SH0ES from 4.4$\sigma$ to $2.6\sigma$ (combining errors in quadrature).
 
 \item LUPE models rely on the sign of the kinetic term ($c_2<0$), causing the scalar field energy density to increase $w_\phi<-1$ and accelerate the universe (figure \ref{fig:canonical_vs_accelerating}).
 By itself LUPE has no impact on $r_s$, but raises $H_0$ for fixed $\theta_*$. Data requires a combination of $\Omega_{\phi,0},  \Omega_{\Lambda,0}\neq 0$ or $\sum m_\nu\sim 0.6$eV if $\Lambda=0$ (appendix \ref{app:uncoupled_lupe_Lambda}).

 \item Planck+BAO allow only LUPE models with $\Lambda\neq0$. The coupling strength is severely restricted in the $\Lambda=0$ case to $|\beta|< 0.05$ 95\% c.l., preventing a coupling from improving the fit for the accelerating cubic Galileon (figure \ref{fig:coupling_constraints}). 
 $\Lambda\neq0$ models reduce $H_0$ tension to $2.5\sigma$ via a combination of LUPE and EEG. 
 
\end{enumerate}

It is remarkable that modified gravity solutions to the Hubble problem require far less fine-tuned initial conditions than other early dark energy models. 
IDEE stems from the initial field velocity and scales only mildly with the dominant matter component.
EEG  stems from the initial field value and its contribution remains constant at early times thanks to the cosmological Vainshtein mechanism. 
Generic initial conditions of the field produce some amount of IDEE and EEG  in coupled cubic Galileons.
To reconcile $H_0$, EEG  requires $\phi_{i}/M_p \sim -0.05/\beta$, a sub-Planckian value of the initial scalar value for typical values of $\beta$.
Early modified gravity solutions require $\Omega_{\phi,i}\sim 0.05$ (EEG), $\Omega_{\phi,i}\sim 10^{-4}$ (IDEE) around BBN to solve the Hubble problem. These are relatively small differences compared with those needed for canonical oscillating fields $\Omega_{{\rm quint},i}\sim \left(T_{\rm eq}/T_{\rm BBN}\right)^4 \sim 10^{-24}$.
While the fine-tuning of initial conditions fares better than in other scenarios, the issues associated to $\Lambda$ and cosmic acceleration remain.

Among the three mechanisms, EEG  is the best candidate to reconcile CMB+BAO and distance ladder, although a combination of EEG  and LUPE remains promising in light of those cosmological datasets. 
These scenarios tension between Planck+BAO and SH0ES to the $\approx 2.5\sigma$ level, comparable to other late DE solutions \cite{DiValentino:2019jae}.
Analyses involving additional datasets are necessary to further constraint these mechanisms. Particularly, late-universe cosmological measurements (e.g. redshift space distortions, weak lensing from galaxy shear and CMB, LSS$\times$CMB cross-correlations or type Ia SNe) will improve the bounds on the coupling $\beta$ for EEG  and $\Omega_\phi$ for LUPE, respectively.
LSS tests, recently used to constrain early quintessence \cite{Hill:2020osr}, will also shed light on early modified gravity.
Such analyses will help clarify whether EEG \& LUPE models can simultaneously alleviate the Hubble and weak lensing tensions, as suggested by figure  \ref{fig:eeg_tensions}.
A variety of additional data can be used to probe these mechanisms further, with precision tests of gravity posing the most outstanding challenge for coupled models.

EEG  models need to match the observed strength of gravity measured in the solar system, which is given by the local field value $\phi_{\rm loc}(t_0,x_0)$ and its derivative (section \ref{sec:challenges_local}). The non-linear nature of the problem (Vainshtein screening) and the hierarchy of scales involved (cosmological background to Solar System) require further modeling to reliably address this issue. 
Two scenarios are plausible: 
1) if the local field velocity is comparable to the cosmological value, then EEG  is severely limited by the time variation of Newton's constant and the stringent bounds from LLR (figure \ref{fig:GNdot}). 
2) If the local field velocity is significantly slower than the cosmological one, a large coupling value is required to recover the correct local strength of gravity today, entering into conflict with cosmology and constraints on scalar forces.
A related issue is the interpretation of SNe and other observations in models in which the strength of gravity depends on redshift and host properties. 

Big-bang nucleosynthesis is sensitive to the early expansion history, allowing the observed abundance of light elements to place bounds on EEG  (section \ref{sec:challenges_bbn}).
These bounds are by themselves weaker than the Planck+BAO, but when combined might improve limits on the initial effective Planck mass. 
Note also that the standard BBN relation between the baryon and helium fraction was assumed and played an important role in constraining EEG  via the damping tail (figure \ref{fig:coupled_triangle}, to right). Varying the helium fraction freely will likely weaken the Planck+BAO limits on EEG.

GW-induced instabilities are sensitive to the late-universe evolution (section \ref{sec:challenges_GWs}). Avoiding the instability limits the value of the coupling beyond CMB+BAO bounds for EEG  and severely limits LUPE, even for $\Lambda\neq0$. 
While these limits are enticing, it is important to remember that instabilities signal a breakdown of the theoretical description, rather than a prediction contradicting known data. A UV-complete theory is needed to establish whether EEG  \& LUPE models can be ruled out by GW-induced instabilities.

EEG, LUPE and IDEE are general mechanisms that can be explored in theories beyond the simple exponentially coupled cubic Galileon.
While none of the models studied here is likely to pass all tests, it is plausible that further model building may overcome these difficulties. 
The notion of IDEE can be generalized beyond the cubic Galileon.%
\footnote{A straightforward IDEE generalization, known as the nKGB model, is specified by the following Horndeski functions
\begin{equation}
 G_2 = -X\,,\quad G_3 \propto X^n\,,
\end{equation}
($n=1$ corresponds to the case studied here). A calculation analogous to the one outlined in section (\ref{sec:dynamics_idee}) shows the following dependence on the equation of state for the scalar field
\begin{equation}
 w_\phi = \frac{1-w_m}{4 n}\,.
\end{equation} 
The conditions for the scalar energy density to dilutes slower than radiation but faster than matter is simply $n>1/2$, approaching the matter scaling in the limit $n\to\infty$. A different value of $n$ may improve the behavior of cosmological perturbations relative to the $n=1$ case studied here. A generalization of this model has been studied in Ref. \cite{Frusciante:2019puu} as LUPE solution.}
In models with a canonical kinetic term one can advance the onset of the kination phase by increasing the hierarchy $c_2/|c_3|$, perhaps even in the pre-recombination era. 
Early modified gravity is also compatible with quartic or quintic Horndeski terms, as long as a kination phase ensures that the speed of GWs is within acceptable bounds at low redshift.
Other simple variations include modifying the coupling function beyond the simple exponential form or adding a potential term. These modifications may help lock up the local value of the scalar field in dense environments, in a manner analogous to the Symmetron model \cite{Hinterbichler:2010es}.
Needless to say, the properties leading to IDEE, EEG and LUPE (and perhaps completely different solutions to the Hubble problem) are likely to exist in extensions of GR other than Horndeski Gravity.
In this sense, this work is only a first systematic exploration of the possibilities of theories beyond Einstein's GR to address the Hubble problem. 

The mechanisms described here are extremely predictive. 
They can be tested using a wide range of observations across vastly different scales and epochs, from precision gravity tests in the laboratory and the Solar System or GW astronomy, all the way to the large-scale structure of the universe, abundance of primordial elements, primary and secondary CMB effects and the cosmic expansion.
Future data on these fronts will be able to determine whether the Hubble problem and other cosmological tensions are due to new physics beyond the $\Lambda$CDM model.
If cosmological tensions endure upcoming scrutiny, combining theoretical and observational insights will be key to illuminate the necessary amendments to the standard model and their fundamental implications for our understanding of nature.

\acknowledgments

I am very grateful for Lam Hui for insightful questions and encouraging me to look closely at the coupled Galileon models, as well as Sunny Vagnozzi and Alex Barreira for detailed comments on a first draft. 
I also thank Carlos Garcia-Garcia, Marius Millea, Iggy Sawicki, Filippo Vernizzi, and Martin White for illuminating discussions that helped me understand different aspects of the analysis.
This work used CLASS \cite{Blas:2011rf}, \texttt{hi\_class} \cite{Zumalacarregui:2016pph,Bellini:2019syt}, MontePython \cite{Audren:2012wb,Brinckmann:2018cvx},  CosmoSlik \cite{2017ascl.soft01004M}, GetDist \cite{Lewis:2019xzd}. 
I acknowledge support from the the Marie Sklodowska-Curie Global Fellowship Project NLO-CO. 
This research used resources of the National Energy Research Scientific Computing Center (NERSC), a U.S. Department of Energy Office of Science User Facility operated under Contract No. DE-AC02-05CH11231.

\appendix

\section{Alternative scalar energy density}

Alternatively to Eq. (\ref{eq:M2_hubble_de}), one can define the Friedmann equation and total scalar energy density as 
\begin{eqnarray}\label{eq:H_total}
 && H^2 = \rho_m + \mathcal{E}\,, \\  
 && \mathcal{E} \equiv \hat{\mathcal{E}}+(1-M_*^2)H^2\,. \label{eq:rho_de_full}
\end{eqnarray}
There are thus two equivalent ways to describe the cosmic expansion:
\begin{itemize}
 \item $\mathcal{E}$ (Eq. \ref{eq:H_total}) is the energy density inferred when interpreting an expansion history within Einstein's GR.
  \item $M_*^2$ and $\hat{\mathcal{E}}$ (Eq. \ref{eq:M2_hubble_de}) separate the effect from the strength of gravity and other contributions from the scalar field ($\propto \dot\phi$). 
\end{itemize}
Both descriptions are equivalent and related by Eq. \ref{eq:rho_de_full}. 
Note that increasing the effective Planck mass, $M_*^2>1$, reduces the expansion rate, which is seen as a negative contribution to $\mathcal{E}$ in Eq. (\ref{eq:rho_de_full}).

\section{Uncoupled LUPE $\Lambda\neq0$ models}\label{app:uncoupled_lupe_Lambda}

I will briefly discuss constraints on uncoupled accelerating cubic Galileon and the role assumptions about the cosmological constant and massive neutrinos. 
The datasets used in these analysis are Planck 2015 and BAO, with the choices and methodology described in Renk \textit{et al.} 2017 \cite{Renk:2017rzu}. Specifically, no IDEE or EEG  is included in these models.
These results are included for completeness, but separate from the main text because they rely on older datasets than the Planck 2018, main analyses presented in section \ref{sec:conclusions}.

\begin{figure}[t]
 \includegraphics[width=0.49\textwidth]{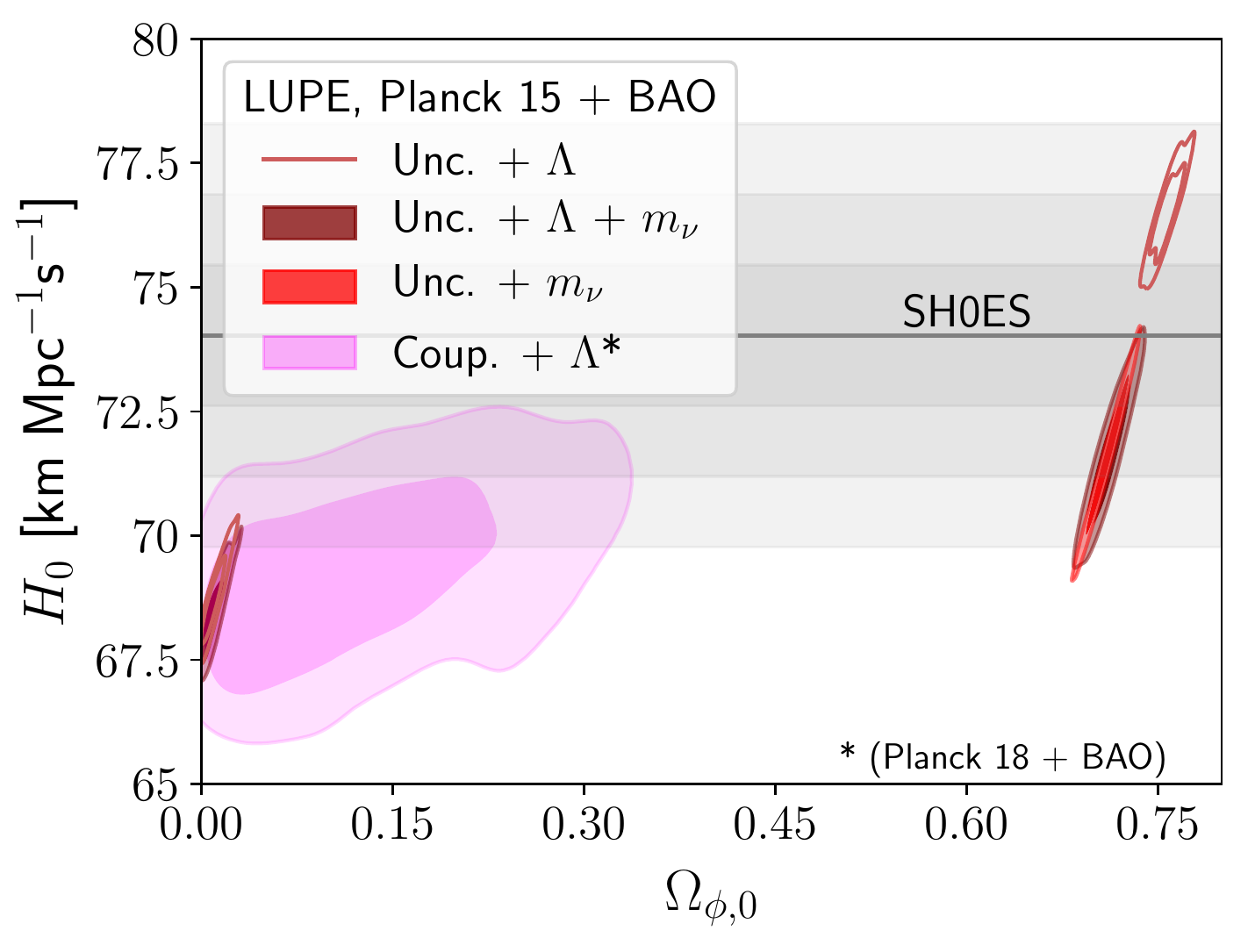}
 \caption{Role of $\Lambda,m_\nu$ on LUPE-only models. Contours show 68, 95 and 99\% c.l. posteriors on uncoupled accelerating cubic Galileon with minimal/variable neutrino mass and cosmological constant for Planck 2015 and BAO data (see \cite{Renk:2017rzu} for details). 
 Accelerating uncoupled $\Lambda\neq0$ Galileons are either dominated by the cosmological constant or the LUPE energy density, but the coupled (EEG) model allows a more flexible combination of both dark energy components. 
\label{fig:lupe_2017_plots}
 }
\end{figure}

I considered the following LUPE scenarios:
\begin{enumerate}
\item[(0)] accelerating uncoupled cubic Galileon with $\Lambda=0$ and free $\sum m_\nu$ (cubic model in Ref. \cite{Renk:2017rzu})
 \item[(1)] model (0) with free $\Lambda$, $\sum m_\nu$ and
 \begin{enumerate}
  \item[(a)]  $\Omega_{\Lambda,0}\sim 0$, $\Omega_{\phi,0}\sim 0.7$
 \item[(b)]   $\Omega_{\Lambda,0}\sim 0.7$, $\Omega_{\phi,0}\sim 0$
 \end{enumerate}
 \item[(2)] model (0) with free $\Lambda$ but fixed $\sum m_\nu = 0.06$eV 
 \begin{enumerate}
  \item[(a)]   $\Omega_{\Lambda,0}\sim 0$, $\Omega_{\phi,0}\sim 0.7$
 \item[(b)]   $\Omega_{\Lambda,0}\sim 0.7$, $\Omega_{\phi,0}\sim 0$
 \end{enumerate}
\end{enumerate}
The values of $\Omega_{\Lambda,0}$ and $\Omega_{\phi,0}$ refer to the initial proposal for the sampling distribution, not to hard priors on the parameters. This distinction was necessary because both regions of the parameter space and had to be explored separately.
The main results are shown in figure \ref{fig:lupe_2017_plots}, where I have also included the coupled accelerating $\Lambda\neq0$ model (with Planck 18) to compare the effects of EEG+LUPE, cf. section \ref{sec:constraints_late}.

The uncoupled Galileon is all or nothing. Comparison between different initial sampling distributions (a vs b) shows that only one form of energy density dominates. The secondary component is limited to $\Omega_{\phi ,0} < 0.022$ in both $\Lambda$-dominated scenarios (b) and $\Omega_{\Lambda,0} < 0.019$ in $\phi$-dominated, free $\sum m_\nu$ (1a) and a more stringent limit $\Omega_{\Lambda,0}<0.003$ for $\phi$-dominated, fixed $\sum m_\nu$ (2a), with all values corresponding to 95\% c.l. exclusions.
The coupled model allows a much wider mixture between the two dark energy components because EEG  lifts the restrictions on the acoustic scale.

These analyses confirm the role of the neutrino mass in $\phi$-dominated scenarios.
Sizeable  $\sum m_\nu$ is necessary both to obtain a better fit and to reconcile the SH0ES value of $H_0$, but this only happens in $\Lambda\sim0$ models. The $\Lambda=0$ (0) and $\phi$-dominated, $\Lambda\sim 0$ (2a) scenarios show only minor differences, consistent with the limits on $\Lambda$ discussed above.
The model with fixed $\sum m_\nu$ (2a) predicts a Hubble parameter above the SH0ES central value, but is excluded by the analysis with variable neutrino mas (1a), of which (2a) is a particular case.
In the $\Lambda$-dominated cases (b) the neutrino mass is constrained at a similar level as in standard $\Lambda$CDM, with negligible differences between variable (1b) and  fixed (2a) $\sum m_\nu$.

\section{Coupling and early field dynamics}\label{sec:dynamics_ic}

Let us examine the effects of a non-zero coupling on the initial conditions of the field. Using the general equations presented in section \ref{sec:dynamics_general} I will discuss the Vainshtein screening mechanism and its effect on three possible sources of initial IDEE: pressure-less matter, particles becoming non-relativistic and a hypothetical kination phase. The high efficiency of the Vainshtein mechanism makes those sources completely negligible for all practical purposes. The same suppression of early dynamics makes EEG  a much more robust and simple mechanism to lower the acoustic scale. 

The Galileon is sourced by the trace of the matter and reduced dark energy momentum tensor (see end of section \ref{sec:dynamics_general})
Several early universe phenomena contribute to the source term-sigma $\Sigma$, cf. Eq. (\ref{eq:sigma_decomposition}), and may affect the initial kinetic energy of the Galileon.
An example is whenever the temperature in the early universe drops below the mass of a particle: for some time that particle remains important in the energy budget, while becoming partially non-relativistic and thus contributing to $\Sigma$.
This phenomenon is known as ``kicks'' in the context of chameleon theories \cite{Erickcek:2013oma,Erickcek:2013dea} (see Ref. \cite{Padilla:2015wlv} for a study of theories with non-canonical kinetic terms of the Dirac-Born-Infeld type and Ref. \cite{Sakstein:2019fmf} for massive neutrino kicks used to set initial conditions for early quintessence).
Phase transitions contribute similarly to the shift-charge density. At the end of this section I will also examine the effects of a hypothetical kination phase, the most favorable situation to overcome the cosmological Vainshtein screening.

We will express the contribution of a kick to the integral in Eq (\ref{eq:charge_sol_explicit}) as 
\begin{equation}\label{eq:kick_simplified}
 {\cal J} \propto \int da H(a)a^2 \Sigma(T(a)) 
 \sim H_0\sqrt{\Omega_R} a_{e}\bar\Sigma\,,
\end{equation}
assuming radiation domination and neglecting the effect of $\Sigma$ on the expansion  (\ref{eq:H_sigma}). The above approximating is equivalent to treating the kick as as a step function with $a_{start}\ll a_{e}$, which is adequate to give an idea of the order of magnitude and time dependence. 
Typical contributions for massive standard model particles are $\Sigma\sim 0.05-0.1$ (See Ref. \cite{Erickcek:2013dea} a detailed computation).

It is possible to express the shift-charge as an energy density fraction for the Galileon using Eqs. (\ref{eq:shift_current_eq},\ref{eq:rho_de_kin}). 
If the cubic term dominates then
\begin{eqnarray}\label{eq:Omega3_kick}
 \hat\Omega_{\phi,3} &=& \frac{1}{4}\sqrt{\frac{a}{|c_3|\Omega_R}}\left(\beta \int_0^a da^\prime \Sigma(T(a^\prime))\right)^{3/2} 
 \\
 &\sim& \frac{1}{4}\frac{1}{\sqrt{|c_3|\Omega_R}}a_{e}^2\left(\beta \bar\Sigma\right)^{3/2} \sqrt{\frac{a}{a_{e}}}\,,
\end{eqnarray}
where the last equality uses the simplified kick expression (\ref{eq:kick_simplified}).
While this contribution dilutes slower than radiation ($\hat\Omega_{\phi,3}\propto\sqrt{a}$), the initial kick is suppressed by $a_{e}^2\ll1$. 
This dependence implies that kicks at an earlier epoch are less important, making it very hard to invoke early universe physics (e.g. new heavy particles with $m>m_\tau$).

The scaling of the cubic Galileon reflects the cosmological Vainshtein screening. This is very different in the case of a quadratic kinetic term, for which
\begin{equation}
\hat\Omega_{\phi,2} 
\sim\frac{3}{8c_2}\beta^2\left(\frac{a_{e}}{a}\right)^2 \bar\Sigma\,. 
\end{equation}
For a canonical kinetic term, a kick contributes a sizeable amount of kinetic energy in the field $\hat\Omega_{\phi,2}\sim \bar\Sigma\beta^2$, which nonetheless kinates away rapidly as $\hat\Omega_{\phi,2}\sim a^{-2}$. 
In contrast, the cubic Galileon is very hard to excite, but any energy injected into the field is persistent, with $\hat\Omega_{\phi,3}$ growing in the radiation era as characteristic of IDEE models, cf section \ref{sec:dynamics_idee}.

Non-luminal Galileons scale more favorably with cosmic expansion, but are equally hard to excite due to the Vainshtein mechanism. If the quartic or quintic term were to dominate the evolution (both the shift-charge and the energy density), the contribution of a kick reads
\begin{eqnarray}
\hat\Omega_{\phi,4} &\sim& \frac{5}{8\left(12 c_4 \Omega_r^2\right)^{1/3}}\left(
 \frac{a}{a_e}\right)^{4/3}a_e^{8/3}\left(\beta\bar\Sigma\right)^{4/3}
 \\
  \hat\Omega_{\phi,5} &\sim& \frac{7}{10}\left(\frac{3}{10c_5 \Omega_r^3}\right)^{1/4}\left(
 \frac{a}{a_e}\right)^{7/4}a_e^{3}\left(\beta\bar\Sigma\right)^{5/4} 
\end{eqnarray}
so early kicks are suppressed by powers of the initial scale factor. Note that while $\hat\Omega_{\phi,4-5}$ grows faster than in the cubic case, this dependence does not compensate for the Vainshtein mechanism, seen here as positive powers of $a_e$, which make kicks at very early times negligible. Note also that the enhancement produced by the small coefficients $c_4,c_5$ (less screening) will not lead to a large kick, but rather to the cubic or canonical term becoming the relevant one.

Just for fun, let us now examine best-case scenario to generate a large IDEE fraction through a coupling.
The best case to generate a large shift-charge density would be a kination phase (e.g. driven by the inflaton) with $w_m\approx -1,\, \Sigma \approx -2,\, H\propto a^{-3}$. 
Note that negative $\Sigma$ requires a negative coupling $\beta<0$ to produce a positive shift-charge. 
Then the integral in Eq. (\ref{eq:charge_sol_explicit}) reads
\begin{equation}\label{eq:kination_shift_contribution}
\int da a^2 H(a) \Sigma 
= -2H_0\sqrt{\Omega_r}a_e\ln\left(a_e/a_i\right)\,,
\end{equation}
where I assumed that kination dominates from $a_i$ to $a_e$ and the universe becomes radiation dominated at $a_e$ (hence relating $H_e = H_0\sqrt{\Omega_r}a_e^2$).
To evaluate the impact of a kination phase on the Galileon density fraction one can substitute $\bar\Sigma\to -2 \ln\left(a_e/a_i\right)$ in the expressions in the previous section.
The logarithmic factor gives a mild dependence on the duration of the kination phase, which can be made arbitrarily large in the limit $a_i\to0$, if the kination phase last long enough.

While possible, imparting a substantial initial energy to the Galileon using a kination phase is extremely unrealistic. The problem is the very rapid scaling of the energy density during a kination phase, with $\rho_i/\rho_e = (a_e/a_i)^6$. The most favorable scenario to affect the acoustic scale via IDEE requires kination to end right before nucleosynthesis, $a_e\sim 10^{10}$, while at the same time producing $\hat\Omega_{\phi}(z_{\rm BBN})\sim 10^{-4}$. 
This would require the kick to be as large as $\bar\Sigma\beta = -2\beta \log(a_e/a_i) \sim 2\cdot 10^9 (\hat\Omega_{\phi,i}/10^{-4})^{2/3}(10^{-10}/a_e)$, corresponding to an initial energy density at the beginning of kination given by $\rho_i/\rho_e = e^{|\Sigma/\beta|}=e^{|\beta|^{-1} 2\cdot 10^9} \sim 6\cdot 10^{868588963}$, where the last value assumes $\beta\sim 1$. 
\footnote{Dear reader: I would love to know if you have ever encountered a larger number in a correct calculation.}
Needless to say, this energy scale is deeply trans-planckian, well beyond the range of validity of the theory as well as the range of validity of classical gravity.

It is clear from the above discussion that the cosmological Vainshtein screening precludes any early universe process to produce a sizeable contribution to IDEE. 
Inflation would dilute the initial energy density of the scalar field very efficiently, requiring a mechanism to produce a sizeable amount of IDEE at reheating or later. 
This necessarily involves physics beyond the classical coupled Galileon theory, possibly through an ultraviolet completion.
This may happen in scenarios of Galilean genesis \cite{Creminelli:2010ba}, a variant of the coupled cubic Galileon in which the scalar field is responsible for setting the initial conditions in the early universe. 
In this scenario, reheating is conjectured to occur when the field configuration exits the effective field theory regime of validity. While a high-energy completion of the theory is necessary for a first principle calculation, it is plausible that the Galileon field producing IDEE might be generated with a sizeable kinetic energy (note that this scalar field might be different from the one causing Galilean genesis).

The Vainshtein mechanism ensures that the initial effective Planck mass $M_*^2(\phi)$ is robust against physical processes in the early universe. 
The smallness of the relative variation of the field $\dot\phi/(H\phi)$ guarantees that $M_*^2(\phi)$ will remain approximately constant until the Hubble rate decreases to a value $H\sim H_0/\sqrt{|c_3|}$. Thus whatever the initial condition $\phi_i$ set in the early universe, its effect on the strength of gravity is robust by virtue of the same physics that prevent the generation of IDEE $\hat\Omega_\phi$. It is interesting that, already at the theoretical level, Enhanced Early Gravity  is much more robust.

\bibliography{galileon_refs}

\begin{thebibliography}{179}
\expandafter\ifx\csname natexlab\endcsname\relax\def\natexlab#1{#1}\fi
\expandafter\ifx\csname bibnamefont\endcsname\relax
  \def\bibnamefont#1{#1}\fi
\expandafter\ifx\csname bibfnamefont\endcsname\relax
  \def\bibfnamefont#1{#1}\fi
\expandafter\ifx\csname citenamefont\endcsname\relax
  \def\citenamefont#1{#1}\fi
\expandafter\ifx\csname url\endcsname\relax
  \def\url#1{\texttt{#1}}\fi
\expandafter\ifx\csname urlprefix\endcsname\relax\def\urlprefix{URL }\fi
\providecommand{\bibinfo}[2]{#2}
\providecommand{\eprint}[2][]{\url{#2}}

\bibitem[{\citenamefont{Aghanim et~al.}(2018)}]{Aghanim:2018eyx}
\bibinfo{author}{\bibfnamefont{N.}~\bibnamefont{Aghanim}} \bibnamefont{et~al.}
  (\bibinfo{collaboration}{Planck}) (\bibinfo{year}{2018}),
  \eprint{1807.06209}.

\bibitem[{\citenamefont{Raveri}(2016)}]{Raveri:2015maa}
\bibinfo{author}{\bibfnamefont{M.}~\bibnamefont{Raveri}},
  \bibinfo{journal}{Phys. Rev.} \textbf{\bibinfo{volume}{D93}},
  \bibinfo{pages}{043522} (\bibinfo{year}{2016}), \eprint{1510.00688}.

\bibitem[{\citenamefont{Riess et~al.}(2019)\citenamefont{Riess, Casertano,
  Yuan, Macri, and Scolnic}}]{Riess:2019cxk}
\bibinfo{author}{\bibfnamefont{A.~G.} \bibnamefont{Riess}},
  \bibinfo{author}{\bibfnamefont{S.}~\bibnamefont{Casertano}},
  \bibinfo{author}{\bibfnamefont{W.}~\bibnamefont{Yuan}},
  \bibinfo{author}{\bibfnamefont{L.~M.} \bibnamefont{Macri}}, \bibnamefont{and}
  \bibinfo{author}{\bibfnamefont{D.}~\bibnamefont{Scolnic}},
  \bibinfo{journal}{Astrophys. J.} \textbf{\bibinfo{volume}{876}},
  \bibinfo{pages}{85} (\bibinfo{year}{2019}), \eprint{1903.07603}.

\bibitem[{\citenamefont{Wong et~al.}(2019)}]{Wong:2019kwg}
\bibinfo{author}{\bibfnamefont{K.~C.} \bibnamefont{Wong}} \bibnamefont{et~al.}
  (\bibinfo{year}{2019}), \eprint{1907.04869}.

\bibitem[{\citenamefont{Shajib et~al.}(2019)}]{Shajib:2019toy}
\bibinfo{author}{\bibfnamefont{A.~J.} \bibnamefont{Shajib}}
  \bibnamefont{et~al.} (\bibinfo{collaboration}{DES}) (\bibinfo{year}{2019}),
  \eprint{1910.06306}.

\bibitem[{\citenamefont{Freedman}(2017)}]{Freedman:2017yms}
\bibinfo{author}{\bibfnamefont{W.~L.} \bibnamefont{Freedman}},
  \bibinfo{journal}{Nat. Astron.} \textbf{\bibinfo{volume}{1}},
  \bibinfo{pages}{0121} (\bibinfo{year}{2017}), \eprint{1706.02739}.

\bibitem[{\citenamefont{Verde et~al.}(2019)\citenamefont{Verde, Treu, and
  Riess}}]{Verde:2019ivm}
\bibinfo{author}{\bibfnamefont{L.}~\bibnamefont{Verde}},
  \bibinfo{author}{\bibfnamefont{T.}~\bibnamefont{Treu}}, \bibnamefont{and}
  \bibinfo{author}{\bibfnamefont{A.~G.} \bibnamefont{Riess}}, in
  \emph{\bibinfo{booktitle}{{Nature Astronomy 2019}}} (\bibinfo{year}{2019}),
  \eprint{1907.10625}.

\bibitem[{\citenamefont{Desmond et~al.}(2019)\citenamefont{Desmond, Jain, and
  Sakstein}}]{Desmond:2019ygn}
\bibinfo{author}{\bibfnamefont{H.}~\bibnamefont{Desmond}},
  \bibinfo{author}{\bibfnamefont{B.}~\bibnamefont{Jain}}, \bibnamefont{and}
  \bibinfo{author}{\bibfnamefont{J.}~\bibnamefont{Sakstein}},
  \bibinfo{journal}{Phys. Rev.} \textbf{\bibinfo{volume}{D100}},
  \bibinfo{pages}{043537} (\bibinfo{year}{2019}), \eprint{1907.03778}.

\bibitem[{\citenamefont{Barreira et~al.}(2013)\citenamefont{Barreira, Li,
  Sanchez, Baugh, and Pascoli}}]{Barreira:2013jma}
\bibinfo{author}{\bibfnamefont{A.}~\bibnamefont{Barreira}},
  \bibinfo{author}{\bibfnamefont{B.}~\bibnamefont{Li}},
  \bibinfo{author}{\bibfnamefont{A.}~\bibnamefont{Sanchez}},
  \bibinfo{author}{\bibfnamefont{C.~M.} \bibnamefont{Baugh}}, \bibnamefont{and}
  \bibinfo{author}{\bibfnamefont{S.}~\bibnamefont{Pascoli}},
  \bibinfo{journal}{Phys. Rev.} \textbf{\bibinfo{volume}{D87}},
  \bibinfo{pages}{103511} (\bibinfo{year}{2013}), \eprint{1302.6241}.

\bibitem[{\citenamefont{Barreira et~al.}(2014)\citenamefont{Barreira, Li,
  Baugh, and Pascoli}}]{Barreira:2014jha}
\bibinfo{author}{\bibfnamefont{A.}~\bibnamefont{Barreira}},
  \bibinfo{author}{\bibfnamefont{B.}~\bibnamefont{Li}},
  \bibinfo{author}{\bibfnamefont{C.}~\bibnamefont{Baugh}}, \bibnamefont{and}
  \bibinfo{author}{\bibfnamefont{S.}~\bibnamefont{Pascoli}},
  \bibinfo{journal}{JCAP} \textbf{\bibinfo{volume}{1408}}, \bibinfo{pages}{059}
  (\bibinfo{year}{2014}), \eprint{1406.0485}.

\bibitem[{\citenamefont{Renk et~al.}(2017)\citenamefont{Renk, Zumalacárregui,
  Montanari, and Barreira}}]{Renk:2017rzu}
\bibinfo{author}{\bibfnamefont{J.}~\bibnamefont{Renk}},
  \bibinfo{author}{\bibfnamefont{M.}~\bibnamefont{Zumalacárregui}},
  \bibinfo{author}{\bibfnamefont{F.}~\bibnamefont{Montanari}},
  \bibnamefont{and} \bibinfo{author}{\bibfnamefont{A.}~\bibnamefont{Barreira}},
  \bibinfo{journal}{JCAP} \textbf{\bibinfo{volume}{1710}}, \bibinfo{pages}{020}
  (\bibinfo{year}{2017}), \eprint{1707.02263}.

\bibitem[{\citenamefont{Di~Valentino et~al.}(2017)\citenamefont{Di~Valentino,
  Melchiorri, and Mena}}]{DiValentino:2017iww}
\bibinfo{author}{\bibfnamefont{E.}~\bibnamefont{Di~Valentino}},
  \bibinfo{author}{\bibfnamefont{A.}~\bibnamefont{Melchiorri}},
  \bibnamefont{and} \bibinfo{author}{\bibfnamefont{O.}~\bibnamefont{Mena}},
  \bibinfo{journal}{Phys. Rev. D} \textbf{\bibinfo{volume}{96}},
  \bibinfo{pages}{043503} (\bibinfo{year}{2017}), \eprint{1704.08342}.

\bibitem[{\citenamefont{Solà et~al.}(2017)\citenamefont{Solà, Gómez-Valent,
  and de~Cruz~Pérez}}]{Sola:2017znb}
\bibinfo{author}{\bibfnamefont{J.}~\bibnamefont{Solà}},
  \bibinfo{author}{\bibfnamefont{A.}~\bibnamefont{Gómez-Valent}},
  \bibnamefont{and}
  \bibinfo{author}{\bibfnamefont{J.}~\bibnamefont{de~Cruz~Pérez}},
  \bibinfo{journal}{Phys. Lett.} \textbf{\bibinfo{volume}{B774}},
  \bibinfo{pages}{317} (\bibinfo{year}{2017}), \eprint{1705.06723}.

\bibitem[{\citenamefont{Poulin et~al.}(2018)\citenamefont{Poulin, Boddy, Bird,
  and Kamionkowski}}]{Poulin:2018zxs}
\bibinfo{author}{\bibfnamefont{V.}~\bibnamefont{Poulin}},
  \bibinfo{author}{\bibfnamefont{K.~K.} \bibnamefont{Boddy}},
  \bibinfo{author}{\bibfnamefont{S.}~\bibnamefont{Bird}}, \bibnamefont{and}
  \bibinfo{author}{\bibfnamefont{M.}~\bibnamefont{Kamionkowski}},
  \bibinfo{journal}{Phys. Rev.} \textbf{\bibinfo{volume}{D97}},
  \bibinfo{pages}{123504} (\bibinfo{year}{2018}), \eprint{1803.02474}.

\bibitem[{\citenamefont{Yang et~al.}(2018)\citenamefont{Yang, Pan,
  Di~Valentino, Nunes, Vagnozzi, and Mota}}]{Yang:2018euj}
\bibinfo{author}{\bibfnamefont{W.}~\bibnamefont{Yang}},
  \bibinfo{author}{\bibfnamefont{S.}~\bibnamefont{Pan}},
  \bibinfo{author}{\bibfnamefont{E.}~\bibnamefont{Di~Valentino}},
  \bibinfo{author}{\bibfnamefont{R.~C.} \bibnamefont{Nunes}},
  \bibinfo{author}{\bibfnamefont{S.}~\bibnamefont{Vagnozzi}}, \bibnamefont{and}
  \bibinfo{author}{\bibfnamefont{D.~F.} \bibnamefont{Mota}},
  \bibinfo{journal}{JCAP} \textbf{\bibinfo{volume}{1809}}, \bibinfo{pages}{019}
  (\bibinfo{year}{2018}), \eprint{1805.08252}.

\bibitem[{\citenamefont{Vattis et~al.}(2019)\citenamefont{Vattis, Koushiappas,
  and Loeb}}]{Vattis:2019efj}
\bibinfo{author}{\bibfnamefont{K.}~\bibnamefont{Vattis}},
  \bibinfo{author}{\bibfnamefont{S.~M.} \bibnamefont{Koushiappas}},
  \bibnamefont{and} \bibinfo{author}{\bibfnamefont{A.}~\bibnamefont{Loeb}},
  \bibinfo{journal}{Phys. Rev.} \textbf{\bibinfo{volume}{D99}},
  \bibinfo{pages}{121302} (\bibinfo{year}{2019}), \eprint{1903.06220}.

\bibitem[{\citenamefont{Di~Valentino et~al.}(2019)\citenamefont{Di~Valentino,
  Melchiorri, Mena, and Vagnozzi}}]{DiValentino:2019ffd}
\bibinfo{author}{\bibfnamefont{E.}~\bibnamefont{Di~Valentino}},
  \bibinfo{author}{\bibfnamefont{A.}~\bibnamefont{Melchiorri}},
  \bibinfo{author}{\bibfnamefont{O.}~\bibnamefont{Mena}}, \bibnamefont{and}
  \bibinfo{author}{\bibfnamefont{S.}~\bibnamefont{Vagnozzi}}
  (\bibinfo{year}{2019}), \eprint{1908.04281}.

\bibitem[{\citenamefont{Bernal et~al.}(2016)\citenamefont{Bernal, Verde, and
  Riess}}]{Bernal:2016gxb}
\bibinfo{author}{\bibfnamefont{J.~L.} \bibnamefont{Bernal}},
  \bibinfo{author}{\bibfnamefont{L.}~\bibnamefont{Verde}}, \bibnamefont{and}
  \bibinfo{author}{\bibfnamefont{A.~G.} \bibnamefont{Riess}},
  \bibinfo{journal}{JCAP} \textbf{\bibinfo{volume}{1610}}, \bibinfo{pages}{019}
  (\bibinfo{year}{2016}), \eprint{1607.05617}.

\bibitem[{\citenamefont{Raveri}(2019)}]{Raveri:2019mxg}
\bibinfo{author}{\bibfnamefont{M.}~\bibnamefont{Raveri}}
  (\bibinfo{year}{2019}), \eprint{1902.01366}.

\bibitem[{\citenamefont{Mörtsell and Dhawan}(2018)}]{Mortsell:2018mfj}
\bibinfo{author}{\bibfnamefont{E.}~\bibnamefont{Mörtsell}} \bibnamefont{and}
  \bibinfo{author}{\bibfnamefont{S.}~\bibnamefont{Dhawan}},
  \bibinfo{journal}{JCAP} \textbf{\bibinfo{volume}{1809}}, \bibinfo{pages}{025}
  (\bibinfo{year}{2018}), \eprint{1801.07260}.

\bibitem[{\citenamefont{Millon et~al.}(2019)}]{Millon:2019slk}
\bibinfo{author}{\bibfnamefont{M.}~\bibnamefont{Millon}} \bibnamefont{et~al.}
  (\bibinfo{year}{2019}), \eprint{1912.08027}.

\bibitem[{\citenamefont{Vagnozzi}(2019)}]{Vagnozzi:2019ezj}
\bibinfo{author}{\bibfnamefont{S.}~\bibnamefont{Vagnozzi}}
  (\bibinfo{year}{2019}), \eprint{1907.07569}.

\bibitem[{\citenamefont{Krishnan et~al.}(2020)\citenamefont{Krishnan, Colgain,
  Ruchika, Sen, Sheikh-Jabbari, and Yang}}]{Krishnan:2020obg}
\bibinfo{author}{\bibfnamefont{C.}~\bibnamefont{Krishnan}},
  \bibinfo{author}{\bibfnamefont{E.~O.} \bibnamefont{Colgain}},
  \bibinfo{author}{\bibnamefont{Ruchika}},
  \bibinfo{author}{\bibfnamefont{A.~A.} \bibnamefont{Sen}},
  \bibinfo{author}{\bibfnamefont{M.~M.} \bibnamefont{Sheikh-Jabbari}},
  \bibnamefont{and} \bibinfo{author}{\bibfnamefont{T.}~\bibnamefont{Yang}}
  (\bibinfo{year}{2020}), \eprint{2002.06044}.

\bibitem[{\citenamefont{Suyu et~al.}(2020)}]{Suyu:2020opl}
\bibinfo{author}{\bibfnamefont{S.~H.} \bibnamefont{Suyu}} \bibnamefont{et~al.}
  (\bibinfo{year}{2020}), \eprint{2002.08378}.

\bibitem[{\citenamefont{Nicolis et~al.}(2009)\citenamefont{Nicolis, Rattazzi,
  and Trincherini}}]{Nicolis:2008in}
\bibinfo{author}{\bibfnamefont{A.}~\bibnamefont{Nicolis}},
  \bibinfo{author}{\bibfnamefont{R.}~\bibnamefont{Rattazzi}}, \bibnamefont{and}
  \bibinfo{author}{\bibfnamefont{E.}~\bibnamefont{Trincherini}},
  \bibinfo{journal}{Phys. Rev.} \textbf{\bibinfo{volume}{D79}},
  \bibinfo{pages}{064036} (\bibinfo{year}{2009}), \eprint{0811.2197}.

\bibitem[{\citenamefont{Abbott et~al.}(2017{\natexlab{a}})}]{Monitor:2017mdv}
\bibinfo{author}{\bibfnamefont{B.~P.} \bibnamefont{Abbott}}
  \bibnamefont{et~al.} (\bibinfo{collaboration}{LIGO Scientific, Virgo,
  Fermi-GBM, INTEGRAL}), \bibinfo{journal}{Astrophys. J.}
  \textbf{\bibinfo{volume}{848}}, \bibinfo{pages}{L13}
  (\bibinfo{year}{2017}{\natexlab{a}}), \eprint{1710.05834}.

\bibitem[{\citenamefont{Ezquiaga and Zumalacárregui}(2017)}]{Ezquiaga:2017ekz}
\bibinfo{author}{\bibfnamefont{J.~M.} \bibnamefont{Ezquiaga}} \bibnamefont{and}
  \bibinfo{author}{\bibfnamefont{M.}~\bibnamefont{Zumalacárregui}},
  \bibinfo{journal}{Phys. Rev. Lett.} \textbf{\bibinfo{volume}{119}},
  \bibinfo{pages}{251304} (\bibinfo{year}{2017}), \eprint{1710.05901}.

\bibitem[{\citenamefont{Creminelli and Vernizzi}(2017)}]{Creminelli:2017sry}
\bibinfo{author}{\bibfnamefont{P.}~\bibnamefont{Creminelli}} \bibnamefont{and}
  \bibinfo{author}{\bibfnamefont{F.}~\bibnamefont{Vernizzi}},
  \bibinfo{journal}{Phys. Rev. Lett.} \textbf{\bibinfo{volume}{119}},
  \bibinfo{pages}{251302} (\bibinfo{year}{2017}), \eprint{1710.05877}.

\bibitem[{\citenamefont{Baker et~al.}(2017)\citenamefont{Baker, Bellini,
  Ferreira, Lagos, Noller, and Sawicki}}]{Baker:2017hug}
\bibinfo{author}{\bibfnamefont{T.}~\bibnamefont{Baker}},
  \bibinfo{author}{\bibfnamefont{E.}~\bibnamefont{Bellini}},
  \bibinfo{author}{\bibfnamefont{P.~G.} \bibnamefont{Ferreira}},
  \bibinfo{author}{\bibfnamefont{M.}~\bibnamefont{Lagos}},
  \bibinfo{author}{\bibfnamefont{J.}~\bibnamefont{Noller}}, \bibnamefont{and}
  \bibinfo{author}{\bibfnamefont{I.}~\bibnamefont{Sawicki}},
  \bibinfo{journal}{Phys. Rev. Lett.} \textbf{\bibinfo{volume}{119}},
  \bibinfo{pages}{251301} (\bibinfo{year}{2017}), \eprint{1710.06394}.

\bibitem[{\citenamefont{Sakstein and Jain}(2017)}]{Sakstein:2017xjx}
\bibinfo{author}{\bibfnamefont{J.}~\bibnamefont{Sakstein}} \bibnamefont{and}
  \bibinfo{author}{\bibfnamefont{B.}~\bibnamefont{Jain}},
  \bibinfo{journal}{Phys. Rev. Lett.} \textbf{\bibinfo{volume}{119}},
  \bibinfo{pages}{251303} (\bibinfo{year}{2017}), \eprint{1710.05893}.

\bibitem[{\citenamefont{Heavens et~al.}(2014)\citenamefont{Heavens, Jimenez,
  and Verde}}]{Heavens:2014rja}
\bibinfo{author}{\bibfnamefont{A.}~\bibnamefont{Heavens}},
  \bibinfo{author}{\bibfnamefont{R.}~\bibnamefont{Jimenez}}, \bibnamefont{and}
  \bibinfo{author}{\bibfnamefont{L.}~\bibnamefont{Verde}},
  \bibinfo{journal}{Phys. Rev. Lett.} \textbf{\bibinfo{volume}{113}},
  \bibinfo{pages}{241302} (\bibinfo{year}{2014}), \eprint{1409.6217}.

\bibitem[{\citenamefont{Aylor et~al.}(2019)\citenamefont{Aylor, Joy, Knox,
  Millea, Raghunathan, and Wu}}]{Aylor:2018drw}
\bibinfo{author}{\bibfnamefont{K.}~\bibnamefont{Aylor}},
  \bibinfo{author}{\bibfnamefont{M.}~\bibnamefont{Joy}},
  \bibinfo{author}{\bibfnamefont{L.}~\bibnamefont{Knox}},
  \bibinfo{author}{\bibfnamefont{M.}~\bibnamefont{Millea}},
  \bibinfo{author}{\bibfnamefont{S.}~\bibnamefont{Raghunathan}},
  \bibnamefont{and} \bibinfo{author}{\bibfnamefont{W.~L.~K.} \bibnamefont{Wu}},
  \bibinfo{journal}{Astrophys. J.} \textbf{\bibinfo{volume}{874}},
  \bibinfo{pages}{4} (\bibinfo{year}{2019}), \eprint{1811.00537}.

\bibitem[{\citenamefont{Knox and Millea}(2019)}]{Knox:2019rjx}
\bibinfo{author}{\bibfnamefont{L.}~\bibnamefont{Knox}} \bibnamefont{and}
  \bibinfo{author}{\bibfnamefont{M.}~\bibnamefont{Millea}}
  (\bibinfo{year}{2019}), \eprint{1908.03663}.

\bibitem[{\citenamefont{Eisenstein and White}(2004)}]{Eisenstein:2004an}
\bibinfo{author}{\bibfnamefont{D.~J.} \bibnamefont{Eisenstein}}
  \bibnamefont{and} \bibinfo{author}{\bibfnamefont{M.~J.} \bibnamefont{White}},
  \bibinfo{journal}{Phys. Rev.} \textbf{\bibinfo{volume}{D70}},
  \bibinfo{pages}{103523} (\bibinfo{year}{2004}), \eprint{astro-ph/0407539}.

\bibitem[{\citenamefont{Barker et~al.}(2020)\citenamefont{Barker, Lasenby,
  Hobson, and Handley}}]{Barker:2020gcp}
\bibinfo{author}{\bibfnamefont{W.~E.~V.} \bibnamefont{Barker}},
  \bibinfo{author}{\bibfnamefont{A.~N.} \bibnamefont{Lasenby}},
  \bibinfo{author}{\bibfnamefont{M.~P.} \bibnamefont{Hobson}},
  \bibnamefont{and} \bibinfo{author}{\bibfnamefont{W.~J.}
  \bibnamefont{Handley}} (\bibinfo{year}{2020}), \eprint{2003.02690}.

\bibitem[{\citenamefont{Di~Valentino et~al.}(2018)\citenamefont{Di~Valentino,
  Bøehm, Hivon, and Bouchet}}]{DiValentino:2017oaw}
\bibinfo{author}{\bibfnamefont{E.}~\bibnamefont{Di~Valentino}},
  \bibinfo{author}{\bibfnamefont{C.}~\bibnamefont{Bøehm}},
  \bibinfo{author}{\bibfnamefont{E.}~\bibnamefont{Hivon}}, \bibnamefont{and}
  \bibinfo{author}{\bibfnamefont{F.~R.} \bibnamefont{Bouchet}},
  \bibinfo{journal}{Phys. Rev.} \textbf{\bibinfo{volume}{D97}},
  \bibinfo{pages}{043513} (\bibinfo{year}{2018}), \eprint{1710.02559}.

\bibitem[{\citenamefont{Kreisch et~al.}(2019)\citenamefont{Kreisch, Cyr-Racine,
  and Doré}}]{Kreisch:2019yzn}
\bibinfo{author}{\bibfnamefont{C.~D.} \bibnamefont{Kreisch}},
  \bibinfo{author}{\bibfnamefont{F.-Y.} \bibnamefont{Cyr-Racine}},
  \bibnamefont{and} \bibinfo{author}{\bibfnamefont{O.}~\bibnamefont{Doré}}
  (\bibinfo{year}{2019}), \eprint{1902.00534}.

\bibitem[{\citenamefont{Escudero and Witte}(2019)}]{Escudero:2019gvw}
\bibinfo{author}{\bibfnamefont{M.}~\bibnamefont{Escudero}} \bibnamefont{and}
  \bibinfo{author}{\bibfnamefont{S.~J.} \bibnamefont{Witte}}
  (\bibinfo{year}{2019}), \eprint{1909.04044}.

\bibitem[{\citenamefont{Blinov et~al.}(2019)\citenamefont{Blinov, Kelly,
  Krnjaic, and McDermott}}]{Blinov:2019gcj}
\bibinfo{author}{\bibfnamefont{N.}~\bibnamefont{Blinov}},
  \bibinfo{author}{\bibfnamefont{K.~J.} \bibnamefont{Kelly}},
  \bibinfo{author}{\bibfnamefont{G.~Z.} \bibnamefont{Krnjaic}},
  \bibnamefont{and} \bibinfo{author}{\bibfnamefont{S.~D.}
  \bibnamefont{McDermott}}, \bibinfo{journal}{Phys. Rev. Lett.}
  \textbf{\bibinfo{volume}{123}}, \bibinfo{pages}{191102}
  (\bibinfo{year}{2019}), \eprint{1905.02727}.

\bibitem[{\citenamefont{Hart and Chluba}(2019)}]{Hart:2019dxi}
\bibinfo{author}{\bibfnamefont{L.}~\bibnamefont{Hart}} \bibnamefont{and}
  \bibinfo{author}{\bibfnamefont{J.}~\bibnamefont{Chluba}}
  (\bibinfo{year}{2019}), \eprint{1912.03986}.

\bibitem[{\citenamefont{Raveri et~al.}(2017)\citenamefont{Raveri, Hu, Hoffman,
  and Wang}}]{Raveri:2017jto}
\bibinfo{author}{\bibfnamefont{M.}~\bibnamefont{Raveri}},
  \bibinfo{author}{\bibfnamefont{W.}~\bibnamefont{Hu}},
  \bibinfo{author}{\bibfnamefont{T.}~\bibnamefont{Hoffman}}, \bibnamefont{and}
  \bibinfo{author}{\bibfnamefont{L.-T.} \bibnamefont{Wang}},
  \bibinfo{journal}{Phys. Rev.} \textbf{\bibinfo{volume}{D96}},
  \bibinfo{pages}{103501} (\bibinfo{year}{2017}), \eprint{1709.04877}.

\bibitem[{\citenamefont{Archidiacono et~al.}(2019)\citenamefont{Archidiacono,
  Hooper, Murgia, Bohr, Lesgourgues, and Viel}}]{Archidiacono:2019wdp}
\bibinfo{author}{\bibfnamefont{M.}~\bibnamefont{Archidiacono}},
  \bibinfo{author}{\bibfnamefont{D.~C.} \bibnamefont{Hooper}},
  \bibinfo{author}{\bibfnamefont{R.}~\bibnamefont{Murgia}},
  \bibinfo{author}{\bibfnamefont{S.}~\bibnamefont{Bohr}},
  \bibinfo{author}{\bibfnamefont{J.}~\bibnamefont{Lesgourgues}},
  \bibnamefont{and} \bibinfo{author}{\bibfnamefont{M.}~\bibnamefont{Viel}},
  \bibinfo{journal}{JCAP} \textbf{\bibinfo{volume}{1910}}, \bibinfo{pages}{055}
  (\bibinfo{year}{2019}), \eprint{1907.01496}.

\bibitem[{\citenamefont{Karwal and Kamionkowski}(2016)}]{Karwal:2016vyq}
\bibinfo{author}{\bibfnamefont{T.}~\bibnamefont{Karwal}} \bibnamefont{and}
  \bibinfo{author}{\bibfnamefont{M.}~\bibnamefont{Kamionkowski}},
  \bibinfo{journal}{Phys. Rev.} \textbf{\bibinfo{volume}{D94}},
  \bibinfo{pages}{103523} (\bibinfo{year}{2016}), \eprint{1608.01309}.

\bibitem[{\citenamefont{Zhao et~al.}(2017)}]{Zhao:2017cud}
\bibinfo{author}{\bibfnamefont{G.-B.} \bibnamefont{Zhao}} \bibnamefont{et~al.},
  \bibinfo{journal}{Nat. Astron.} \textbf{\bibinfo{volume}{1}},
  \bibinfo{pages}{627} (\bibinfo{year}{2017}), \eprint{1701.08165}.

\bibitem[{\citenamefont{Doran and Robbers}(2006)}]{Doran:2006kp}
\bibinfo{author}{\bibfnamefont{M.}~\bibnamefont{Doran}} \bibnamefont{and}
  \bibinfo{author}{\bibfnamefont{G.}~\bibnamefont{Robbers}},
  \bibinfo{journal}{JCAP} \textbf{\bibinfo{volume}{0606}}, \bibinfo{pages}{026}
  (\bibinfo{year}{2006}), \eprint{astro-ph/0601544}.

\bibitem[{\citenamefont{Pettorino et~al.}(2013)\citenamefont{Pettorino,
  Amendola, and Wetterich}}]{Pettorino:2013ia}
\bibinfo{author}{\bibfnamefont{V.}~\bibnamefont{Pettorino}},
  \bibinfo{author}{\bibfnamefont{L.}~\bibnamefont{Amendola}}, \bibnamefont{and}
  \bibinfo{author}{\bibfnamefont{C.}~\bibnamefont{Wetterich}},
  \bibinfo{journal}{Phys. Rev.} \textbf{\bibinfo{volume}{D87}},
  \bibinfo{pages}{083009} (\bibinfo{year}{2013}), \eprint{1301.5279}.

\bibitem[{\citenamefont{Lin et~al.}(2019{\natexlab{a}})\citenamefont{Lin,
  Raveri, and Hu}}]{Lin:2018nxe}
\bibinfo{author}{\bibfnamefont{M.-X.} \bibnamefont{Lin}},
  \bibinfo{author}{\bibfnamefont{M.}~\bibnamefont{Raveri}}, \bibnamefont{and}
  \bibinfo{author}{\bibfnamefont{W.}~\bibnamefont{Hu}}, \bibinfo{journal}{Phys.
  Rev.} \textbf{\bibinfo{volume}{D99}}, \bibinfo{pages}{043514}
  (\bibinfo{year}{2019}{\natexlab{a}}), \eprint{1810.02333}.

\bibitem[{\citenamefont{Poulin et~al.}(2019)\citenamefont{Poulin, Smith,
  Karwal, and Kamionkowski}}]{Poulin:2018cxd}
\bibinfo{author}{\bibfnamefont{V.}~\bibnamefont{Poulin}},
  \bibinfo{author}{\bibfnamefont{T.~L.} \bibnamefont{Smith}},
  \bibinfo{author}{\bibfnamefont{T.}~\bibnamefont{Karwal}}, \bibnamefont{and}
  \bibinfo{author}{\bibfnamefont{M.}~\bibnamefont{Kamionkowski}},
  \bibinfo{journal}{Phys. Rev. Lett.} \textbf{\bibinfo{volume}{122}},
  \bibinfo{pages}{221301} (\bibinfo{year}{2019}), \eprint{1811.04083}.

\bibitem[{\citenamefont{Agrawal et~al.}(2019)\citenamefont{Agrawal, Cyr-Racine,
  Pinner, and Randall}}]{Agrawal:2019lmo}
\bibinfo{author}{\bibfnamefont{P.}~\bibnamefont{Agrawal}},
  \bibinfo{author}{\bibfnamefont{F.-Y.} \bibnamefont{Cyr-Racine}},
  \bibinfo{author}{\bibfnamefont{D.}~\bibnamefont{Pinner}}, \bibnamefont{and}
  \bibinfo{author}{\bibfnamefont{L.}~\bibnamefont{Randall}}
  (\bibinfo{year}{2019}), \eprint{1904.01016}.

\bibitem[{\citenamefont{Smith et~al.}(2019)\citenamefont{Smith, Poulin, and
  Amin}}]{Smith:2019ihp}
\bibinfo{author}{\bibfnamefont{T.~L.} \bibnamefont{Smith}},
  \bibinfo{author}{\bibfnamefont{V.}~\bibnamefont{Poulin}}, \bibnamefont{and}
  \bibinfo{author}{\bibfnamefont{M.~A.} \bibnamefont{Amin}}
  (\bibinfo{year}{2019}), \eprint{1908.06995}.

\bibitem[{\citenamefont{Caldwell and Linder}(2005)}]{Caldwell:2005tm}
\bibinfo{author}{\bibfnamefont{R.~R.} \bibnamefont{Caldwell}} \bibnamefont{and}
  \bibinfo{author}{\bibfnamefont{E.~V.} \bibnamefont{Linder}},
  \bibinfo{journal}{Phys. Rev. Lett.} \textbf{\bibinfo{volume}{95}},
  \bibinfo{pages}{141301} (\bibinfo{year}{2005}), \eprint{astro-ph/0505494}.

\bibitem[{\citenamefont{Turner}(1983)}]{Turner:1983he}
\bibinfo{author}{\bibfnamefont{M.~S.} \bibnamefont{Turner}},
  \bibinfo{journal}{Phys. Rev.} \textbf{\bibinfo{volume}{D28}},
  \bibinfo{pages}{1243} (\bibinfo{year}{1983}).

\bibitem[{\citenamefont{Sakstein and Trodden}(2019)}]{Sakstein:2019fmf}
\bibinfo{author}{\bibfnamefont{J.}~\bibnamefont{Sakstein}} \bibnamefont{and}
  \bibinfo{author}{\bibfnamefont{M.}~\bibnamefont{Trodden}}
  (\bibinfo{year}{2019}), \eprint{1911.11760}.

\bibitem[{\citenamefont{Lin et~al.}(2019{\natexlab{b}})\citenamefont{Lin,
  Benevento, Hu, and Raveri}}]{Lin:2019qug}
\bibinfo{author}{\bibfnamefont{M.-X.} \bibnamefont{Lin}},
  \bibinfo{author}{\bibfnamefont{G.}~\bibnamefont{Benevento}},
  \bibinfo{author}{\bibfnamefont{W.}~\bibnamefont{Hu}}, \bibnamefont{and}
  \bibinfo{author}{\bibfnamefont{M.}~\bibnamefont{Raveri}},
  \bibinfo{journal}{Phys. Rev.} \textbf{\bibinfo{volume}{D100}},
  \bibinfo{pages}{063542} (\bibinfo{year}{2019}{\natexlab{b}}),
  \eprint{1905.12618}.

\bibitem[{\citenamefont{Niedermann and Sloth}(2019)}]{Niedermann:2019olb}
\bibinfo{author}{\bibfnamefont{F.}~\bibnamefont{Niedermann}} \bibnamefont{and}
  \bibinfo{author}{\bibfnamefont{M.~S.} \bibnamefont{Sloth}}
  (\bibinfo{year}{2019}), \eprint{1910.10739}.

\bibitem[{\citenamefont{Ezquiaga and Zumalacárregui}(2018)}]{Ezquiaga:2018btd}
\bibinfo{author}{\bibfnamefont{J.~M.} \bibnamefont{Ezquiaga}} \bibnamefont{and}
  \bibinfo{author}{\bibfnamefont{M.}~\bibnamefont{Zumalacárregui}},
  \bibinfo{journal}{Front. Astron. Space Sci.} \textbf{\bibinfo{volume}{5}},
  \bibinfo{pages}{44} (\bibinfo{year}{2018}), \eprint{1807.09241}.

\bibitem[{\citenamefont{Horndeski}(1974)}]{Horndeski:1974wa}
\bibinfo{author}{\bibfnamefont{G.~W.} \bibnamefont{Horndeski}},
  \bibinfo{journal}{Int.J.Theor.Phys.} \textbf{\bibinfo{volume}{10}},
  \bibinfo{pages}{363} (\bibinfo{year}{1974}).

\bibitem[{\citenamefont{Deffayet et~al.}(2011)\citenamefont{Deffayet, Gao,
  Steer, and Zahariade}}]{Deffayet:2011gz}
\bibinfo{author}{\bibfnamefont{C.}~\bibnamefont{Deffayet}},
  \bibinfo{author}{\bibfnamefont{X.}~\bibnamefont{Gao}},
  \bibinfo{author}{\bibfnamefont{D.~A.} \bibnamefont{Steer}}, \bibnamefont{and}
  \bibinfo{author}{\bibfnamefont{G.}~\bibnamefont{Zahariade}},
  \bibinfo{journal}{Phys. Rev.} \textbf{\bibinfo{volume}{D84}},
  \bibinfo{pages}{064039} (\bibinfo{year}{2011}), \eprint{1103.3260}.

\bibitem[{\citenamefont{Kobayashi et~al.}(2011)\citenamefont{Kobayashi,
  Yamaguchi, and Yokoyama}}]{Kobayashi:2011nu}
\bibinfo{author}{\bibfnamefont{T.}~\bibnamefont{Kobayashi}},
  \bibinfo{author}{\bibfnamefont{M.}~\bibnamefont{Yamaguchi}},
  \bibnamefont{and} \bibinfo{author}{\bibfnamefont{J.}~\bibnamefont{Yokoyama}},
  \bibinfo{journal}{Prog. Theor. Phys.} \textbf{\bibinfo{volume}{126}},
  \bibinfo{pages}{511} (\bibinfo{year}{2011}), \eprint{1105.5723}.

\bibitem[{\citenamefont{Zumalacárregui
  et~al.}(2017)\citenamefont{Zumalacárregui, Bellini, Sawicki, Lesgourgues,
  and Ferreira}}]{Zumalacarregui:2016pph}
\bibinfo{author}{\bibfnamefont{M.}~\bibnamefont{Zumalacárregui}},
  \bibinfo{author}{\bibfnamefont{E.}~\bibnamefont{Bellini}},
  \bibinfo{author}{\bibfnamefont{I.}~\bibnamefont{Sawicki}},
  \bibinfo{author}{\bibfnamefont{J.}~\bibnamefont{Lesgourgues}},
  \bibnamefont{and} \bibinfo{author}{\bibfnamefont{P.~G.}
  \bibnamefont{Ferreira}}, \bibinfo{journal}{JCAP}
  \textbf{\bibinfo{volume}{1708}}, \bibinfo{pages}{019} (\bibinfo{year}{2017}),
  \eprint{1605.06102}.

\bibitem[{\citenamefont{Bellini et~al.}(2019)\citenamefont{Bellini, Sawicki,
  and Zumalacárregui}}]{Bellini:2019syt}
\bibinfo{author}{\bibfnamefont{E.}~\bibnamefont{Bellini}},
  \bibinfo{author}{\bibfnamefont{I.}~\bibnamefont{Sawicki}}, \bibnamefont{and}
  \bibinfo{author}{\bibfnamefont{M.}~\bibnamefont{Zumalacárregui}}
  (\bibinfo{year}{2019}), \eprint{1909.01828}.

\bibitem[{\citenamefont{Deffayet et~al.}(2009)\citenamefont{Deffayet,
  Esposito-Farese, and Vikman}}]{Deffayet:2009wt}
\bibinfo{author}{\bibfnamefont{C.}~\bibnamefont{Deffayet}},
  \bibinfo{author}{\bibfnamefont{G.}~\bibnamefont{Esposito-Farese}},
  \bibnamefont{and} \bibinfo{author}{\bibfnamefont{A.}~\bibnamefont{Vikman}},
  \bibinfo{journal}{Phys. Rev.} \textbf{\bibinfo{volume}{D79}},
  \bibinfo{pages}{084003} (\bibinfo{year}{2009}), \eprint{0901.1314}.

\bibitem[{\citenamefont{De~Felice and Tsujikawa}(2010)}]{DeFelice:2010pv}
\bibinfo{author}{\bibfnamefont{A.}~\bibnamefont{De~Felice}} \bibnamefont{and}
  \bibinfo{author}{\bibfnamefont{S.}~\bibnamefont{Tsujikawa}},
  \bibinfo{journal}{Phys. Rev. Lett.} \textbf{\bibinfo{volume}{105}},
  \bibinfo{pages}{111301} (\bibinfo{year}{2010}), \eprint{1007.2700}.

\bibitem[{\citenamefont{Bellini and Jimenez}(2013)}]{Bellini:2013hea}
\bibinfo{author}{\bibfnamefont{E.}~\bibnamefont{Bellini}} \bibnamefont{and}
  \bibinfo{author}{\bibfnamefont{R.}~\bibnamefont{Jimenez}},
  \bibinfo{journal}{Phys. Dark Univ.} \textbf{\bibinfo{volume}{2}},
  \bibinfo{pages}{179} (\bibinfo{year}{2013}), \eprint{1306.1262}.

\bibitem[{\citenamefont{Aker et~al.}(2019)}]{Aker:2019uuj}
\bibinfo{author}{\bibfnamefont{M.}~\bibnamefont{Aker}} \bibnamefont{et~al.}
  (\bibinfo{collaboration}{KATRIN}), \bibinfo{journal}{Phys. Rev. Lett.}
  \textbf{\bibinfo{volume}{123}}, \bibinfo{pages}{221802}
  (\bibinfo{year}{2019}), \eprint{1909.06048}.

\bibitem[{\citenamefont{Ferraro et~al.}(2015)\citenamefont{Ferraro, Sherwin,
  and Spergel}}]{Ferraro:2014msa}
\bibinfo{author}{\bibfnamefont{S.}~\bibnamefont{Ferraro}},
  \bibinfo{author}{\bibfnamefont{B.~D.} \bibnamefont{Sherwin}},
  \bibnamefont{and} \bibinfo{author}{\bibfnamefont{D.~N.}
  \bibnamefont{Spergel}}, \bibinfo{journal}{Phys. Rev.}
  \textbf{\bibinfo{volume}{D91}}, \bibinfo{pages}{083533}
  (\bibinfo{year}{2015}), \eprint{1401.1193}.

\bibitem[{\citenamefont{Lombriser and Taylor}(2016)}]{Lombriser:2015sxa}
\bibinfo{author}{\bibfnamefont{L.}~\bibnamefont{Lombriser}} \bibnamefont{and}
  \bibinfo{author}{\bibfnamefont{A.}~\bibnamefont{Taylor}},
  \bibinfo{journal}{JCAP} \textbf{\bibinfo{volume}{1603}}, \bibinfo{pages}{031}
  (\bibinfo{year}{2016}), \eprint{1509.08458}.

\bibitem[{\citenamefont{Brax et~al.}(2016)\citenamefont{Brax, Burrage, and
  Davis}}]{Brax:2015dma}
\bibinfo{author}{\bibfnamefont{P.}~\bibnamefont{Brax}},
  \bibinfo{author}{\bibfnamefont{C.}~\bibnamefont{Burrage}}, \bibnamefont{and}
  \bibinfo{author}{\bibfnamefont{A.-C.} \bibnamefont{Davis}},
  \bibinfo{journal}{JCAP} \textbf{\bibinfo{volume}{1603}}, \bibinfo{pages}{004}
  (\bibinfo{year}{2016}), \eprint{1510.03701}.

\bibitem[{\citenamefont{Bettoni et~al.}(2017)\citenamefont{Bettoni, Ezquiaga,
  Hinterbichler, and Zumalacárregui}}]{Bettoni:2016mij}
\bibinfo{author}{\bibfnamefont{D.}~\bibnamefont{Bettoni}},
  \bibinfo{author}{\bibfnamefont{J.~M.} \bibnamefont{Ezquiaga}},
  \bibinfo{author}{\bibfnamefont{K.}~\bibnamefont{Hinterbichler}},
  \bibnamefont{and}
  \bibinfo{author}{\bibfnamefont{M.}~\bibnamefont{Zumalacárregui}},
  \bibinfo{journal}{Phys. Rev.} \textbf{\bibinfo{volume}{D95}},
  \bibinfo{pages}{084029} (\bibinfo{year}{2017}), \eprint{1608.01982}.

\bibitem[{\citenamefont{Zumalacárregui and
  García-Bellido}(2014)}]{Zumalacarregui:2013pma}
\bibinfo{author}{\bibfnamefont{M.}~\bibnamefont{Zumalacárregui}}
  \bibnamefont{and}
  \bibinfo{author}{\bibfnamefont{J.}~\bibnamefont{García-Bellido}},
  \bibinfo{journal}{Phys. Rev.} \textbf{\bibinfo{volume}{D89}},
  \bibinfo{pages}{064046} (\bibinfo{year}{2014}), \eprint{1308.4685}.

\bibitem[{\citenamefont{Gleyzes et~al.}(2015)\citenamefont{Gleyzes, Langlois,
  Piazza, and Vernizzi}}]{Gleyzes:2014dya}
\bibinfo{author}{\bibfnamefont{J.}~\bibnamefont{Gleyzes}},
  \bibinfo{author}{\bibfnamefont{D.}~\bibnamefont{Langlois}},
  \bibinfo{author}{\bibfnamefont{F.}~\bibnamefont{Piazza}}, \bibnamefont{and}
  \bibinfo{author}{\bibfnamefont{F.}~\bibnamefont{Vernizzi}},
  \bibinfo{journal}{Phys. Rev. Lett.} \textbf{\bibinfo{volume}{114}},
  \bibinfo{pages}{211101} (\bibinfo{year}{2015}), \eprint{1404.6495}.

\bibitem[{\citenamefont{Langlois and Noui}(2016)}]{Langlois:2015cwa}
\bibinfo{author}{\bibfnamefont{D.}~\bibnamefont{Langlois}} \bibnamefont{and}
  \bibinfo{author}{\bibfnamefont{K.}~\bibnamefont{Noui}},
  \bibinfo{journal}{JCAP} \textbf{\bibinfo{volume}{1602}}, \bibinfo{pages}{034}
  (\bibinfo{year}{2016}), \eprint{1510.06930}.

\bibitem[{\citenamefont{Creminelli et~al.}(2018)\citenamefont{Creminelli,
  Lewandowski, Tambalo, and Vernizzi}}]{Creminelli:2018xsv}
\bibinfo{author}{\bibfnamefont{P.}~\bibnamefont{Creminelli}},
  \bibinfo{author}{\bibfnamefont{M.}~\bibnamefont{Lewandowski}},
  \bibinfo{author}{\bibfnamefont{G.}~\bibnamefont{Tambalo}}, \bibnamefont{and}
  \bibinfo{author}{\bibfnamefont{F.}~\bibnamefont{Vernizzi}},
  \bibinfo{journal}{JCAP} \textbf{\bibinfo{volume}{1812}}, \bibinfo{pages}{025}
  (\bibinfo{year}{2018}), \eprint{1809.03484}.

\bibitem[{\citenamefont{Creminelli
  et~al.}(2019{\natexlab{a}})\citenamefont{Creminelli, Tambalo, Vernizzi, and
  Yingcharoenrat}}]{Creminelli:2019nok}
\bibinfo{author}{\bibfnamefont{P.}~\bibnamefont{Creminelli}},
  \bibinfo{author}{\bibfnamefont{G.}~\bibnamefont{Tambalo}},
  \bibinfo{author}{\bibfnamefont{F.}~\bibnamefont{Vernizzi}}, \bibnamefont{and}
  \bibinfo{author}{\bibfnamefont{V.}~\bibnamefont{Yingcharoenrat}},
  \bibinfo{journal}{JCAP} \textbf{\bibinfo{volume}{1910}}, \bibinfo{pages}{072}
  (\bibinfo{year}{2019}{\natexlab{a}}), \eprint{1906.07015}.

\bibitem[{\citenamefont{Peirone et~al.}(2019)\citenamefont{Peirone, Benevento,
  Frusciante, and Tsujikawa}}]{Peirone:2019yjs}
\bibinfo{author}{\bibfnamefont{S.}~\bibnamefont{Peirone}},
  \bibinfo{author}{\bibfnamefont{G.}~\bibnamefont{Benevento}},
  \bibinfo{author}{\bibfnamefont{N.}~\bibnamefont{Frusciante}},
  \bibnamefont{and}
  \bibinfo{author}{\bibfnamefont{S.}~\bibnamefont{Tsujikawa}},
  \bibinfo{journal}{Phys. Rev.} \textbf{\bibinfo{volume}{D100}},
  \bibinfo{pages}{063509} (\bibinfo{year}{2019}), \eprint{1905.11364}.

\bibitem[{\citenamefont{Babichev et~al.}(2018)\citenamefont{Babichev,
  Charmousis, Esposito-Farèse, and Lehébel}}]{Babichev:2017lmw}
\bibinfo{author}{\bibfnamefont{E.}~\bibnamefont{Babichev}},
  \bibinfo{author}{\bibfnamefont{C.}~\bibnamefont{Charmousis}},
  \bibinfo{author}{\bibfnamefont{G.}~\bibnamefont{Esposito-Farèse}},
  \bibnamefont{and} \bibinfo{author}{\bibfnamefont{A.}~\bibnamefont{Lehébel}},
  \bibinfo{journal}{Phys. Rev. Lett.} \textbf{\bibinfo{volume}{120}},
  \bibinfo{pages}{241101} (\bibinfo{year}{2018}), \eprint{1712.04398}.

\bibitem[{\citenamefont{Vainshtein}(1972)}]{Vainshtein:1972sx}
\bibinfo{author}{\bibfnamefont{A.}~\bibnamefont{Vainshtein}},
  \bibinfo{journal}{Phys.Lett.B} \textbf{\bibinfo{volume}{39}},
  \bibinfo{pages}{393} (\bibinfo{year}{1972}).

\bibitem[{\citenamefont{de~Rham et~al.}(2013)\citenamefont{de~Rham, Tolley, and
  Wesley}}]{deRham:2012fw}
\bibinfo{author}{\bibfnamefont{C.}~\bibnamefont{de~Rham}},
  \bibinfo{author}{\bibfnamefont{A.~J.} \bibnamefont{Tolley}},
  \bibnamefont{and} \bibinfo{author}{\bibfnamefont{D.~H.}
  \bibnamefont{Wesley}}, \bibinfo{journal}{Phys. Rev.}
  \textbf{\bibinfo{volume}{D87}}, \bibinfo{pages}{044025}
  (\bibinfo{year}{2013}), \eprint{1208.0580}.

\bibitem[{\citenamefont{Chu and Trodden}(2013)}]{Chu:2012kz}
\bibinfo{author}{\bibfnamefont{Y.-Z.} \bibnamefont{Chu}} \bibnamefont{and}
  \bibinfo{author}{\bibfnamefont{M.}~\bibnamefont{Trodden}},
  \bibinfo{journal}{Phys. Rev.} \textbf{\bibinfo{volume}{D87}},
  \bibinfo{pages}{024011} (\bibinfo{year}{2013}), \eprint{1210.6651}.

\bibitem[{\citenamefont{Dar et~al.}(2019)\citenamefont{Dar, De~Rham, Deskins,
  Giblin, and Tolley}}]{Dar:2018dra}
\bibinfo{author}{\bibfnamefont{F.}~\bibnamefont{Dar}},
  \bibinfo{author}{\bibfnamefont{C.}~\bibnamefont{De~Rham}},
  \bibinfo{author}{\bibfnamefont{J.~T.} \bibnamefont{Deskins}},
  \bibinfo{author}{\bibfnamefont{J.~T.} \bibnamefont{Giblin}},
  \bibnamefont{and} \bibinfo{author}{\bibfnamefont{A.~J.}
  \bibnamefont{Tolley}}, \bibinfo{journal}{Class. Quant. Grav.}
  \textbf{\bibinfo{volume}{36}}, \bibinfo{pages}{025008}
  (\bibinfo{year}{2019}), \eprint{1808.02165}.

\bibitem[{\citenamefont{Brax et~al.}(2020)\citenamefont{Brax, Heisenberg, and
  Kuntz}}]{Brax:2020ujo}
\bibinfo{author}{\bibfnamefont{P.}~\bibnamefont{Brax}},
  \bibinfo{author}{\bibfnamefont{L.}~\bibnamefont{Heisenberg}},
  \bibnamefont{and} \bibinfo{author}{\bibfnamefont{A.}~\bibnamefont{Kuntz}}
  (\bibinfo{year}{2020}), \eprint{2002.12590}.

\bibitem[{\citenamefont{Creminelli
  et~al.}(2019{\natexlab{b}})\citenamefont{Creminelli, Tambalo, Vernizzi, and
  Yingcharoenrat}}]{Creminelli:2019kjy}
\bibinfo{author}{\bibfnamefont{P.}~\bibnamefont{Creminelli}},
  \bibinfo{author}{\bibfnamefont{G.}~\bibnamefont{Tambalo}},
  \bibinfo{author}{\bibfnamefont{F.}~\bibnamefont{Vernizzi}}, \bibnamefont{and}
  \bibinfo{author}{\bibfnamefont{V.}~\bibnamefont{Yingcharoenrat}}
  (\bibinfo{year}{2019}{\natexlab{b}}), \eprint{1910.14035}.

\bibitem[{\citenamefont{Chow and Khoury}(2009)}]{Chow:2009fm}
\bibinfo{author}{\bibfnamefont{N.}~\bibnamefont{Chow}} \bibnamefont{and}
  \bibinfo{author}{\bibfnamefont{J.}~\bibnamefont{Khoury}},
  \bibinfo{journal}{Phys. Rev.} \textbf{\bibinfo{volume}{D80}},
  \bibinfo{pages}{024037} (\bibinfo{year}{2009}), \eprint{0905.1325}.

\bibitem[{\citenamefont{Silva and Koyama}(2009)}]{Silva:2009km}
\bibinfo{author}{\bibfnamefont{F.~P.} \bibnamefont{Silva}} \bibnamefont{and}
  \bibinfo{author}{\bibfnamefont{K.}~\bibnamefont{Koyama}},
  \bibinfo{journal}{Phys. Rev. D} \textbf{\bibinfo{volume}{80}},
  \bibinfo{pages}{121301} (\bibinfo{year}{2009}), \eprint{0909.4538}.

\bibitem[{\citenamefont{Appleby and
  Linder}(2012{\natexlab{a}})}]{Appleby:2011aa}
\bibinfo{author}{\bibfnamefont{S.}~\bibnamefont{Appleby}} \bibnamefont{and}
  \bibinfo{author}{\bibfnamefont{E.~V.} \bibnamefont{Linder}},
  \bibinfo{journal}{JCAP} \textbf{\bibinfo{volume}{1203}}, \bibinfo{pages}{043}
  (\bibinfo{year}{2012}{\natexlab{a}}), \eprint{1112.1981}.

\bibitem[{\citenamefont{Appleby and
  Linder}(2012{\natexlab{b}})}]{Appleby:2012ba}
\bibinfo{author}{\bibfnamefont{S.~A.} \bibnamefont{Appleby}} \bibnamefont{and}
  \bibinfo{author}{\bibfnamefont{E.~V.} \bibnamefont{Linder}},
  \bibinfo{journal}{JCAP} \textbf{\bibinfo{volume}{1208}}, \bibinfo{pages}{026}
  (\bibinfo{year}{2012}{\natexlab{b}}), \eprint{1204.4314}.

\bibitem[{\citenamefont{Neveu et~al.}(2014{\natexlab{a}})\citenamefont{Neveu,
  Ruhlmann-Kleider, Astier, Besançon, Conley, Guy, Möller,
  Palanque-Delabrouille, and Babichev}}]{Neveu:2014vua}
\bibinfo{author}{\bibfnamefont{J.}~\bibnamefont{Neveu}},
  \bibinfo{author}{\bibfnamefont{V.}~\bibnamefont{Ruhlmann-Kleider}},
  \bibinfo{author}{\bibfnamefont{P.}~\bibnamefont{Astier}},
  \bibinfo{author}{\bibfnamefont{M.}~\bibnamefont{Besançon}},
  \bibinfo{author}{\bibfnamefont{A.}~\bibnamefont{Conley}},
  \bibinfo{author}{\bibfnamefont{J.}~\bibnamefont{Guy}},
  \bibinfo{author}{\bibfnamefont{A.}~\bibnamefont{Möller}},
  \bibinfo{author}{\bibfnamefont{N.}~\bibnamefont{Palanque-Delabrouille}},
  \bibnamefont{and} \bibinfo{author}{\bibfnamefont{E.}~\bibnamefont{Babichev}},
  \bibinfo{journal}{Astron. Astrophys.} \textbf{\bibinfo{volume}{569}},
  \bibinfo{pages}{A90} (\bibinfo{year}{2014}{\natexlab{a}}),
  \eprint{1403.0854}.

\bibitem[{\citenamefont{Bhattacharya et~al.}(2016)\citenamefont{Bhattacharya,
  Dialektopoulos, and Tomaras}}]{Bhattacharya:2015chc}
\bibinfo{author}{\bibfnamefont{S.}~\bibnamefont{Bhattacharya}},
  \bibinfo{author}{\bibfnamefont{K.~F.} \bibnamefont{Dialektopoulos}},
  \bibnamefont{and} \bibinfo{author}{\bibfnamefont{T.~N.}
  \bibnamefont{Tomaras}}, \bibinfo{journal}{JCAP}
  \textbf{\bibinfo{volume}{1605}}, \bibinfo{pages}{036} (\bibinfo{year}{2016}),
  \eprint{1512.08856}.

\bibitem[{\citenamefont{Burrage et~al.}(2017)\citenamefont{Burrage, Parkinson,
  and Seery}}]{Burrage:2015lla}
\bibinfo{author}{\bibfnamefont{C.}~\bibnamefont{Burrage}},
  \bibinfo{author}{\bibfnamefont{D.}~\bibnamefont{Parkinson}},
  \bibnamefont{and} \bibinfo{author}{\bibfnamefont{D.}~\bibnamefont{Seery}},
  \bibinfo{journal}{Phys. Rev.} \textbf{\bibinfo{volume}{D96}},
  \bibinfo{pages}{043509} (\bibinfo{year}{2017}), \eprint{1502.03710}.

\bibitem[{\citenamefont{Burrage et~al.}(2019)\citenamefont{Burrage, Dombrowski,
  and Saadeh}}]{Burrage:2019afs}
\bibinfo{author}{\bibfnamefont{C.}~\bibnamefont{Burrage}},
  \bibinfo{author}{\bibfnamefont{J.}~\bibnamefont{Dombrowski}},
  \bibnamefont{and} \bibinfo{author}{\bibfnamefont{D.}~\bibnamefont{Saadeh}},
  \bibinfo{journal}{JCAP} \textbf{\bibinfo{volume}{1910}}, \bibinfo{pages}{023}
  (\bibinfo{year}{2019}), \eprint{1905.06260}.

\bibitem[{\citenamefont{Tsujikawa}(2019)}]{Tsujikawa:2019pih}
\bibinfo{author}{\bibfnamefont{S.}~\bibnamefont{Tsujikawa}},
  \bibinfo{journal}{Phys. Rev.} \textbf{\bibinfo{volume}{D100}},
  \bibinfo{pages}{043510} (\bibinfo{year}{2019}), \eprint{1903.07092}.

\bibitem[{\citenamefont{Ezquiaga et~al.}(2017)\citenamefont{Ezquiaga,
  García-Bellido, and Zumalacárregui}}]{Ezquiaga:2017ner}
\bibinfo{author}{\bibfnamefont{J.~M.} \bibnamefont{Ezquiaga}},
  \bibinfo{author}{\bibfnamefont{J.}~\bibnamefont{García-Bellido}},
  \bibnamefont{and}
  \bibinfo{author}{\bibfnamefont{M.}~\bibnamefont{Zumalacárregui}},
  \bibinfo{journal}{Phys. Rev.} \textbf{\bibinfo{volume}{D95}},
  \bibinfo{pages}{084039} (\bibinfo{year}{2017}), \eprint{1701.05476}.

\bibitem[{\citenamefont{Blas et~al.}(2011)\citenamefont{Blas, Lesgourgues, and
  Tram}}]{Blas:2011rf}
\bibinfo{author}{\bibfnamefont{D.}~\bibnamefont{Blas}},
  \bibinfo{author}{\bibfnamefont{J.}~\bibnamefont{Lesgourgues}},
  \bibnamefont{and} \bibinfo{author}{\bibfnamefont{T.}~\bibnamefont{Tram}},
  \bibinfo{journal}{JCAP} \textbf{\bibinfo{volume}{1107}}, \bibinfo{pages}{034}
  (\bibinfo{year}{2011}), \eprint{1104.2933}.

\bibitem[{\citenamefont{Bellini and Sawicki}(2014)}]{Bellini:2014fua}
\bibinfo{author}{\bibfnamefont{E.}~\bibnamefont{Bellini}} \bibnamefont{and}
  \bibinfo{author}{\bibfnamefont{I.}~\bibnamefont{Sawicki}},
  \bibinfo{journal}{JCAP} \textbf{\bibinfo{volume}{1407}}, \bibinfo{pages}{050}
  (\bibinfo{year}{2014}), \eprint{1404.3713}.

\bibitem[{\citenamefont{Bettoni and Zumalacárregui}(2015)}]{Bettoni:2015wta}
\bibinfo{author}{\bibfnamefont{D.}~\bibnamefont{Bettoni}} \bibnamefont{and}
  \bibinfo{author}{\bibfnamefont{M.}~\bibnamefont{Zumalacárregui}},
  \bibinfo{journal}{Phys. Rev.} \textbf{\bibinfo{volume}{D91}},
  \bibinfo{pages}{104009} (\bibinfo{year}{2015}), \eprint{1502.02666}.

\bibitem[{\citenamefont{Pujolas et~al.}(2011)\citenamefont{Pujolas, Sawicki,
  and Vikman}}]{Pujolas:2011he}
\bibinfo{author}{\bibfnamefont{O.}~\bibnamefont{Pujolas}},
  \bibinfo{author}{\bibfnamefont{I.}~\bibnamefont{Sawicki}}, \bibnamefont{and}
  \bibinfo{author}{\bibfnamefont{A.}~\bibnamefont{Vikman}},
  \bibinfo{journal}{JHEP} \textbf{\bibinfo{volume}{11}}, \bibinfo{pages}{156}
  (\bibinfo{year}{2011}), \eprint{1103.5360}.

\bibitem[{\citenamefont{Sawicki et~al.}(2013)\citenamefont{Sawicki, Saltas,
  Amendola, and Kunz}}]{Sawicki:2012re}
\bibinfo{author}{\bibfnamefont{I.}~\bibnamefont{Sawicki}},
  \bibinfo{author}{\bibfnamefont{I.~D.} \bibnamefont{Saltas}},
  \bibinfo{author}{\bibfnamefont{L.}~\bibnamefont{Amendola}}, \bibnamefont{and}
  \bibinfo{author}{\bibfnamefont{M.}~\bibnamefont{Kunz}},
  \bibinfo{journal}{JCAP} \textbf{\bibinfo{volume}{1301}}, \bibinfo{pages}{004}
  (\bibinfo{year}{2013}), \eprint{1208.4855}.

\bibitem[{\citenamefont{Deffayet et~al.}(2010)\citenamefont{Deffayet, Pujolas,
  Sawicki, and Vikman}}]{Deffayet:2010qz}
\bibinfo{author}{\bibfnamefont{C.}~\bibnamefont{Deffayet}},
  \bibinfo{author}{\bibfnamefont{O.}~\bibnamefont{Pujolas}},
  \bibinfo{author}{\bibfnamefont{I.}~\bibnamefont{Sawicki}}, \bibnamefont{and}
  \bibinfo{author}{\bibfnamefont{A.}~\bibnamefont{Vikman}},
  \bibinfo{journal}{JCAP} \textbf{\bibinfo{volume}{1010}}, \bibinfo{pages}{026}
  (\bibinfo{year}{2010}), \eprint{1008.0048}.

\bibitem[{\citenamefont{Barreira et~al.}(2012)\citenamefont{Barreira, Li,
  Baugh, and Pascoli}}]{Barreira:2012kk}
\bibinfo{author}{\bibfnamefont{A.}~\bibnamefont{Barreira}},
  \bibinfo{author}{\bibfnamefont{B.}~\bibnamefont{Li}},
  \bibinfo{author}{\bibfnamefont{C.~M.} \bibnamefont{Baugh}}, \bibnamefont{and}
  \bibinfo{author}{\bibfnamefont{S.}~\bibnamefont{Pascoli}},
  \bibinfo{journal}{Phys. Rev.} \textbf{\bibinfo{volume}{D86}},
  \bibinfo{pages}{124016} (\bibinfo{year}{2012}), \eprint{1208.0600}.

\bibitem[{\citenamefont{Neveu et~al.}(2013)\citenamefont{Neveu,
  Ruhlmann-Kleider, Conley, Palanque-Delabrouille, Astier, Guy, and
  Babichev}}]{Neveu:2013mfa}
\bibinfo{author}{\bibfnamefont{J.}~\bibnamefont{Neveu}},
  \bibinfo{author}{\bibfnamefont{V.}~\bibnamefont{Ruhlmann-Kleider}},
  \bibinfo{author}{\bibfnamefont{A.}~\bibnamefont{Conley}},
  \bibinfo{author}{\bibfnamefont{N.}~\bibnamefont{Palanque-Delabrouille}},
  \bibinfo{author}{\bibfnamefont{P.}~\bibnamefont{Astier}},
  \bibinfo{author}{\bibfnamefont{J.}~\bibnamefont{Guy}}, \bibnamefont{and}
  \bibinfo{author}{\bibfnamefont{E.}~\bibnamefont{Babichev}},
  \bibinfo{journal}{Astron. Astrophys.} \textbf{\bibinfo{volume}{555}},
  \bibinfo{pages}{A53} (\bibinfo{year}{2013}), \eprint{1302.2786}.

\bibitem[{\citenamefont{Neveu et~al.}(2014{\natexlab{b}})\citenamefont{Neveu,
  Ruhlmann-Kleider, and Besançon}}]{Neveu:2014kba}
\bibinfo{author}{\bibfnamefont{J.}~\bibnamefont{Neveu}},
  \bibinfo{author}{\bibfnamefont{V.}~\bibnamefont{Ruhlmann-Kleider}},
  \bibnamefont{and}
  \bibinfo{author}{\bibfnamefont{M.}~\bibnamefont{Besançon}}, in
  \emph{\bibinfo{booktitle}{{Proceedings, 49th Rencontres de Moriond on
  Cosmology: La Thuile, Italy, March 15-22, 2014}}}
  (\bibinfo{year}{2014}{\natexlab{b}}), pp. \bibinfo{pages}{293--296}.

\bibitem[{\citenamefont{Neveu et~al.}(2017)\citenamefont{Neveu,
  Ruhlmann-Kleider, Astier, Besançon, Guy, Möller, and
  Babichev}}]{Neveu:2016gxp}
\bibinfo{author}{\bibfnamefont{J.}~\bibnamefont{Neveu}},
  \bibinfo{author}{\bibfnamefont{V.}~\bibnamefont{Ruhlmann-Kleider}},
  \bibinfo{author}{\bibfnamefont{P.}~\bibnamefont{Astier}},
  \bibinfo{author}{\bibfnamefont{M.}~\bibnamefont{Besançon}},
  \bibinfo{author}{\bibfnamefont{J.}~\bibnamefont{Guy}},
  \bibinfo{author}{\bibfnamefont{A.}~\bibnamefont{Möller}}, \bibnamefont{and}
  \bibinfo{author}{\bibfnamefont{E.}~\bibnamefont{Babichev}},
  \bibinfo{journal}{Astron. Astrophys.} \textbf{\bibinfo{volume}{600}},
  \bibinfo{pages}{A40} (\bibinfo{year}{2017}), \eprint{1605.02627}.

\bibitem[{\citenamefont{Leloup et~al.}(2019)\citenamefont{Leloup,
  Ruhlmann-Kleider, Neveu, and De~Mattia}}]{Leloup:2019fas}
\bibinfo{author}{\bibfnamefont{C.}~\bibnamefont{Leloup}},
  \bibinfo{author}{\bibfnamefont{V.}~\bibnamefont{Ruhlmann-Kleider}},
  \bibinfo{author}{\bibfnamefont{J.}~\bibnamefont{Neveu}}, \bibnamefont{and}
  \bibinfo{author}{\bibfnamefont{A.}~\bibnamefont{De~Mattia}},
  \bibinfo{journal}{JCAP} \textbf{\bibinfo{volume}{1905}}, \bibinfo{pages}{011}
  (\bibinfo{year}{2019}), \eprint{1902.07065}.

\bibitem[{\citenamefont{Umiltà et~al.}(2015)\citenamefont{Umiltà, Ballardini,
  Finelli, and Paoletti}}]{Umilta:2015cta}
\bibinfo{author}{\bibfnamefont{C.}~\bibnamefont{Umiltà}},
  \bibinfo{author}{\bibfnamefont{M.}~\bibnamefont{Ballardini}},
  \bibinfo{author}{\bibfnamefont{F.}~\bibnamefont{Finelli}}, \bibnamefont{and}
  \bibinfo{author}{\bibfnamefont{D.}~\bibnamefont{Paoletti}},
  \bibinfo{journal}{JCAP} \textbf{\bibinfo{volume}{1508}}, \bibinfo{pages}{017}
  (\bibinfo{year}{2015}), \eprint{1507.00718}.

\bibitem[{\citenamefont{Ballardini et~al.}(2016)\citenamefont{Ballardini,
  Finelli, Umiltà, and Paoletti}}]{Ballardini:2016cvy}
\bibinfo{author}{\bibfnamefont{M.}~\bibnamefont{Ballardini}},
  \bibinfo{author}{\bibfnamefont{F.}~\bibnamefont{Finelli}},
  \bibinfo{author}{\bibfnamefont{C.}~\bibnamefont{Umiltà}}, \bibnamefont{and}
  \bibinfo{author}{\bibfnamefont{D.}~\bibnamefont{Paoletti}},
  \bibinfo{journal}{JCAP} \textbf{\bibinfo{volume}{1605}}, \bibinfo{pages}{067}
  (\bibinfo{year}{2016}), \eprint{1601.03387}.

\bibitem[{\citenamefont{Rossi et~al.}(2019)\citenamefont{Rossi, Ballardini,
  Braglia, Finelli, Paoletti, Starobinsky, and Umiltà}}]{Rossi:2019lgt}
\bibinfo{author}{\bibfnamefont{M.}~\bibnamefont{Rossi}},
  \bibinfo{author}{\bibfnamefont{M.}~\bibnamefont{Ballardini}},
  \bibinfo{author}{\bibfnamefont{M.}~\bibnamefont{Braglia}},
  \bibinfo{author}{\bibfnamefont{F.}~\bibnamefont{Finelli}},
  \bibinfo{author}{\bibfnamefont{D.}~\bibnamefont{Paoletti}},
  \bibinfo{author}{\bibfnamefont{A.~A.} \bibnamefont{Starobinsky}},
  \bibnamefont{and} \bibinfo{author}{\bibfnamefont{C.}~\bibnamefont{Umiltà}},
  \bibinfo{journal}{Phys. Rev.} \textbf{\bibinfo{volume}{D100}},
  \bibinfo{pages}{103524} (\bibinfo{year}{2019}), \eprint{1906.10218}.

\bibitem[{\citenamefont{Ballesteros et~al.}(2020)\citenamefont{Ballesteros,
  Notari, and Rompineve}}]{Ballesteros:2020sik}
\bibinfo{author}{\bibfnamefont{G.}~\bibnamefont{Ballesteros}},
  \bibinfo{author}{\bibfnamefont{A.}~\bibnamefont{Notari}}, \bibnamefont{and}
  \bibinfo{author}{\bibfnamefont{F.}~\bibnamefont{Rompineve}}
  (\bibinfo{year}{2020}), \eprint{2004.05049}.

\bibitem[{\citenamefont{Braglia et~al.}(2020)\citenamefont{Braglia, Ballardini,
  Emond, Finelli, Gumrukcuoglu, Koyama, and Paoletti}}]{Braglia:2020iik}
\bibinfo{author}{\bibfnamefont{M.}~\bibnamefont{Braglia}},
  \bibinfo{author}{\bibfnamefont{M.}~\bibnamefont{Ballardini}},
  \bibinfo{author}{\bibfnamefont{W.~T.} \bibnamefont{Emond}},
  \bibinfo{author}{\bibfnamefont{F.}~\bibnamefont{Finelli}},
  \bibinfo{author}{\bibfnamefont{A.~E.} \bibnamefont{Gumrukcuoglu}},
  \bibinfo{author}{\bibfnamefont{K.}~\bibnamefont{Koyama}}, \bibnamefont{and}
  \bibinfo{author}{\bibfnamefont{D.}~\bibnamefont{Paoletti}}
  (\bibinfo{year}{2020}), \eprint{2004.11161}.

\bibitem[{\citenamefont{Giusarma et~al.}(2016)\citenamefont{Giusarma, Gerbino,
  Mena, Vagnozzi, Ho, and Freese}}]{Giusarma:2016phn}
\bibinfo{author}{\bibfnamefont{E.}~\bibnamefont{Giusarma}},
  \bibinfo{author}{\bibfnamefont{M.}~\bibnamefont{Gerbino}},
  \bibinfo{author}{\bibfnamefont{O.}~\bibnamefont{Mena}},
  \bibinfo{author}{\bibfnamefont{S.}~\bibnamefont{Vagnozzi}},
  \bibinfo{author}{\bibfnamefont{S.}~\bibnamefont{Ho}}, \bibnamefont{and}
  \bibinfo{author}{\bibfnamefont{K.}~\bibnamefont{Freese}},
  \bibinfo{journal}{Phys. Rev.} \textbf{\bibinfo{volume}{D94}},
  \bibinfo{pages}{083522} (\bibinfo{year}{2016}), \eprint{1605.04320}.

\bibitem[{\citenamefont{Vagnozzi et~al.}(2017)\citenamefont{Vagnozzi, Giusarma,
  Mena, Freese, Gerbino, Ho, and Lattanzi}}]{Vagnozzi:2017ovm}
\bibinfo{author}{\bibfnamefont{S.}~\bibnamefont{Vagnozzi}},
  \bibinfo{author}{\bibfnamefont{E.}~\bibnamefont{Giusarma}},
  \bibinfo{author}{\bibfnamefont{O.}~\bibnamefont{Mena}},
  \bibinfo{author}{\bibfnamefont{K.}~\bibnamefont{Freese}},
  \bibinfo{author}{\bibfnamefont{M.}~\bibnamefont{Gerbino}},
  \bibinfo{author}{\bibfnamefont{S.}~\bibnamefont{Ho}}, \bibnamefont{and}
  \bibinfo{author}{\bibfnamefont{M.}~\bibnamefont{Lattanzi}},
  \bibinfo{journal}{Phys. Rev.} \textbf{\bibinfo{volume}{D96}},
  \bibinfo{pages}{123503} (\bibinfo{year}{2017}), \eprint{1701.08172}.

\bibitem[{\citenamefont{Peirone et~al.}(2018)\citenamefont{Peirone, Frusciante,
  Hu, Raveri, and Silvestri}}]{Peirone:2017vcq}
\bibinfo{author}{\bibfnamefont{S.}~\bibnamefont{Peirone}},
  \bibinfo{author}{\bibfnamefont{N.}~\bibnamefont{Frusciante}},
  \bibinfo{author}{\bibfnamefont{B.}~\bibnamefont{Hu}},
  \bibinfo{author}{\bibfnamefont{M.}~\bibnamefont{Raveri}}, \bibnamefont{and}
  \bibinfo{author}{\bibfnamefont{A.}~\bibnamefont{Silvestri}},
  \bibinfo{journal}{Phys. Rev.} \textbf{\bibinfo{volume}{D97}},
  \bibinfo{pages}{063518} (\bibinfo{year}{2018}), \eprint{1711.04760}.

\bibitem[{\citenamefont{Aghanim et~al.}(2019)}]{Aghanim:2019ame}
\bibinfo{author}{\bibfnamefont{N.}~\bibnamefont{Aghanim}} \bibnamefont{et~al.}
  (\bibinfo{collaboration}{Planck}) (\bibinfo{year}{2019}),
  \eprint{1907.12875}.

\bibitem[{\citenamefont{Alam et~al.}(2017)}]{Alam:2016hwk}
\bibinfo{author}{\bibfnamefont{S.}~\bibnamefont{Alam}} \bibnamefont{et~al.}
  (\bibinfo{collaboration}{BOSS}), \bibinfo{journal}{Mon. Not. Roy. Astron.
  Soc.} \textbf{\bibinfo{volume}{470}}, \bibinfo{pages}{2617}
  (\bibinfo{year}{2017}), \eprint{1607.03155}.

\bibitem[{\citenamefont{Beutler et~al.}(2011)\citenamefont{Beutler, Blake,
  Colless, Jones, Staveley-Smith, Campbell, Parker, Saunders, and
  Watson}}]{Beutler:2011hx}
\bibinfo{author}{\bibfnamefont{F.}~\bibnamefont{Beutler}},
  \bibinfo{author}{\bibfnamefont{C.}~\bibnamefont{Blake}},
  \bibinfo{author}{\bibfnamefont{M.}~\bibnamefont{Colless}},
  \bibinfo{author}{\bibfnamefont{D.~H.} \bibnamefont{Jones}},
  \bibinfo{author}{\bibfnamefont{L.}~\bibnamefont{Staveley-Smith}},
  \bibinfo{author}{\bibfnamefont{L.}~\bibnamefont{Campbell}},
  \bibinfo{author}{\bibfnamefont{Q.}~\bibnamefont{Parker}},
  \bibinfo{author}{\bibfnamefont{W.}~\bibnamefont{Saunders}}, \bibnamefont{and}
  \bibinfo{author}{\bibfnamefont{F.}~\bibnamefont{Watson}},
  \bibinfo{journal}{Mon. Not. Roy. Astron. Soc.}
  \textbf{\bibinfo{volume}{416}}, \bibinfo{pages}{3017} (\bibinfo{year}{2011}),
  \eprint{1106.3366}.

\bibitem[{\citenamefont{Ross et~al.}(2015)\citenamefont{Ross, Samushia,
  Howlett, Percival, Burden, and Manera}}]{Ross:2014qpa}
\bibinfo{author}{\bibfnamefont{A.~J.} \bibnamefont{Ross}},
  \bibinfo{author}{\bibfnamefont{L.}~\bibnamefont{Samushia}},
  \bibinfo{author}{\bibfnamefont{C.}~\bibnamefont{Howlett}},
  \bibinfo{author}{\bibfnamefont{W.~J.} \bibnamefont{Percival}},
  \bibinfo{author}{\bibfnamefont{A.}~\bibnamefont{Burden}}, \bibnamefont{and}
  \bibinfo{author}{\bibfnamefont{M.}~\bibnamefont{Manera}},
  \bibinfo{journal}{Mon. Not. Roy. Astron. Soc.}
  \textbf{\bibinfo{volume}{449}}, \bibinfo{pages}{835} (\bibinfo{year}{2015}),
  \eprint{1409.3242}.

\bibitem[{\citenamefont{Sherwin and White}(2019)}]{Sherwin:2018wbu}
\bibinfo{author}{\bibfnamefont{B.~D.} \bibnamefont{Sherwin}} \bibnamefont{and}
  \bibinfo{author}{\bibfnamefont{M.}~\bibnamefont{White}},
  \bibinfo{journal}{JCAP} \textbf{\bibinfo{volume}{1902}}, \bibinfo{pages}{027}
  (\bibinfo{year}{2019}), \eprint{1808.04384}.

\bibitem[{\citenamefont{Carter et~al.}(2019)\citenamefont{Carter, Beutler,
  Percival, DeRose, Wechsler, and Zhao}}]{Carter:2019ulk}
\bibinfo{author}{\bibfnamefont{P.}~\bibnamefont{Carter}},
  \bibinfo{author}{\bibfnamefont{F.}~\bibnamefont{Beutler}},
  \bibinfo{author}{\bibfnamefont{W.~J.} \bibnamefont{Percival}},
  \bibinfo{author}{\bibfnamefont{J.}~\bibnamefont{DeRose}},
  \bibinfo{author}{\bibfnamefont{R.~H.} \bibnamefont{Wechsler}},
  \bibnamefont{and} \bibinfo{author}{\bibfnamefont{C.}~\bibnamefont{Zhao}}
  (\bibinfo{year}{2019}), \eprint{1906.03035}.

\bibitem[{\citenamefont{Bellini and Zumalacarregui}(2015)}]{Bellini:2015oua}
\bibinfo{author}{\bibfnamefont{E.}~\bibnamefont{Bellini}} \bibnamefont{and}
  \bibinfo{author}{\bibfnamefont{M.}~\bibnamefont{Zumalacarregui}},
  \bibinfo{journal}{Phys. Rev.} \textbf{\bibinfo{volume}{D92}},
  \bibinfo{pages}{063522} (\bibinfo{year}{2015}), \eprint{1505.03839}.

\bibitem[{\citenamefont{Cardona et~al.}(2017)\citenamefont{Cardona, Kunz, and
  Pettorino}}]{Cardona:2016ems}
\bibinfo{author}{\bibfnamefont{W.}~\bibnamefont{Cardona}},
  \bibinfo{author}{\bibfnamefont{M.}~\bibnamefont{Kunz}}, \bibnamefont{and}
  \bibinfo{author}{\bibfnamefont{V.}~\bibnamefont{Pettorino}},
  \bibinfo{journal}{JCAP} \textbf{\bibinfo{volume}{1703}}, \bibinfo{pages}{056}
  (\bibinfo{year}{2017}), \eprint{1611.06088}.

\bibitem[{\citenamefont{Zhang et~al.}(2017)\citenamefont{Zhang, Childress,
  Davis, Karpenka, Lidman, Schmidt, and Smith}}]{Zhang:2017aqn}
\bibinfo{author}{\bibfnamefont{B.~R.} \bibnamefont{Zhang}},
  \bibinfo{author}{\bibfnamefont{M.~J.} \bibnamefont{Childress}},
  \bibinfo{author}{\bibfnamefont{T.~M.} \bibnamefont{Davis}},
  \bibinfo{author}{\bibfnamefont{N.~V.} \bibnamefont{Karpenka}},
  \bibinfo{author}{\bibfnamefont{C.}~\bibnamefont{Lidman}},
  \bibinfo{author}{\bibfnamefont{B.~P.} \bibnamefont{Schmidt}},
  \bibnamefont{and} \bibinfo{author}{\bibfnamefont{M.}~\bibnamefont{Smith}},
  \bibinfo{journal}{Mon. Not. Roy. Astron. Soc.}
  \textbf{\bibinfo{volume}{471}}, \bibinfo{pages}{2254} (\bibinfo{year}{2017}),
  \eprint{1706.07573}.

\bibitem[{\citenamefont{Follin and Knox}(2018)}]{Follin:2017ljs}
\bibinfo{author}{\bibfnamefont{B.}~\bibnamefont{Follin}} \bibnamefont{and}
  \bibinfo{author}{\bibfnamefont{L.}~\bibnamefont{Knox}},
  \bibinfo{journal}{Mon. Not. Roy. Astron. Soc.}
  \textbf{\bibinfo{volume}{477}}, \bibinfo{pages}{4534} (\bibinfo{year}{2018}),
  \eprint{1707.01175}.

\bibitem[{\citenamefont{Feeney et~al.}(2018)\citenamefont{Feeney, Mortlock, and
  Dalmasso}}]{Feeney:2017sgx}
\bibinfo{author}{\bibfnamefont{S.~M.} \bibnamefont{Feeney}},
  \bibinfo{author}{\bibfnamefont{D.~J.} \bibnamefont{Mortlock}},
  \bibnamefont{and} \bibinfo{author}{\bibfnamefont{N.}~\bibnamefont{Dalmasso}},
  \bibinfo{journal}{Mon. Not. Roy. Astron. Soc.}
  \textbf{\bibinfo{volume}{476}}, \bibinfo{pages}{3861} (\bibinfo{year}{2018}),
  \eprint{1707.00007}.

\bibitem[{\citenamefont{Dhawan et~al.}(2020)\citenamefont{Dhawan, Brout,
  Scolnic, Goobar, Riess, and Miranda}}]{Dhawan:2020xmp}
\bibinfo{author}{\bibfnamefont{S.}~\bibnamefont{Dhawan}},
  \bibinfo{author}{\bibfnamefont{D.}~\bibnamefont{Brout}},
  \bibinfo{author}{\bibfnamefont{D.}~\bibnamefont{Scolnic}},
  \bibinfo{author}{\bibfnamefont{A.}~\bibnamefont{Goobar}},
  \bibinfo{author}{\bibfnamefont{A.~G.} \bibnamefont{Riess}}, \bibnamefont{and}
  \bibinfo{author}{\bibfnamefont{V.}~\bibnamefont{Miranda}}
  (\bibinfo{year}{2020}), \eprint{2001.09260}.

\bibitem[{\citenamefont{Abbott et~al.}(2017{\natexlab{b}})}]{Abbott:2017xzu}
\bibinfo{author}{\bibfnamefont{B.~P.} \bibnamefont{Abbott}}
  \bibnamefont{et~al.} (\bibinfo{collaboration}{LIGO Scientific, Virgo, 1M2H,
  Dark Energy Camera GW-E, DES, DLT40, Las Cumbres Observatory, VINROUGE,
  MASTER}), \bibinfo{journal}{Nature} \textbf{\bibinfo{volume}{551}},
  \bibinfo{pages}{85} (\bibinfo{year}{2017}{\natexlab{b}}),
  \eprint{1710.05835}.

\bibitem[{\citenamefont{Freedman et~al.}(2019)}]{Freedman:2019jwv}
\bibinfo{author}{\bibfnamefont{W.~L.} \bibnamefont{Freedman}}
  \bibnamefont{et~al.} (\bibinfo{year}{2019}), \eprint{1907.05922}.

\bibitem[{\citenamefont{Audren et~al.}(2013)\citenamefont{Audren, Lesgourgues,
  Benabed, and Prunet}}]{Audren:2012wb}
\bibinfo{author}{\bibfnamefont{B.}~\bibnamefont{Audren}},
  \bibinfo{author}{\bibfnamefont{J.}~\bibnamefont{Lesgourgues}},
  \bibinfo{author}{\bibfnamefont{K.}~\bibnamefont{Benabed}}, \bibnamefont{and}
  \bibinfo{author}{\bibfnamefont{S.}~\bibnamefont{Prunet}},
  \bibinfo{journal}{JCAP} \textbf{\bibinfo{volume}{1302}}, \bibinfo{pages}{001}
  (\bibinfo{year}{2013}), \eprint{1210.7183}.

\bibitem[{\citenamefont{Brinckmann and Lesgourgues}(2018)}]{Brinckmann:2018cvx}
\bibinfo{author}{\bibfnamefont{T.}~\bibnamefont{Brinckmann}} \bibnamefont{and}
  \bibinfo{author}{\bibfnamefont{J.}~\bibnamefont{Lesgourgues}}
  (\bibinfo{year}{2018}), \eprint{1804.07261}.

\bibitem[{\citenamefont{Lewis}(2019)}]{Lewis:2019xzd}
\bibinfo{author}{\bibfnamefont{A.}~\bibnamefont{Lewis}} (\bibinfo{year}{2019}),
  \eprint{1910.13970}.

\bibitem[{\citenamefont{{Millea}}(2017)}]{2017ascl.soft01004M}
\bibinfo{author}{\bibfnamefont{M.}~\bibnamefont{{Millea}}},
  \emph{\bibinfo{title}{{CosmoSlik: Cosmology sampler of likelihoods}}},
  \bibinfo{howpublished}{Astrophysics Source Code Library}
  (\bibinfo{year}{2017}), \eprint{1701.004}.

\bibitem[{\citenamefont{Kimura et~al.}(2012)\citenamefont{Kimura, Kobayashi,
  and Yamamoto}}]{Kimura:2011td}
\bibinfo{author}{\bibfnamefont{R.}~\bibnamefont{Kimura}},
  \bibinfo{author}{\bibfnamefont{T.}~\bibnamefont{Kobayashi}},
  \bibnamefont{and} \bibinfo{author}{\bibfnamefont{K.}~\bibnamefont{Yamamoto}},
  \bibinfo{journal}{Phys. Rev.} \textbf{\bibinfo{volume}{D85}},
  \bibinfo{pages}{123503} (\bibinfo{year}{2012}), \eprint{1110.3598}.

\bibitem[{\citenamefont{Abbott et~al.}(2018)}]{Abbott:2017wau}
\bibinfo{author}{\bibfnamefont{T.}~\bibnamefont{Abbott}} \bibnamefont{et~al.}
  (\bibinfo{collaboration}{DES}), \bibinfo{journal}{Phys. Rev. D}
  \textbf{\bibinfo{volume}{98}}, \bibinfo{pages}{043526}
  (\bibinfo{year}{2018}), \eprint{1708.01530}.

\bibitem[{\citenamefont{Hildebrandt et~al.}(2017)}]{Hildebrandt:2016iqg}
\bibinfo{author}{\bibfnamefont{H.}~\bibnamefont{Hildebrandt}}
  \bibnamefont{et~al.}, \bibinfo{journal}{Mon. Not. Roy. Astron. Soc.}
  \textbf{\bibinfo{volume}{465}}, \bibinfo{pages}{1454} (\bibinfo{year}{2017}),
  \eprint{1606.05338}.

\bibitem[{\citenamefont{Cyburt et~al.}(2016)\citenamefont{Cyburt, Fields,
  Olive, and Yeh}}]{Cyburt:2015mya}
\bibinfo{author}{\bibfnamefont{R.~H.} \bibnamefont{Cyburt}},
  \bibinfo{author}{\bibfnamefont{B.~D.} \bibnamefont{Fields}},
  \bibinfo{author}{\bibfnamefont{K.~A.} \bibnamefont{Olive}}, \bibnamefont{and}
  \bibinfo{author}{\bibfnamefont{T.-H.} \bibnamefont{Yeh}},
  \bibinfo{journal}{Rev. Mod. Phys.} \textbf{\bibinfo{volume}{88}},
  \bibinfo{pages}{015004} (\bibinfo{year}{2016}), \eprint{1505.01076}.

\bibitem[{\citenamefont{de~Salas and Pastor}(2016)}]{deSalas:2016ztq}
\bibinfo{author}{\bibfnamefont{P.~F.} \bibnamefont{de~Salas}} \bibnamefont{and}
  \bibinfo{author}{\bibfnamefont{S.}~\bibnamefont{Pastor}},
  \bibinfo{journal}{JCAP} \textbf{\bibinfo{volume}{1607}}, \bibinfo{pages}{051}
  (\bibinfo{year}{2016}), \eprint{1606.06986}.

\bibitem[{\citenamefont{Bashinsky and Seljak}(2004)}]{Bashinsky:2003tk}
\bibinfo{author}{\bibfnamefont{S.}~\bibnamefont{Bashinsky}} \bibnamefont{and}
  \bibinfo{author}{\bibfnamefont{U.}~\bibnamefont{Seljak}},
  \bibinfo{journal}{Phys. Rev.} \textbf{\bibinfo{volume}{D69}},
  \bibinfo{pages}{083002} (\bibinfo{year}{2004}), \eprint{astro-ph/0310198}.

\bibitem[{\citenamefont{Baumann et~al.}(2016)\citenamefont{Baumann, Green,
  Meyers, and Wallisch}}]{Baumann:2015rya}
\bibinfo{author}{\bibfnamefont{D.}~\bibnamefont{Baumann}},
  \bibinfo{author}{\bibfnamefont{D.}~\bibnamefont{Green}},
  \bibinfo{author}{\bibfnamefont{J.}~\bibnamefont{Meyers}}, \bibnamefont{and}
  \bibinfo{author}{\bibfnamefont{B.}~\bibnamefont{Wallisch}},
  \bibinfo{journal}{JCAP} \textbf{\bibinfo{volume}{1601}}, \bibinfo{pages}{007}
  (\bibinfo{year}{2016}), \eprint{1508.06342}.

\bibitem[{\citenamefont{Follin et~al.}(2015)\citenamefont{Follin, Knox, Millea,
  and Pan}}]{Follin:2015hya}
\bibinfo{author}{\bibfnamefont{B.}~\bibnamefont{Follin}},
  \bibinfo{author}{\bibfnamefont{L.}~\bibnamefont{Knox}},
  \bibinfo{author}{\bibfnamefont{M.}~\bibnamefont{Millea}}, \bibnamefont{and}
  \bibinfo{author}{\bibfnamefont{Z.}~\bibnamefont{Pan}},
  \bibinfo{journal}{Phys. Rev. Lett.} \textbf{\bibinfo{volume}{115}},
  \bibinfo{pages}{091301} (\bibinfo{year}{2015}), \eprint{1503.07863}.

\bibitem[{\citenamefont{Aver et~al.}(2015)\citenamefont{Aver, Olive, and
  Skillman}}]{Aver:2015iza}
\bibinfo{author}{\bibfnamefont{E.}~\bibnamefont{Aver}},
  \bibinfo{author}{\bibfnamefont{K.~A.} \bibnamefont{Olive}}, \bibnamefont{and}
  \bibinfo{author}{\bibfnamefont{E.~D.} \bibnamefont{Skillman}},
  \bibinfo{journal}{JCAP} \textbf{\bibinfo{volume}{1507}}, \bibinfo{pages}{011}
  (\bibinfo{year}{2015}), \eprint{1503.08146}.

\bibitem[{\citenamefont{Tanabashi et~al.}(2018)}]{Tanabashi:2018oca}
\bibinfo{author}{\bibfnamefont{M.}~\bibnamefont{Tanabashi}}
  \bibnamefont{et~al.} (\bibinfo{collaboration}{Particle Data Group}),
  \bibinfo{journal}{Phys. Rev.} \textbf{\bibinfo{volume}{D98}},
  \bibinfo{pages}{030001} (\bibinfo{year}{2018}).

\bibitem[{\citenamefont{Bean et~al.}(2001)\citenamefont{Bean, Hansen, and
  Melchiorri}}]{Bean:2001wt}
\bibinfo{author}{\bibfnamefont{R.}~\bibnamefont{Bean}},
  \bibinfo{author}{\bibfnamefont{S.~H.} \bibnamefont{Hansen}},
  \bibnamefont{and}
  \bibinfo{author}{\bibfnamefont{A.}~\bibnamefont{Melchiorri}},
  \bibinfo{journal}{Phys. Rev.} \textbf{\bibinfo{volume}{D64}},
  \bibinfo{pages}{103508} (\bibinfo{year}{2001}), \eprint{astro-ph/0104162}.

\bibitem[{\citenamefont{Izotov et~al.}(2014)\citenamefont{Izotov, Thuan, and
  Guseva}}]{Izotov:2014fga}
\bibinfo{author}{\bibfnamefont{Y.~I.} \bibnamefont{Izotov}},
  \bibinfo{author}{\bibfnamefont{T.~X.} \bibnamefont{Thuan}}, \bibnamefont{and}
  \bibinfo{author}{\bibfnamefont{N.~G.} \bibnamefont{Guseva}},
  \bibinfo{journal}{Mon. Not. Roy. Astron. Soc.}
  \textbf{\bibinfo{volume}{445}}, \bibinfo{pages}{778} (\bibinfo{year}{2014}),
  \eprint{1408.6953}.

\bibitem[{\citenamefont{Coc et~al.}(2006)\citenamefont{Coc, Olive, Uzan, and
  Vangioni}}]{Coc:2006rt}
\bibinfo{author}{\bibfnamefont{A.}~\bibnamefont{Coc}},
  \bibinfo{author}{\bibfnamefont{K.~A.} \bibnamefont{Olive}},
  \bibinfo{author}{\bibfnamefont{J.-P.} \bibnamefont{Uzan}}, \bibnamefont{and}
  \bibinfo{author}{\bibfnamefont{E.}~\bibnamefont{Vangioni}},
  \bibinfo{journal}{Phys. Rev.} \textbf{\bibinfo{volume}{D73}},
  \bibinfo{pages}{083525} (\bibinfo{year}{2006}), \eprint{astro-ph/0601299}.

\bibitem[{\citenamefont{Bambi et~al.}(2005)\citenamefont{Bambi, Giannotti, and
  Villante}}]{Bambi:2005fi}
\bibinfo{author}{\bibfnamefont{C.}~\bibnamefont{Bambi}},
  \bibinfo{author}{\bibfnamefont{M.}~\bibnamefont{Giannotti}},
  \bibnamefont{and} \bibinfo{author}{\bibfnamefont{F.~L.}
  \bibnamefont{Villante}}, \bibinfo{journal}{Phys. Rev.}
  \textbf{\bibinfo{volume}{D71}}, \bibinfo{pages}{123524}
  (\bibinfo{year}{2005}), \eprint{astro-ph/0503502}.

\bibitem[{\citenamefont{Erickcek et~al.}(2014)\citenamefont{Erickcek, Barnaby,
  Burrage, and Huang}}]{Erickcek:2013dea}
\bibinfo{author}{\bibfnamefont{A.~L.} \bibnamefont{Erickcek}},
  \bibinfo{author}{\bibfnamefont{N.}~\bibnamefont{Barnaby}},
  \bibinfo{author}{\bibfnamefont{C.}~\bibnamefont{Burrage}}, \bibnamefont{and}
  \bibinfo{author}{\bibfnamefont{Z.}~\bibnamefont{Huang}},
  \bibinfo{journal}{Phys. Rev.} \textbf{\bibinfo{volume}{D89}},
  \bibinfo{pages}{084074} (\bibinfo{year}{2014}), \eprint{1310.5149}.

\bibitem[{\citenamefont{Iocco et~al.}(2009)\citenamefont{Iocco, Mangano, Miele,
  Pisanti, and Serpico}}]{Iocco:2008va}
\bibinfo{author}{\bibfnamefont{F.}~\bibnamefont{Iocco}},
  \bibinfo{author}{\bibfnamefont{G.}~\bibnamefont{Mangano}},
  \bibinfo{author}{\bibfnamefont{G.}~\bibnamefont{Miele}},
  \bibinfo{author}{\bibfnamefont{O.}~\bibnamefont{Pisanti}}, \bibnamefont{and}
  \bibinfo{author}{\bibfnamefont{P.~D.} \bibnamefont{Serpico}},
  \bibinfo{journal}{Phys. Rept.} \textbf{\bibinfo{volume}{472}},
  \bibinfo{pages}{1} (\bibinfo{year}{2009}), \eprint{0809.0631}.

\bibitem[{\citenamefont{Dvali et~al.}(2003)\citenamefont{Dvali, Gruzinov, and
  Zaldarriaga}}]{Dvali:2002vf}
\bibinfo{author}{\bibfnamefont{G.}~\bibnamefont{Dvali}},
  \bibinfo{author}{\bibfnamefont{A.}~\bibnamefont{Gruzinov}}, \bibnamefont{and}
  \bibinfo{author}{\bibfnamefont{M.}~\bibnamefont{Zaldarriaga}},
  \bibinfo{journal}{Phys. Rev.} \textbf{\bibinfo{volume}{D68}},
  \bibinfo{pages}{024012} (\bibinfo{year}{2003}), \eprint{hep-ph/0212069}.

\bibitem[{\citenamefont{Brax et~al.}(2011)\citenamefont{Brax, Burrage, and
  Davis}}]{Brax:2011sv}
\bibinfo{author}{\bibfnamefont{P.}~\bibnamefont{Brax}},
  \bibinfo{author}{\bibfnamefont{C.}~\bibnamefont{Burrage}}, \bibnamefont{and}
  \bibinfo{author}{\bibfnamefont{A.-C.} \bibnamefont{Davis}},
  \bibinfo{journal}{JCAP} \textbf{\bibinfo{volume}{1109}}, \bibinfo{pages}{020}
  (\bibinfo{year}{2011}), \eprint{1106.1573}.

\bibitem[{\citenamefont{Clifton et~al.}(2012)\citenamefont{Clifton, Ferreira,
  Padilla, and Skordis}}]{Clifton:2011jh}
\bibinfo{author}{\bibfnamefont{T.}~\bibnamefont{Clifton}},
  \bibinfo{author}{\bibfnamefont{P.~G.} \bibnamefont{Ferreira}},
  \bibinfo{author}{\bibfnamefont{A.}~\bibnamefont{Padilla}}, \bibnamefont{and}
  \bibinfo{author}{\bibfnamefont{C.}~\bibnamefont{Skordis}},
  \bibinfo{journal}{Phys. Rept.} \textbf{\bibinfo{volume}{513}},
  \bibinfo{pages}{1} (\bibinfo{year}{2012}), \eprint{1106.2476}.

\bibitem[{\citenamefont{Belgacem
  et~al.}(2019{\natexlab{a}})\citenamefont{Belgacem, Finke, Frassino, and
  Maggiore}}]{Belgacem:2018wtb}
\bibinfo{author}{\bibfnamefont{E.}~\bibnamefont{Belgacem}},
  \bibinfo{author}{\bibfnamefont{A.}~\bibnamefont{Finke}},
  \bibinfo{author}{\bibfnamefont{A.}~\bibnamefont{Frassino}}, \bibnamefont{and}
  \bibinfo{author}{\bibfnamefont{M.}~\bibnamefont{Maggiore}},
  \bibinfo{journal}{JCAP} \textbf{\bibinfo{volume}{1902}}, \bibinfo{pages}{035}
  (\bibinfo{year}{2019}{\natexlab{a}}), \eprint{1812.11181}.

\bibitem[{\citenamefont{Afshordi et~al.}(2014)\citenamefont{Afshordi,
  Fontanini, and Guariento}}]{Afshordi:2014qaa}
\bibinfo{author}{\bibfnamefont{N.}~\bibnamefont{Afshordi}},
  \bibinfo{author}{\bibfnamefont{M.}~\bibnamefont{Fontanini}},
  \bibnamefont{and} \bibinfo{author}{\bibfnamefont{D.~C.}
  \bibnamefont{Guariento}}, \bibinfo{journal}{Phys. Rev.}
  \textbf{\bibinfo{volume}{D90}}, \bibinfo{pages}{084012}
  (\bibinfo{year}{2014}), \eprint{1408.5538}.

\bibitem[{\citenamefont{Brito et~al.}(2014)\citenamefont{Brito, Terrana,
  Johnson, and Cardoso}}]{Brito:2014ifa}
\bibinfo{author}{\bibfnamefont{R.}~\bibnamefont{Brito}},
  \bibinfo{author}{\bibfnamefont{A.}~\bibnamefont{Terrana}},
  \bibinfo{author}{\bibfnamefont{M.}~\bibnamefont{Johnson}}, \bibnamefont{and}
  \bibinfo{author}{\bibfnamefont{V.}~\bibnamefont{Cardoso}},
  \bibinfo{journal}{Phys. Rev.} \textbf{\bibinfo{volume}{D90}},
  \bibinfo{pages}{124035} (\bibinfo{year}{2014}), \eprint{1409.0886}.

\bibitem[{\citenamefont{Lagos and Zhu}(2020)}]{Lagos:2020mzy}
\bibinfo{author}{\bibfnamefont{M.}~\bibnamefont{Lagos}} \bibnamefont{and}
  \bibinfo{author}{\bibfnamefont{H.}~\bibnamefont{Zhu}} (\bibinfo{year}{2020}),
  \eprint{2003.01038}.

\bibitem[{\citenamefont{Babichev et~al.}(2011)\citenamefont{Babichev, Deffayet,
  and Esposito-Farese}}]{Babichev:2011iz}
\bibinfo{author}{\bibfnamefont{E.}~\bibnamefont{Babichev}},
  \bibinfo{author}{\bibfnamefont{C.}~\bibnamefont{Deffayet}}, \bibnamefont{and}
  \bibinfo{author}{\bibfnamefont{G.}~\bibnamefont{Esposito-Farese}},
  \bibinfo{journal}{Phys. Rev. Lett.} \textbf{\bibinfo{volume}{107}},
  \bibinfo{pages}{251102} (\bibinfo{year}{2011}), \eprint{1107.1569}.

\bibitem[{\citenamefont{Burrage and Dombrowski}(2020)}]{Burrage:2020jkj}
\bibinfo{author}{\bibfnamefont{C.}~\bibnamefont{Burrage}} \bibnamefont{and}
  \bibinfo{author}{\bibfnamefont{J.}~\bibnamefont{Dombrowski}}
  (\bibinfo{year}{2020}), \eprint{2004.14260}.

\bibitem[{\citenamefont{Amendola et~al.}(1999)\citenamefont{Amendola,
  Corasaniti, and Occhionero}}]{Amendola:1999vu}
\bibinfo{author}{\bibfnamefont{L.}~\bibnamefont{Amendola}},
  \bibinfo{author}{\bibfnamefont{P.~S.} \bibnamefont{Corasaniti}},
  \bibnamefont{and}
  \bibinfo{author}{\bibfnamefont{F.}~\bibnamefont{Occhionero}}
  (\bibinfo{year}{1999}), \eprint{astro-ph/9907222}.

\bibitem[{\citenamefont{Garcia-Berro et~al.}(1999)\citenamefont{Garcia-Berro,
  Gaztanaga, Isern, Benvenuto, and Althaus}}]{GarciaBerro:1999bq}
\bibinfo{author}{\bibfnamefont{E.}~\bibnamefont{Garcia-Berro}},
  \bibinfo{author}{\bibfnamefont{E.}~\bibnamefont{Gaztanaga}},
  \bibinfo{author}{\bibfnamefont{J.}~\bibnamefont{Isern}},
  \bibinfo{author}{\bibfnamefont{O.}~\bibnamefont{Benvenuto}},
  \bibnamefont{and} \bibinfo{author}{\bibfnamefont{L.}~\bibnamefont{Althaus}}
  (\bibinfo{year}{1999}), \eprint{astro-ph/9907440}.

\bibitem[{\citenamefont{Riazuelo and Uzan}(2002)}]{Riazuelo:2001mg}
\bibinfo{author}{\bibfnamefont{A.}~\bibnamefont{Riazuelo}} \bibnamefont{and}
  \bibinfo{author}{\bibfnamefont{J.-P.} \bibnamefont{Uzan}},
  \bibinfo{journal}{Phys. Rev.} \textbf{\bibinfo{volume}{D66}},
  \bibinfo{pages}{023525} (\bibinfo{year}{2002}), \eprint{astro-ph/0107386}.

\bibitem[{\citenamefont{Wright and Li}(2018)}]{Wright:2017rsu}
\bibinfo{author}{\bibfnamefont{B.~S.} \bibnamefont{Wright}} \bibnamefont{and}
  \bibinfo{author}{\bibfnamefont{B.}~\bibnamefont{Li}}, \bibinfo{journal}{Phys.
  Rev.} \textbf{\bibinfo{volume}{D97}}, \bibinfo{pages}{083505}
  (\bibinfo{year}{2018}), \eprint{1710.07018}.

\bibitem[{\citenamefont{Sakstein et~al.}(2019)\citenamefont{Sakstein, Desmond,
  and Jain}}]{Sakstein:2019qgn}
\bibinfo{author}{\bibfnamefont{J.}~\bibnamefont{Sakstein}},
  \bibinfo{author}{\bibfnamefont{H.}~\bibnamefont{Desmond}}, \bibnamefont{and}
  \bibinfo{author}{\bibfnamefont{B.}~\bibnamefont{Jain}},
  \bibinfo{journal}{Phys. Rev.} \textbf{\bibinfo{volume}{D100}},
  \bibinfo{pages}{104035} (\bibinfo{year}{2019}), \eprint{1907.03775}.

\bibitem[{\citenamefont{Hanımeli et~al.}(2019)\citenamefont{Hanımeli, Lamine,
  Tutusaus, and Blanchard}}]{Hanimeli:2019wrt}
\bibinfo{author}{\bibfnamefont{E.~T.} \bibnamefont{Hanımeli}},
  \bibinfo{author}{\bibfnamefont{B.}~\bibnamefont{Lamine}},
  \bibinfo{author}{\bibfnamefont{I.}~\bibnamefont{Tutusaus}}, \bibnamefont{and}
  \bibinfo{author}{\bibfnamefont{A.}~\bibnamefont{Blanchard}}
  (\bibinfo{year}{2019}), \eprint{1910.08325}.

\bibitem[{\citenamefont{Linden et~al.}(2009)\citenamefont{Linden, Virey, and
  Tilquin}}]{Linden:2009vh}
\bibinfo{author}{\bibfnamefont{S.}~\bibnamefont{Linden}},
  \bibinfo{author}{\bibfnamefont{J.-M.} \bibnamefont{Virey}}, \bibnamefont{and}
  \bibinfo{author}{\bibfnamefont{A.}~\bibnamefont{Tilquin}},
  \bibinfo{journal}{Astron. Astrophys.} \textbf{\bibinfo{volume}{50}},
  \bibinfo{pages}{1095} (\bibinfo{year}{2009}), \eprint{0907.4495}.

\bibitem[{\citenamefont{Zumalacarregui and
  Seljak}(2018)}]{Zumalacarregui:2017qqd}
\bibinfo{author}{\bibfnamefont{M.}~\bibnamefont{Zumalacarregui}}
  \bibnamefont{and} \bibinfo{author}{\bibfnamefont{U.}~\bibnamefont{Seljak}},
  \bibinfo{journal}{Phys. Rev. Lett.} \textbf{\bibinfo{volume}{121}},
  \bibinfo{pages}{141101} (\bibinfo{year}{2018}), \eprint{1712.02240}.

\bibitem[{\citenamefont{Liao et~al.}(2020)\citenamefont{Liao, Shafieloo,
  Keeley, and Linder}}]{Liao:2020zko}
\bibinfo{author}{\bibfnamefont{K.}~\bibnamefont{Liao}},
  \bibinfo{author}{\bibfnamefont{A.}~\bibnamefont{Shafieloo}},
  \bibinfo{author}{\bibfnamefont{R.~E.} \bibnamefont{Keeley}},
  \bibnamefont{and} \bibinfo{author}{\bibfnamefont{E.~V.} \bibnamefont{Linder}}
  (\bibinfo{year}{2020}), \eprint{2002.10605}.

\bibitem[{\citenamefont{Hofmann and Müller}(2018)}]{Hofmann:2018myc}
\bibinfo{author}{\bibfnamefont{F.}~\bibnamefont{Hofmann}} \bibnamefont{and}
  \bibinfo{author}{\bibfnamefont{J.}~\bibnamefont{Müller}},
  \bibinfo{journal}{Class. Quant. Grav.} \textbf{\bibinfo{volume}{35}},
  \bibinfo{pages}{035015} (\bibinfo{year}{2018}).

\bibitem[{\citenamefont{Noller}(2020)}]{Noller:2020afd}
\bibinfo{author}{\bibfnamefont{J.}~\bibnamefont{Noller}}
  (\bibinfo{year}{2020}), \eprint{2001.05469}.

\bibitem[{\citenamefont{Lagos et~al.}(2019)\citenamefont{Lagos, Fishbach,
  Landry, and Holz}}]{Lagos:2019kds}
\bibinfo{author}{\bibfnamefont{M.}~\bibnamefont{Lagos}},
  \bibinfo{author}{\bibfnamefont{M.}~\bibnamefont{Fishbach}},
  \bibinfo{author}{\bibfnamefont{P.}~\bibnamefont{Landry}}, \bibnamefont{and}
  \bibinfo{author}{\bibfnamefont{D.~E.} \bibnamefont{Holz}},
  \bibinfo{journal}{Phys. Rev.} \textbf{\bibinfo{volume}{D99}},
  \bibinfo{pages}{083504} (\bibinfo{year}{2019}), \eprint{1901.03321}.

\bibitem[{\citenamefont{Belgacem
  et~al.}(2019{\natexlab{b}})}]{Belgacem:2019pkk}
\bibinfo{author}{\bibfnamefont{E.}~\bibnamefont{Belgacem}} \bibnamefont{et~al.}
  (\bibinfo{collaboration}{LISA Cosmology Working Group}),
  \bibinfo{journal}{JCAP} \textbf{\bibinfo{volume}{1907}}, \bibinfo{pages}{024}
  (\bibinfo{year}{2019}{\natexlab{b}}), \eprint{1906.01593}.

\bibitem[{\citenamefont{Maggiore et~al.}(2019)}]{Maggiore:2019uih}
\bibinfo{author}{\bibfnamefont{M.}~\bibnamefont{Maggiore}} \bibnamefont{et~al.}
  (\bibinfo{year}{2019}), \eprint{1912.02622}.

\bibitem[{\citenamefont{Dalang and Lombriser}(2019)}]{Dalang:2019fma}
\bibinfo{author}{\bibfnamefont{C.}~\bibnamefont{Dalang}} \bibnamefont{and}
  \bibinfo{author}{\bibfnamefont{L.}~\bibnamefont{Lombriser}},
  \bibinfo{journal}{JCAP} \textbf{\bibinfo{volume}{1910}}, \bibinfo{pages}{013}
  (\bibinfo{year}{2019}), \eprint{1906.12333}.

\bibitem[{\citenamefont{Dalang et~al.}(2019)\citenamefont{Dalang, Fleury, and
  Lombriser}}]{Dalang:2019rke}
\bibinfo{author}{\bibfnamefont{C.}~\bibnamefont{Dalang}},
  \bibinfo{author}{\bibfnamefont{P.}~\bibnamefont{Fleury}}, \bibnamefont{and}
  \bibinfo{author}{\bibfnamefont{L.}~\bibnamefont{Lombriser}}
  (\bibinfo{year}{2019}), \eprint{1912.06117}.

\bibitem[{\citenamefont{Garoffolo et~al.}(2019)\citenamefont{Garoffolo,
  Tasinato, Carbone, Bertacca, and Matarrese}}]{Garoffolo:2019mna}
\bibinfo{author}{\bibfnamefont{A.}~\bibnamefont{Garoffolo}},
  \bibinfo{author}{\bibfnamefont{G.}~\bibnamefont{Tasinato}},
  \bibinfo{author}{\bibfnamefont{C.}~\bibnamefont{Carbone}},
  \bibinfo{author}{\bibfnamefont{D.}~\bibnamefont{Bertacca}}, \bibnamefont{and}
  \bibinfo{author}{\bibfnamefont{S.}~\bibnamefont{Matarrese}}
  (\bibinfo{year}{2019}), \eprint{1912.08093}.

\bibitem[{\citenamefont{Wolf and Lagos}(2020)}]{Wolf:2019hun}
\bibinfo{author}{\bibfnamefont{W.~J.} \bibnamefont{Wolf}} \bibnamefont{and}
  \bibinfo{author}{\bibfnamefont{M.}~\bibnamefont{Lagos}},
  \bibinfo{journal}{Phys. Rev. Lett.} \textbf{\bibinfo{volume}{124}},
  \bibinfo{pages}{061101} (\bibinfo{year}{2020}), \eprint{1910.10580}.

\bibitem[{\citenamefont{Di~Valentino et~al.}(2020)\citenamefont{Di~Valentino,
  Melchiorri, Mena, and Vagnozzi}}]{DiValentino:2019jae}
\bibinfo{author}{\bibfnamefont{E.}~\bibnamefont{Di~Valentino}},
  \bibinfo{author}{\bibfnamefont{A.}~\bibnamefont{Melchiorri}},
  \bibinfo{author}{\bibfnamefont{O.}~\bibnamefont{Mena}}, \bibnamefont{and}
  \bibinfo{author}{\bibfnamefont{S.}~\bibnamefont{Vagnozzi}},
  \bibinfo{journal}{Phys. Rev.} \textbf{\bibinfo{volume}{D101}},
  \bibinfo{pages}{063502} (\bibinfo{year}{2020}), \eprint{1910.09853}.

\bibitem[{\citenamefont{Hill et~al.}(2020)\citenamefont{Hill, McDonough,
  Toomey, and Alexander}}]{Hill:2020osr}
\bibinfo{author}{\bibfnamefont{J.~C.} \bibnamefont{Hill}},
  \bibinfo{author}{\bibfnamefont{E.}~\bibnamefont{McDonough}},
  \bibinfo{author}{\bibfnamefont{M.~W.} \bibnamefont{Toomey}},
  \bibnamefont{and} \bibinfo{author}{\bibfnamefont{S.}~\bibnamefont{Alexander}}
  (\bibinfo{year}{2020}), \eprint{2003.07355}.

\bibitem[{\citenamefont{Frusciante et~al.}(2019)\citenamefont{Frusciante,
  Peirone, Atayde, and De~Felice}}]{Frusciante:2019puu}
\bibinfo{author}{\bibfnamefont{N.}~\bibnamefont{Frusciante}},
  \bibinfo{author}{\bibfnamefont{S.}~\bibnamefont{Peirone}},
  \bibinfo{author}{\bibfnamefont{L.}~\bibnamefont{Atayde}}, \bibnamefont{and}
  \bibinfo{author}{\bibfnamefont{A.}~\bibnamefont{De~Felice}}
  (\bibinfo{year}{2019}), \eprint{1912.07586}.

\bibitem[{\citenamefont{Hinterbichler and Khoury}(2010)}]{Hinterbichler:2010es}
\bibinfo{author}{\bibfnamefont{K.}~\bibnamefont{Hinterbichler}}
  \bibnamefont{and} \bibinfo{author}{\bibfnamefont{J.}~\bibnamefont{Khoury}},
  \bibinfo{journal}{Phys. Rev. Lett.} \textbf{\bibinfo{volume}{104}},
  \bibinfo{pages}{231301} (\bibinfo{year}{2010}), \eprint{1001.4525}.

\bibitem[{\citenamefont{Erickcek et~al.}(2013)\citenamefont{Erickcek, Barnaby,
  Burrage, and Huang}}]{Erickcek:2013oma}
\bibinfo{author}{\bibfnamefont{A.~L.} \bibnamefont{Erickcek}},
  \bibinfo{author}{\bibfnamefont{N.}~\bibnamefont{Barnaby}},
  \bibinfo{author}{\bibfnamefont{C.}~\bibnamefont{Burrage}}, \bibnamefont{and}
  \bibinfo{author}{\bibfnamefont{Z.}~\bibnamefont{Huang}},
  \bibinfo{journal}{Phys. Rev. Lett.} \textbf{\bibinfo{volume}{110}},
  \bibinfo{pages}{171101} (\bibinfo{year}{2013}), \eprint{1304.0009}.

\bibitem[{\citenamefont{Padilla et~al.}(2016)\citenamefont{Padilla, Platts,
  Stefanyszyn, Walters, Weltman, and Wilson}}]{Padilla:2015wlv}
\bibinfo{author}{\bibfnamefont{A.}~\bibnamefont{Padilla}},
  \bibinfo{author}{\bibfnamefont{E.}~\bibnamefont{Platts}},
  \bibinfo{author}{\bibfnamefont{D.}~\bibnamefont{Stefanyszyn}},
  \bibinfo{author}{\bibfnamefont{A.}~\bibnamefont{Walters}},
  \bibinfo{author}{\bibfnamefont{A.}~\bibnamefont{Weltman}}, \bibnamefont{and}
  \bibinfo{author}{\bibfnamefont{T.}~\bibnamefont{Wilson}},
  \bibinfo{journal}{JCAP} \textbf{\bibinfo{volume}{1603}}, \bibinfo{pages}{058}
  (\bibinfo{year}{2016}), \eprint{1511.05761}.

\bibitem[{\citenamefont{Creminelli et~al.}(2010)\citenamefont{Creminelli,
  Nicolis, and Trincherini}}]{Creminelli:2010ba}
\bibinfo{author}{\bibfnamefont{P.}~\bibnamefont{Creminelli}},
  \bibinfo{author}{\bibfnamefont{A.}~\bibnamefont{Nicolis}}, \bibnamefont{and}
  \bibinfo{author}{\bibfnamefont{E.}~\bibnamefont{Trincherini}},
  \bibinfo{journal}{JCAP} \textbf{\bibinfo{volume}{1011}}, \bibinfo{pages}{021}
  (\bibinfo{year}{2010}), \eprint{1007.0027}.

\end{thebibliography}

\end{document}